\documentclass[aps,prx,twocolumn,showpacs,floatfix,nobibnotes,nofootinbib,superscriptaddress,amsmath,amssymb,longbibliography]{revtex4-1}
\usepackage[english]{babel}
\usepackage[colorlinks,urlcolor={blue}]{hyperref}
\usepackage{eurosym}
\usepackage[usenames]{color}
\usepackage{graphicx}
\usepackage{amsmath}
\usepackage{amsfonts}
\usepackage{amssymb}
\usepackage{mathrsfs}
\usepackage{bm}
\usepackage{verbatim}
\usepackage{ulem}
\usepackage{siunitx}
\usepackage{upgreek}
\usepackage[textsize=small]{todonotes}

\newcommand{\beq}{\begin{equation}}
\newcommand{\eeq}{\end{equation}}
\newcommand{\bea}{\begin{eqnarray}}
\newcommand{\eea}{\end{eqnarray}}

\begin{document}

\title{Multi-photon stimulated grasers assisted by laser-plasma interactions}

\author{C.-J.~Yang}
\affiliation{ELI-NP, ``Horia Hulubei" National Institute for Physics and Nuclear Engineering, 30 Reactorului Street, RO-077125, Bucharest-Măgurele, Romania}
\email{chieh.jen@eli-np.ro (corresponding author)}

\author{K.~M. Spohr}
\affiliation{ELI-NP, ``Horia Hulubei" National Institute for Physics and Nuclear Engineering, 30 Reactorului Street, RO-077125, Bucharest-Măgurele, Romania}
\affiliation{School of Computing, Engineering and Physical Sciences, University of the West of Scotland, High Street, PA1 2BE, Paisley, Scotland}
\email{klaus.spohr@eli-np.ro (corresponding author)}
\author{D. Doria}
\affiliation{ELI-NP, ``Horia Hulubei" National Institute for Physics and Nuclear Engineering, 30 Reactorului Street, RO-077125, Bucharest-Măgurele, Romania}
\date{\today }

\begin{abstract}
We investigate theoretically the possibility of achieving the stimulated amplification of $\gamma $-rays. Herein, our approach circumvents the so-called ``graser dilemma" through a non-linear, multi-photon mechanism. Our work foresees the combination of a high-intensity $\gamma-$flash generated by the interaction of a high-intensity laser pulse with plasma and intensive photons supplied by an additional laser. We show that multi-photon stimulated emission processes can have a larger effective cross-section compared to a one-photon process. The bandwidth of the supplied photons can also be tuned to curtail linewidth broadening. Naturally, M\"{o}{\ss}bauer transitions can be chosen to apply the scheme in the first instance. Furthermore, we derive that even multi-photon stimulated emission in the form of an anti-Stokes type could allow our scheme to be applied to non-M\"{o}{\ss}bauer nuclei, provided that the supplied photon energy can be tuned to compensate for the recoil and other broadening induced losses. The graser development can be spearheaded using multi-PW class high-power laser systems such as the \SI{10}{PW} installation at Extreme Light Infrastructure - Nuclear Physics (ELI-NP) in Romania.
\end{abstract}

\maketitle

\vspace{10mm}

\section{Introduction}
In the dawn of the 21\textsuperscript{st} century, the pursuit of a coherent $\gamma-$ray source, known as a graser, has proven elusive despite decades of relentless efforts summarized e.g. in~\cite{RevModPhys.53.687,RevModPhys.69.1085}. The technical realization of this endeavor has remained a formidable challenge, casting a shadow over the gamut of potential applications it could unlock for advancing fundamental science and technology. Theoretical and technical obstacles, seemingly insurmountable, have impeded progress in this quest.
In this challenging landscape, a beacon of optimism emerges from recent breakthroughs in multi-petawatt high-power laser systems (HPLS)~\footnote{See https://www.icuil.org/ for a comprehesive overview of current facilities.}, exemplified by the \SI{10}{PW} ELI-NP installation~\cite{Tanaka2020,Doria2020,Radier2022}. These cutting-edge systems not only represent a milestone in laser technology but also open exciting possibilities in the burgeoning realm of nuclear photonics. Building on insights gleaned from the work of peers ~\cite{PhysRevLett.127.052501,PhysRevC.100.064610,PhysRevC.101.044304,von2020theory,PhysRevC.106.024606,PhysRevC.105.054001,PhysRevC.99.044610,PhysRevC.100.041601,Li2021,PhysRevC.104.044614,Lv2022,cheng2023laserassisted,PhysRevA.108.L021502,Spohr:2023koj,Wu:2023vti}, our inspiration is further fueled by a previous study~~\cite{our_iso}, where we demonstrated that the sheer intensity of $\gamma-$photons, generated through laser-matter interactions, could offer a transformative avenue for overcoming the significant angular momentum gaps inherent in many nuclear systems. Extending this paradigm, our current work delves deeper into the realm of multi-photon absorption and emission. We will exploit non-linear effects by combining a high-brilliance $\gamma-$flash from laser-matter interaction and multiple high-intensity photon sources. This strategy provides a general avenue to circumvent the obstacles preventing the conception of nuclear $\gamma$-ray laser systems so far. Our approach is new as in previous works depicting graser concepts in the past decades, HPLS systems in the PW-regime that can provide flashes of intensive $\gamma$-rays coincident with intensive optical beams to initiate multi-photon effects could not have been foreseen to be in existence. This includes the aforementioned seminal works by Baldwin \textit{et al.}  ~\cite{RevModPhys.53.687,RevModPhys.69.1085}.\\

The structure of our work is as follows. First, we discuss pathways to overcome the graser dilemma in Sec.~\ref{sec2}. This comprises a review of the low cross-sections for stimulated emission for isomers, the broadening of the narrow absorption line breadth, photon removal effects, and the recoil energy losses that prevent re-absorption. Then, in Sec.~\ref{sec3}, we investigate several possibilities for designing the first graser. In Sec.~\ref{sec4}, we discuss how to control the multi-photon process to achieve different objectives. Finally, our findings are summarized in Sec.~\ref{sec5}.

\section{Circumventing the graser dilemma}
\label{sec2}
Einstein's theory utilizing detailed balancing dictates the following relation between the spontaneous emission coefficient $\mathcal{A}$ and stimulated emission coefficient $\mathcal{B}$~\cite{1916DPhyG..18..318E}:
\begin{equation}
\mathcal{A}=\mathcal{B}\,\frac{8\pi h \upsilon^3}{c^3} ,\label{eq1}
\end{equation}
where $h$ is the Planck constant, $\upsilon$ is the frequency of the photon, and $c$ is the speed of light.
Thus, the probability of spontaneous emission grows quickly with the increase of $\upsilon$, leading to a shorter window of time for the system to accumulate a sufficient amount of lasing states, i.e. the excited state where the $\gamma$-ray is stimulated from. Such a build-up is, however, necessary for grasing to happen. For nuclei, long-lived isomer states exist, and the spontaneous emission coefficient $\mathcal{A}$ can be made small by special spin-isospin arrangements. Nevertheless, Eq.~(\ref{eq1}) indicates that the stimulated emission coefficient $\mathcal{B}$ must be even more suppressed, which prohibits the $\gamma$-ray from being amplified. Realistic calculations indicate that one encounters the graser dilemma: the incompatibility between available pumping power against spontaneous decay and the minimum required stimulated emission cross-section for grasing. 

Consequently, proposals to overcome the above dilemma can be categorized into two main categories: Firstly, those that adopt alternative pumping that does not directly involve photons to avoid Eq.~(\ref{eq1}), which include nuclear reactions~\cite{baldwin1962,Mughabghab1973,1444442,r1963a,g1973a,l1973b,1444212} or charged particles pumping~\cite{Byrne1974,Walker1999,karamian2007prospects,PhysRevLett.96.042505,rex2006first,Feng2022,Feng2023} and secondly those that utilize the non-linear effects with intense optical laser pumping~\cite{Winterberg1986,PhysRevLett.42.1397,PhysRevC.20.1942,PhysRevC.23.50,PhysRevC.23.1007,Collins1982}. In this work, we follow and extend the spirit of the latter and utilize the non-linearity of multi-photon processes. In detail, a temporal coincidence of $\gamma-$flash with a secondary beam of very dense photons.
This circumvents the dilemma by enlarging the effective $\gamma$-ray absorption cross-section and compensating any homogeneous shift due to recoil or other mechanisms.

\subsection{The smallness of stimulated emission cross-sections: Problems and solutions}
\label{sec2a}
From Eq.~(\ref{eq1}) the asymptotic stimulated emission cross-section is given by~\cite{RevModPhys.53.687},
\begin{equation}
\sigma_\mathrm{asy}=\frac{1}{8\pi\upsilon^2}\frac{\Gamma_{\gamma}}{\Gamma} ,\label{eq2}
\end{equation}
where $\Gamma_{\gamma}$ is the natural width of the excited state with its relationship to the half-life governed solely by the Uncertainty Principle, and $\Gamma$ represents the total width after considering all broadening effects, $\Gamma_{\gamma}\leq\Gamma$. Eq.~(\ref{eq2}) is a ``hard" limit which dictates that any individual photo-absorption or stimulated emission cross-section cannot exceed $\sigma_\mathrm{asy}\approx\frac{10^{-22}}{2\pi E_{\gamma}^2}$ cm$^2$, with $E_{\gamma}$ being the energy of $\gamma$-photons given in MeV. 
However, the ansatz leading to Eqs.~(\ref{eq1}) and (\ref{eq2}), assumes the yields of being \textit{linear} to the radiation density of the incident field. In other words, it is based on the approximation that the incident field has low intensity, so each scattering center will interact with one incoming particle per time. Events concerning two or more particles interacting with one scattering center are no longer small and eventually dominate over single-processes as the intensity of the incoming field grows, which was first investigated in terms of two-photon absorption~\cite{GM}, and later generalized to n-photon process (nPP)~\cite{LAMBROPOULOS197687,Friedrich2017,Delone2000,Grechukhin1963,PhysRevA.9.1762,Friar1974,Friar1975,Kramp1987,PhysRevLett.7.170,Alburger1964,Harihar1970,Nakayama1973,Vanderleeden:1970zn,PhysRevLett.53.1897,Soderstrom:2020iaz,freirefernández2023measurement}.
Note that this additional ingredient is of combinatorial origin. Thus, the enhanced effective cross-section of this ingredient does not violate any physical limit deduced from the two-body (one-to-one) scattering process. In some senses, the original limit is circumvented instead of violated. A similar analog of this combinatorial enhancement is found in nuclear many-body interactions~\cite{Yang:2019hkn,Yang:2021vxa}. 

Meanwhile, it turns out that the ingredient of supplying intense optical photons (as proposed in Refs.~\cite{PhysRevLett.42.1397,PhysRevC.20.1942,PhysRevC.23.50,PhysRevC.23.1007,Collins1982}) alone is not sufficient to overcome the dilemma. Generating sufficient populations of short-lived excited states requires large amounts of $\gamma$-photons plus at least a graser-level pumping power. Since there is no graser to start with, isomers are likely to be the only choice to build the first graser. However, the angular momentum quantum number difference ($\Delta J^{\pi}$) between any isomer state and the final state (the state after the release of $\gamma $ rays) must be large (in general $\gtrsim3$) to allow the existence of isomers. Since a photon has intrinsic $J^{\pi}=1^{-}$, it has a low probability of triggering a transition with multipolarity higher than E1. Therefore, one must resort to multi-photon processes. For the effect to be pronounced, it then requires each virtual state to resonate with a nuclear intermediate state, which is usually separated by at least $\gtrsim \SI{1}{keV}$, making the desired multi-photon transition incompatible with the optical photon energy. Previous proposals~\cite{PhysRevLett.42.1397,PhysRevC.20.1942,PhysRevC.23.50,PhysRevC.23.1007,Collins1982} then rely on spontaneous decays to go over transitions with gaps larger than the optical photon energy. Unfortunately, calculations showed that this process is not efficient enough, even enhanced by optical photons~\cite{BECKER1984441}.  

Recent progress in HPLS could become a game-changer for the above situation, as intensive $\gamma$-photons can be produced.
In Ref.~\cite{our_iso}, we proposed an improved isomer-pumping and depletion scheme combining the ingredient of supplying optical or infrared photons and the $\gamma$-photons generated from HPLS via laser-matter interaction. The transition rate through the multi-photon process (nPP) with favorable intermediate states reads
 \begin{equation}
R_\mathrm{nPP}=\frac{e^{2n} \mathcal{E}^2_1 \cdots \mathcal{E}^2_n}{4^n\hbar^{2n}}|\mathcal{M}^{(n)}|^2 2\pi\delta_t(\omega-\omega_1\cdots-\omega_n), \label{prn}
\end{equation}
where $e$ is the charge, $\omega_\mathrm{i}=E_\mathrm{i}/\hbar$, with $E_\mathrm{i}$ the energy of the
$i^{th}$ photon with amplitude $\mathcal{E}_\mathrm{i}$, $\hbar$
the reduced Planck constant.
The interaction kernel $\mathcal{M}^{(n)}$ sums over eigenstates $\langle m_{2}|\sim\langle m_n|$, i.e.,
\begin{widetext}
\begin{equation}
\mathcal{M}^{(n)}=\sum_{m_2}\cdots\sum_{m_n}\left[\frac{ \langle f|\hat{H}_{n}|m_n\rangle \cdots \langle m_3|\hat{H}_{2}|m_2\rangle\langle m_2|\hat{H}_{1}|i\rangle}{(\omega_1-\omega_{m_{2}i})(\omega_2-\omega_{m_3 m_2})\cdots(\omega_{n-1}-\omega_{m_{n}m_{n-1}})}+(\text{all permutation})\right], \label{mn}
\end{equation}
\end{widetext}
where $\langle \mathrm{f}|$ and $\langle \mathrm{i}|$ are the final and initial state, $\hat{H_\mathrm{i}}$ is the transition operator due to the i$^\mathrm{th}$ photon. $\omega_{m\mathrm{i}m_\mathrm{j}}=(E_\mathrm{i}-E_\mathrm{j})/\hbar$. 
Note that here we have generalized the previous n-photon-absorption (nPA) in Ref.~\cite{our_iso} to nPP with $\omega_\mathrm{i}>0$ ($<0$) indicating absorption (emission) of photons, which then gives upward (downward) transitions across virtual states.  
Under Weisskopf estimates~\cite{Blatt:1952ije}, the resulting effective cross-section (given in [\SI{}{cm^2}]) experienced by the last $\gamma-$photon (labelled here as the lasing photon with energy $E_{\gamma}$) reads
\begin{align}
\sigma^{\mathrm{nPP},|\Delta J|\leq n}_\mathrm{eff}&\approx 10^{-25}\frac{E_{\gamma}}{w_>}\prod_{i=1}^{i=n-1}\left[3\cdot 10^{-24}\mathcal{P}_\mathrm{i} \,\frac{X_\mathrm{i}}{(\Delta E_\mathrm{i})^2}\right] , \label{eff_e1} 
\end{align}
where $X_\mathrm{i}=A^{2/3}$ (with $A$ the total number of nucleons) if the transition carried by the $i^\mathrm{th}$-photon is of E1 type. For M1 and E2 type transitions, $X_\mathrm{i}$ will be suppressed by $\gtrsim25$ and $\gtrsim100$ times concerning the E1 value.  
The superscript $|\Delta J|\leq n$ indicates the largest angular momentum difference that nPP can achieve. $E_{\gamma}$ and $w_>$ must have the same unit, with $w_>$ the larger of the total width of $\langle f|$ or the laser bandwidth. $\mathcal{P}_\mathrm{i}$ is the power density of photons in \SI{}{Wcm^{-2}}. $\Delta E_\mathrm{i}$ (in eV) represents the energy difference between each virtual state and the state it resonates with.

For $\sigma^{\mathrm{nPP},|\Delta J|\leq n}_\mathrm{eff}$ in Eq.~(\ref{eff_e1}) to be appreciable, all adjacent virtual/intermediate states must be separated within an E2 transition. Note that the main consideration in arranging an nPP sequence is to connect states with favorable transition multipolarities. An adjacent pair $\langle m_\mathrm{i}|$ and $\langle m_\mathrm{i+1}|$ need not be the two states closest in energy. 
An ideal scenario is that every virtual state is separated from an intermediate state by the natural width of the intermediate state, i.e., $\Delta E_\mathrm{i}\lesssim \SI{e-3}{eV}$ to \SI{e-4}{eV}. 
Thus, if one could optimize the laser-plasma interaction to maintain $\mathcal{P}_{\gamma}\gtrsim \SI{e17}{Wcm^{-2}}$ per eV interval over $\SI{1}{keV} \lesssim E_{\gamma}\lesssim \SI{5}{MeV}$, nPP will be of the same order of magnitude as 1PP, which has $\sigma_\mathrm{eff}\approx \SI{e-19}{cm^2}$ for $A\approx200$ nuclei. In this case, supplying extra photon beams is unnecessary, at least in retaining a large effective nPP cross-section. Otherwise, providing photons at other energies could compensate for the missing intensity. 
The supplied photons (defined as $\omega$-photons in this work) mostly\footnote{One exception would be that the intermediate states are separated within E2 transitions and energy reachable by the supplied photon.} carry out M1 transitions and merely act as a means of boosting the combinatorial choices together with the $\gamma$-photons. In this instance, half of the denominator would become the energy of the supplied photons instead of the natural width of the intermediate state. An averaged $\Delta E_\mathrm{i}\lesssim10^{-(2+a/2)}$\,eV in Eq.~(\ref{eff_e1}) is expected if the supplied photons have an energy $E_{\omega}\sim10^{-a}$\,eV (with $a\leq4$) and a bandwidth $10^{-b}$\,eV. To retain $\sigma^\mathrm{nPP}_\mathrm{eff}\approx \SI{e-19}{cm^{2}}$, it then demands the supplied photons to have $\mathcal{P_{\omega}}\approx10^{18-a+x}$\,\SI{}{Wcm^{-2}} with $x$ the order of magnitude missing in $\mathcal{P}_{\gamma}$ with respect to $10^{18-a+b}$ \SI{}{Wcm^{-2}}.\footnote{The estimate assumes that at each step the $\gamma$-photon carries out an E1 transition, unless under the presence of the helical $\gamma$ beam~\cite{Taira2017,Maruyama2022,Shen2019,Peele2002,Terhalle2011,PhysRevLett.113.153901,Hemsing2013,Ababekri:2022mob,PhysRevA.98.052130,Zhu2018,Feng2019,PhysRevApplied.14.014094,Zhang2023,Liu2016,Zhang2021,Liu2020,PhysRevLett.106.013001,PhysRevLett.121.074801,Hu2021,Younis2022,PhysRevLett.131.202502,Balabanski2024}.} 
In any case, a continuous $\gamma$ spectrum will automatically enable combinations with supplied photons to maximize the resonance with each intermediate state. The only requirement is their intensity and neither the $\gamma$-photons nor the supplied photons must be coherent.

Particle-in-cell (PIC) simulations~\cite{PhysRevApplied.13.054024,Xue2020,PhysRevE.104.045206,Heppe2022,Shen2024} indicate that a continuous $\gamma-$photon spectrum can be generated by an HPLS. Existing calculations suggest that photons accumulated within \SI{1}{eV} interval anywhere from $E=\SI{1}{keV}$ to \SI{5}{MeV} have an intensity $\mathcal{P}_{\gamma}\approx \SI{e9}{Wcm^{-2}}$ to \SI{e12}{Wcm^{-2}}. Thus, aside from a few special cases, an $\sigma_\mathrm{eff}^{\mathrm{nPP},|\Delta J|\leq n} \approx \SI{e-19}{cm^2}$  can only be retained by increasing the intensity of the supplied photons $\mathcal{P_{\omega}}\gtrsim \SI{e23}{Wcm^{-2}}$, under the condition that both the energy and the bandwidth of the supplied photons $\approx \SI{1}{eV}$. 
Pulses from such HPLS, though available today~\cite{Yoon:19,Yoon:21,Li:18,Yu:18,ur2015eli,Tanaka2020}, trigger the laser-plasma mechanism to destroy solid targets~\cite{Borghesi2006,roth2002energetic,logan2006assessment,Roth2008}. However, it is not necessary to have the nPP cross-section comparable to 1PP for an arbitrarily large $n$, nor does the effective cross-section need to be held at \SI{e-19}{cm^{2}} to enable grasing to occur. 
The range of $|\Delta J|\leq 4$ already covers a wide variety of isomers, and manipulating them with $\sigma_\mathrm{eff}^{\mathrm{4pp},|\Delta J|\leq 4}\gtrsim$$\SI{e-25}{cm^2}$ would be sufficient for grasing to occur, as will be discussed in Sec.~\ref{sec2c}.


\begin{figure}[t]
\includegraphics[width=0.47\textwidth,clip=true]{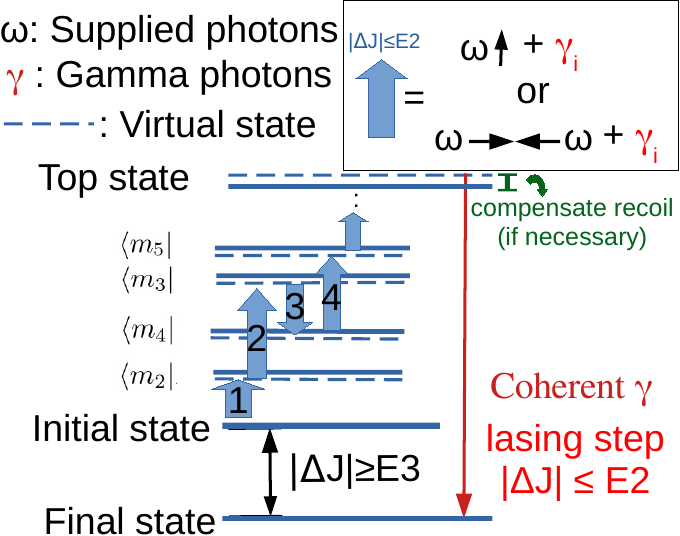}
\caption{Illustration of an nPP graser example. The number inside each arrow labels the order of the process at each step, which can be photo-absorption or stimulated emission. Two possible arrangements of the supplied photons are listed: One in which the enhancement is carried out by one photon beam and the second one in which two beams with equal but opposite momentum are used. Optical or infrared photons with tunable energy $E_{\omega}$ are supplied by intensive lasers with a duration that covers the entire lasing process. }
\label{fig1}
\end{figure}

Fig.~\ref{fig1} illustrates our general scheme, with the two scenarios to be discussed further in Section~\ref{sec3}. The entire process is a multi-photon process ending in an anti-Stokes form. Due to the multi-step nature of nPP, some clarifications regarding the commonly used terminology are needed. In the 1PP case, one refers to the state where the photons are stimulated as the lasing state, and there could be storage states that supply the population of the lasing state. The definitions could become tenebrous once the process involves multi-photons. In this work, we denote the entire lasing process, starting with an initial state and ending with a final state. 
The initial state must have a number density $N_\mathrm{i}$ greater than $N_f$, which labels the number density of the final state after the release of $\gamma$-photons\footnote{In this sense, one could identify the initial state as the ``lasing state".}.
The transition from the initial to the final state is carried out by nPP, which goes through various virtual states. Each virtual state will resonate with an intermediate state to maximize the nPP amplitude. We denote the physical state that resonates with the last virtual state before the release of $\gamma$ as the ``top state". 
Transitions between virtual states are carried out by combining the $\gamma$-photons generated from laser-matter interactions with one or two $\omega$-photon sources provided continuously throughout the lasing process by intensive lasers. If the energy gap $E_\mathrm{gap}$ between adjacent states is reachable by the $\omega$-photon, the transition can be completed by the $\omega$-photon alone. Otherwise, $\gamma$-photons need to be employed. In the latter case, each $\gamma_\mathrm{i}$ could carry out up to an E2-transition\footnote{In this work ``up to" means the inclusion of all transitions which have equal or larger probabilities than the specified multipolarities. E.g., up to E2 means either an E1, M1 or E2 transitions.}, while the $\omega$-photons carry out M1-transition~\cite{our_iso}. Note that although alternatives have been suggested~\cite{1444212}, our scheme requires a population inversion, which can be achieved by nuclear reactions~\cite{PhysRevC.83.024604}, separation of isotopes~\cite{sharp,1443607,billings1953photochemical,Gunning1963,liuti1966isotopic}, or HPLS assisted single-~\cite{PhysRevC.61.044303} or multi-photon absorption mechanism~\cite{our_iso}. 
The preferred path (e.g., the nPP route from initial state through $\langle m_2|$, $\langle m_3|$, ..., toward the ``top state") will depend on the detailed structure of the nuclei. In general, the top state is a state which can decay easily to a state other than the initial state. 
The final state is not necessarily the ground state. It is advantageous to have a short-lived final state so that population inversion can persist for a longer time against the initial state's spontaneous decay (or any depletion). 

In summary, with the ingredient of an HPLS produced $\gamma-$flash, it is possible that the nPP effective cross-section of transitions across multipolarities can already be enhanced up to $\approx \SI{e-19}{cm^2}$ utilizing the most powerful HPLS available presently~\cite{Tanaka2020}. This tremendous value of $\sigma_\mathrm{eff}^\mathrm{nPP}$ is comparable to the single-photon absorption cross-section of E1 type and enables stimulated emissions for a wide variety of nuclear isomer states. Although the opposite/competing mechanism exists, it is only significant after the population is reversed between the initial state and the final state. Most importantly, the nPP-induced transitions are entirely decoupled from spontaneous emission, which is dominated by single-photon processes, therein belies a strength of our ansatz. Thus, the limitation imposed by Eq.~(\ref{eq1}) is levitated, and the initial state can have a longer half-life $t_{1/2}$ to accumulate its population without hindering its stimulated emission.

Two additional profound advantages can be realized through the nPP-stimulated graser. First, the supplied photons can be tuned to compensate for the recoil or any shift of the absorption line, which makes non-M\"{o}{\ss}bauer graser possible. Second, the bandwidth of the supplied photons provides a natural way to levitate the broadening of the narrow line breadth issue, which will be discussed in the next section.

\subsection{Broadening of the narrow absorption line breadth: problems and solutions}
\label{sec2b}
Another major issue of grasers, which becomes severe when long-lived isomers are adopted as the initial state, concerns the huge difference between the natural absorption width and the total width after broadening. 
Note that even without the pumping versus spontaneous decay dilemma, the $\Gamma_{\gamma}/\Gamma$ factor of Eq.~(\ref{eq2}) is still presented.
To achieve population inversion, it is preferable if one could adopt isomers as the initial state. Meanwhile, this also means the initial state is associated with a narrower width $\Gamma_{\gamma}=ln(2)\hbar/t_{1/2}$. For example, a \SI{10}{s} half-life corresponds to a natural line width \SI{3.7e-17}{eV}. Meanwhile, various broadening mechanisms disturb nuclei in the target, and even in the most preferable cases of selected M\"{o}{\ss}bauer isomers, the resulting widths are at best of the same order and are almost always much larger than the natural width $\Gamma_{\gamma}$~\cite{RevModPhys.53.687}. For non-M\"{o}{\ss}bauer nuclei, the broadening is dominated by the ``Doppler breadth", i.e., $\Gamma\approx\Gamma_\mathrm{D}\sim \frac{3.3}{\hbar}\sqrt{\mathcal{R}_\mathrm{loss}\mathrm{k_{B}}\theta}$, where $\mathcal{R}_\mathrm{loss}$ is the recoil energy loss, $\mathrm{k_{B}}$ is the Boltzmann constant, and $\theta$ is the absolute temperature. For $\mathcal{R}_\mathrm{loss}=\SI{1}{eV}$ and room temperature, $\Gamma_\mathrm{D}\sim \SI{0.3}{eV}$. Thus, for a $t_{1/2}=\SI{10}{s}$ non-M\"{o}{\ss}bauer isomer, its absorption line is broadened by $10^{16}$ times. Line broadening can sometimes be seen as desirable, especially if $\Gamma_\mathrm{D}>\mathcal{R}_\mathrm{loss}$ happens (mostly at high temperature), which will, in principle, allow the stimulated $\gamma$-photon to have a capacity covering the recoil-loss. However, as with broadening the condition $\Gamma_\mathrm{D}\gg \Gamma_{\gamma}$ generally weakens the amplification process, there is, de-facto, no progress with respect to a practical graser solution to be expected from broadening effects in general~\cite{Spohr2006}. For graser setups that solely rely on a single-pulse amplified linearly, the above means a one per $10^{16}$ chance that a stimulated-$\gamma$ can find a nucleus with the correct vibration velocity which is suitable for the next stimulation process. Thus, even with a 99\% isomer purity, the broadening will completely negate the possibility of grasing. Various mechanisms to narrow $\Gamma_\mathrm{D}$ have been proposed several decades ago concerning selected M\"{o}{\ss}bauer nuclei~\cite{kho1973,ill1974a,ill1974b,go1974a,go1974b,na1973,kagan1975,kar1976,kar1977}.

On the other hand, the supplied photons in the nPP scheme alleviate this difficulty significantly. The $\omega$-photons provided by intensive lasers come with a bandwidth that is typically of order $w_{\omega}\sim \SI{e-3}{eV} - \SI{e-4}{eV}$ and could be easily enlarged at least up to 100 times by shortening the duration of the pulse. Powerful large-bandwidth lasers with $\sim \SI{0.1}{eV}$ for both the central photon energy and bandwidth are also possible~\cite{Lesko2021,Ozawa2021}. We will use the term ``spectrum width" to describe this maximum achievable coverage in energy. In practice, as long as a spectrum width $w_{\omega}\approx\Gamma_\mathrm{D}\approx \SI{0.1}{eV}$ to \SI{1}{eV} is available, most of the $\gamma$-photons resulting from the amplification process will have intense enough $\omega$-photons with suitable energy to combine with (via nPP) so that the stimulating process can carry on\footnote{An even smaller $w_{\omega}\approx 10^{-2}\Gamma_\mathrm{D}$ is acceptable provided that the minimum concentration of initial state (to be discussed in the following subsection) can be increased accordingly.}. The trade-off in that scheme is that the resulting $\gamma$-ray will have an energy-spread $\sim w_{\omega}$.

\subsection{Photon removal by Compton scattering and pair production: problems and solutions}
\label{sec2c}
Now, we consider the photon removal effects and discuss to what degree we can reduce them in our nPP scheme.
Previous estimates~\cite{Baldwin1974} indicate only lasing photons with energy $\lesssim1$ MeV will be favorable. This is mainly due to the photon removal effect by Compton scattering and pair production for $E_{\gamma}\geq 
\SI{1.022}{MeV}$, in combination with the limitation given by Eq.~(\ref{eq2}). As shown above, the limitation in the cross-section can now be removed through nPP. Here, we focus on the remaining photon removal mechanism.

First, since the low-energy photon beam will be supplied continuously (compared to the duration of the entire lasing process) with an intensity $\mathcal{P_{\omega}}\gg \SI{e10}{Wcm^{-2}}$ its loss due to various interactions with electrons, though significant in magnitude, is still negligible compared to its original intensity. The main concern is the energy/momentum shift of the stimulated $\gamma$-photons caused by Compton scattering with electrons in the material. For $E_{\gamma}\geq \SI{1.022}{MeV}$, $\gamma$-photons will start to experience positron-electron pairs produced in the electric field, which further reduces their chance of triggering another stimulated emission. 
\begin{figure}[t]
\includegraphics[width=0.5\textwidth,clip=true]{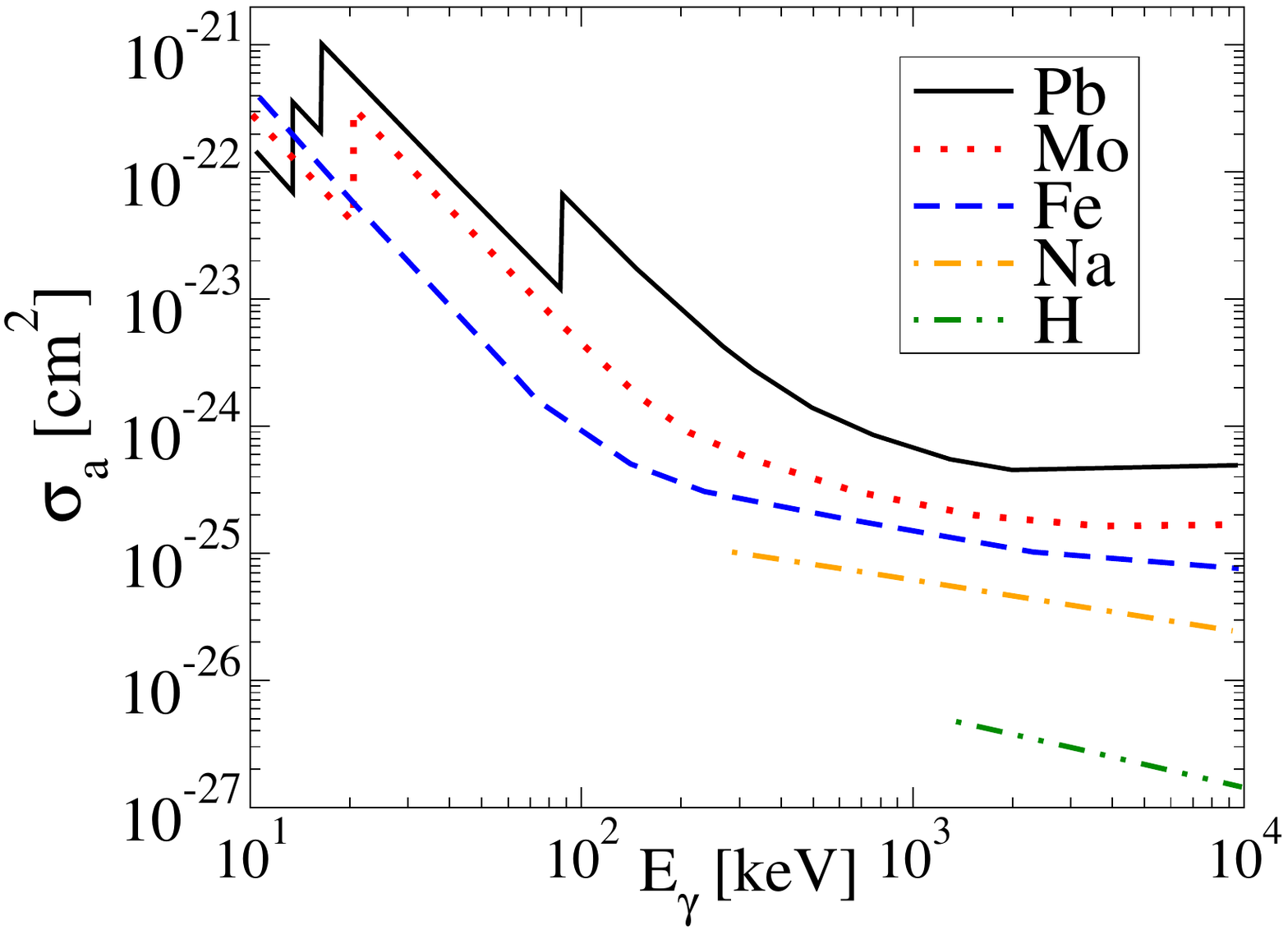}
\caption{Total non-nuclear photon removal cross
section $\sigma_a$ of several substances in the
\SI{10}{keV} to \SI{10}{MeV} energy range (converted from Fig.~5 of Ref.~\cite{Baldwin1974}; or Fig.~2 of Ref.~\cite{RevModPhys.53.687}).}
\label{fig2}
\end{figure}

Fig.~\ref{fig2} illustrates the total non-nuclear photon removal cross-section $\sigma_a$ as a function of energy.
One can see that the nonnuclear photon removal cross-section decreases with $E_{\gamma}$ but reaches a minimum and becomes constant after $E_{\gamma}\gtrsim \SI{1}{MeV}$ because electrons and positions interacting with the electric fields produced by the nucleus persist after $E_{\gamma}\geq \SI{1.022}{MeV}$.

For grasing to happen, it requires~\cite{RevModPhys.53.687,allen1971amplified}
\begin{equation}
N_\mathrm{i}\,\sigma_\mathrm{eff}^\mathrm{nPP}\geq(N_\mathrm{tot}\,\sigma_a+1/L), \label{equal}
\end{equation}
where $N_\mathrm{tot}$ is the number density of total atoms in the target, and $L$ is the length of the target in the direction of the graser propagation. Therefore, the absolute minimum of the effective cross-section required corresponds to $\sigma_\mathrm{eff}^\mathrm{nPP}\gtrsim \SI{e-25}{cm^2}$ for most of the nuclei, assuming that $1/L$ can be made small compared to other terms in Eq.~(\ref{equal}) and one could populate $N_\mathrm{i}\approx N_\mathrm{tot}$.

On the other hand, assuming that the effective cross-section of $\gamma$-photons can be enhanced to $\sigma_\mathrm{eff}^\mathrm{nPP}\approx \SI{e-20}{cm^2}$ for $n\leq4$, it is likely that the required minimum concentration of the initial state is $N_\mathrm{i}/N_\mathrm{tot}\geq \SI{e-5}{}$. 

Any shift in $E_{\gamma}$ larger than the absorption width is counted as photon removal effects in Fig.~\ref{fig2}. In our scheme, the criterion is less stringent, as the bandwidth of the supplied optical or infrared photons replaces the absorption width. However, this is of minimal help, as the significant sources of photon removal above \SI{1}{MeV}, namely Compton scattering and pair production, create shifts in wavelength at the order of the Compton wavelength of the scattered particle (electron or nucleus), which corresponds to an energy $\gtrsim \SI{0.1}{keV}$ and cannot be compensated by the $\omega$-photons supplied by typical lasers. Nevertheless, it is worth mentioning that intense $1\sim5$ keV photons can be generated from betatron radiation in specific laser-matter interaction schemes. The recently suggested intensity $\mathcal{P}\approx\SI{e10}{Wcm^{-2}}$~\cite{Shen2024} is just two orders of magnitude\footnote{This source needs to participate nPP together with other photons, and need to be larger than $\gamma$-photons' intensity $\mathcal{P}_{\gamma}\approx\SI{e12}{Wcm^{-2}}$ in order to not lower $\sigma^{nPP}_{eff}$ further.} away from starting significantly compensating the above-mentioned photon removal effect.  

Neither the pair production rate nor the charge of a specific nucleus can be changed. Thus, without further increasing the effective cross-section, the only option for lowering the minimum concentration would be to choose a lower-$Z$ nucleus such as Na or H in Fig.~\ref{fig2}. Adopting gas targets will not change the concentration requirement but will lower the total number of isomers required by $\sim \SI{e-3}{}$ times due to the lower $N_\mathrm{tot}$, and therefore potentially shorten the pumping time further\footnote{The recoil issue can in principle be solved by the supplied optical photons and will be discussed later.}. 

\begin{figure}[t]
\includegraphics[width=0.5\textwidth,clip=true]{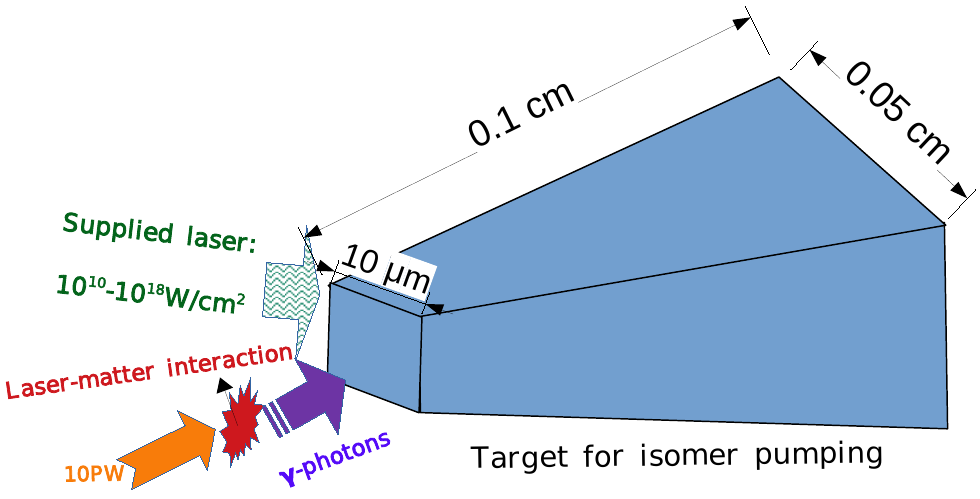}
\caption{Illustration (not to scale) of a possible target design. The waist of the supplied laser is tunable and could cover the entire entry area of $\sim  \SI{0.05}{cm} \times \SI{0.05}{cm}$.}
\label{isotar}
\end{figure}

Typical $\gamma$-photons generated by an HPLS through laser-plasma interaction are confined within an initial area $\sim  \SI{10}{\upmu m} \times \SI{10}{\upmu m}$, at production with an angular divergence $\theta\lesssim \SI{30}{^\circ}$ for the majority of the beam. Unlike the case of grasers, for isomer pumping, nPP is not required to cover all the $|\Delta J|$ between the initial and the isomer state, as there could be favorable decay paths from the final state to the desired isomer state. Thus, 2PP could be sufficient to cover $|\Delta J|\gtrsim4$. It is then reasonable to assume that pumping with a $\sigma^\mathrm{2PP}_\mathrm{eff}\approx \SI{e-21}{cm^{2}}$ can be done without destroying the target. In this scenario, most isomers will be populated within a volume, as illustrated in Fig.~\ref{isotar}. Assuming a solid number density $N_\mathrm{tot}\sim \SI{e22}{cm^{-3}}$ and $N_\mathrm{i}/N_\mathrm{tot}=\SI{e-5}{}$, the minimum number of isomers required within this volume is $\approx \SI{e13}{}$.   
In the pumping scheme, which utilizes $\gamma-$flash from laser-plasma interactions with an \SI{10}{PW} HPLS and with $\sigma^{2pa}_\mathrm{eff}$ enhanced to $\approx \SI{e-21}{cm^2}$ via the assisted optical photons~\cite{our_iso}, the yields of isomers per laser shot in the target illustrated in Fig.~\ref{isotar} is mainly limited by the number of $\gamma$ photons produced per eV interval. The corresponding value is $\sim \SI{e9}{}$ based on the PIC simulations based on the optimal setup currently available in the literature~\cite{PhysRevApplied.13.054024}. Thus, the minimum concentration is likely to be reached with $\sim \SI{e4}{}$ shots of an HPLS, which will be in the reach of a few days of beamtime at a state-of-the-art \SI{10}{PW} facility such as ELI-NP.  

Populating a sufficient concentration of isomers, though it can be time-consuming depending on the HPLS's repetition rate, does not suffer from any theoretical problem as long as the $t_{1/2}$ of the chosen isomer is greater than the total pumping time. On the other hand, the strict requirement on $\sigma_\mathrm{eff}^\mathrm{nPP}$ to enable grasing can lead to a situation that supplying optical photons with $\mathcal{P_{\omega}}\geq \SI{e18}{Wcm^{-2}}$ becomes necessary, which then destroys solid targets. Furthermore, inverse bremsstrahlung (aka inverse Compton scattering) in the corresponding plasma environment could potentially alter the total photon removal cross-section. However, even in this case, the target-fragmentation process or the mechanical shock-wave propagation is much slower than the speed of light and the entire stimulated emission process. Thus, a one-time-use graser with a nano-wired-like or other special target designs might be favorable for the first experimental realization.

We note that further enhancing the $\gamma-$flash by a factor of $10^{1}\sim 10^3$, either by optimizing the target design with new laser-plasma interaction schemes or simply improving the HPLS setup itself\footnote{One can increase the yields of isomers per laser shot by $n$ times by supplying $n$ additional $\omega$-photon beams (each differs by $\gtrsim \SI{1}{eV}$ in energy) so that a wider energy domain in the $\gamma$ spectrum can be used in the pumping process~\cite{our_iso}.}, is likely. If the enhancement is reflected in the intensity of the $\gamma-$flash, i.e., $\mathcal{P_{\gamma}}$ itself, then $\sigma_\mathrm{eff}^\mathrm{nPP}$ is also enhanced accordingly (up to a threshold value to be determined experimentally, where a further increase in $\mathcal{P_{\gamma}}$ will play a significant role to the target with unfavorable outcomes, like the optical photons after $\mathcal{P_{\omega}}\geq \SI{e18}{Wcm^{-2}}$). Thus, each order of magnitude in the form of such enhancement will shorten the total pumping time by up to two orders of magnitude for the same amount of yields, as both the effective cross-section and the actual number of $\gamma$ photons are increased ten times. At the same time, the required $N_\mathrm{i}/N_\mathrm{tot}$ is lowered by one order of magnitude due to the increase of $\sigma_\mathrm{eff}^\mathrm{nPP}$, leading to a total reduction between $\SI{e2}{}\sim \SI{e3}{}$ for the graser pumping time. 

\subsection{Recoil: Problem and solution}
One well-known problem in the realization of grasers concerns recoil. Unlike typical lasers, where the photon involves transitions between different atomic/molecular states and is of eV-scale in energy, the energy of $\gamma$-photons involves transitions between nuclear states and is of the order of MeV-scale. Thus, the recoil is no longer negligible, and the stimulated emitted photon could lose up to $\sim \SI{10}{eV}$ in energy, which is at least $3-4$ orders greater than the width of any state usable for lasing. With this recoil energy loss, the emitted photon cannot stimulate another emission. In this regard, grasers can only occur in special cases where individual nuclei are tightly bound in the lattice structure of solids, i.e., the M\"{o}{\ss}bauer nuclei~\cite{Mssbauer1958}. 

However, the nPP mechanism provides an extra benefit such that the supplied photons can be used to compensate for these recoil losses. Herein, we discuss two of the simplest possibilities that could be used to overcome the recoil problem for non-M\"{o}{\ss}bauer nuclei.
In the past, various recoil-compensation mechanisms have been proposed~\cite{Krasnov1975,PhysRevLett.42.1397,PhysRevC.20.1942,PhysRevC.23.50,PhysRevC.23.1007,Collins1982}.
As a general rule, since recoiling costs energy, nPP must involve at least one anti-Stokes process where the supplied photon(s) absorption is followed by a stimulated emission of the $\gamma-$photon.
In the following, we demonstrate that the recoil losses, even in the most profound case by using gas targets, can be compensated easily by the supplied photons. 
Denoting $\omega_{in}$ the energy of the incoming photon, $M_A$ the mass of the nucleus, then by momentum- and energy-conservation,
\begin{align}
\omega_\mathrm{in}+M_AV_\mathrm{i}&=M_A V, \label{recoil} \\ 
\omega_\mathrm{in}+\frac{1}{2}M_AV_\mathrm{i}^2&=\frac{1}{2}M_A V^2+\omega, \label{recoil1}
\end{align}
where $V$ ($_\mathrm{i}$) is the nucleus's recoil (initial) velocity and $\omega$ is the energy gap between the top state and the final state. Here, we have adopted natural unit $\hbar=c=1$. One can then obtain the energy loss per absorption or emission\footnote{The same equations apply to the case of photon-emission, i.e., just replaces $\omega_{in}$ by $\omega_{out}$, where $\omega_{out}$ is the energy of the emitted photon.}, i.e.,
\begin{equation}
\mathcal{R}_\mathrm{loss}\equiv \omega_\mathrm{in}-\omega=\frac{\omega_{in}^2}{2M_A}+\omega_\mathrm{in}V_\mathrm{i}. \label{recoil2}
\end{equation}
For gases, $V_{\mathrm{i}}\approx\sqrt{3RT/M_\mathrm{mol}}$, with $R=\SI{8.31}{Jk^{-1}mol^{-1}}$, $M_{mol}$ the molar mass in \SI{}{kg/mol} and $T$ the temperature in Kelvin. At room temperature $V_\mathrm{i}\approx O(10^{-6})$. Note that in our nPP scheme, the top virtual state of the nucleus before releasing the desired $\gamma$-ray can have a larger value of $V_\mathrm{i}$, as it has absorbed the momentum from those $(n-1)$ photons.  
Thus, the absorption line will be shifted by \SI{3}{eV} to \SI{6}{eV} for a nucleus with $A\approx100-200$ nucleus with a $\gamma-$photon energy $\omega_\mathrm{in} \approx \SI{1}{MeV}$. The standard spread of $V_\mathrm{i}$ will also broaden the line, which is of order $3.3\sqrt{\mathcal{R}_\mathrm{loss}k\theta}\approx \SI{0.1}{eV}$ as discussed before.

\begin{figure}[t]
\includegraphics[width=0.47\textwidth,clip=true]{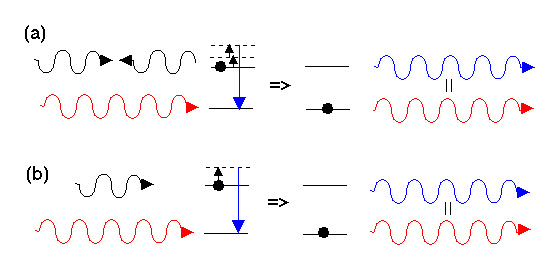}
\caption{Illustration of two simplest recoil-compensating schemes via nPP. The black curved lines represent the supplied photons, and the incoming (stimulated) $\gamma-$photon is represented by the red (blue) curved lines. Solid (dashed) lines represent physical (virtual) states.}
\label{fignpp}
\end{figure}

The most straightforward way to compensate for the recoil energy loss is to supply two head-on photon beams with opposite momentum $\omega_1$, which then leaves Eq.~(\ref{recoil}) unchanged but replaces $\omega_{in}$ in Eq.~(\ref{recoil1}) by $\omega_{in}+2\omega_1$. Thus, from Eq.~(\ref{recoil2}), one can solve for $\omega_\mathrm{in}$ as a function of $\omega$ and $V_\mathrm{i}$, then the photon energy of the supplied lasers needs to be tuned to $\omega_1=\mathcal{R}_\mathrm{loss}/2$ to compensate for the recoil. This compensation can be realized in the pumping process (a) of Fig.~\ref{fignpp}, where the two opposite photons provide zero net momentum but pump the nucleus to a virtual state slightly above the top state so that the stimulated-emitted $\gamma-$photon can have an energy $\omega_\mathrm{in}$ which enables it to re-combine with the supplied $2\omega_1$ in the amplification cycle.\\

A second scheme is to compensate for the recoil by one supplied photon, as illustrated in (b) of Fig.~\ref{fignpp}. As the total momentum increases, the required $\omega_1$ will be slightly larger than $\mathcal{R}_\mathrm{loss}$. 

Note that the energy of the supplied beams \textit{does not} need to match the recoil loss exactly; a tuning until the energy plus its spectrum width covers the required $\omega_1$ would be sufficient. A caveat is that the supplied optical photons are absorbed in the anti-Stokes processes. Thus, they needed to be supplied continuously, and their number-intensity ($\equiv I_n$) will serve as an upper limit of the graser number-intensity (but \textit{not} the graser intensity which is the product of $I_n\times E_{\gamma}$). 

The supplied photons generally resonate with the top state via the M1 transition. Before pumping toward the top state, the main contribution consists of $\omega_1$-photons in combination with $\gamma$-photons at suitable energies that resonate with each intermediate state. Then, at the last pumping step before lasing occurs, the virtual state must be $2\omega_1$ (or $\omega_1$ for the one-photon-compensate scheme) above the top state. In the beginning, this will lead to a slightly smaller nPP cross-section compared to the maximum Eq.~(\ref{eff_e1}) could have, but this choice will have $\sigma_\mathrm{eff}^\mathrm{nPP}$ be amplified quickly as the intensity of the $\gamma$ flux grows exponentially. 

\section{Designing the first single pulse graser}
\label{sec3}
In the above discussion, we only focus on the general solutions to alleviate the graser dilemma. To overcome technical limitations at present, specific choices of target nuclei need to be made.
As a general rule, both PIC simulations and experiments suggested that the $\gamma-$flash generated by HPLS has more photons located at the lower end of the energy spectrum, i.e., $\SI{1}{keV}\lesssim E_{\gamma}\lesssim \SI{1}{MeV}$. This favors low-lying isomers or those with small pumping/depletion gaps between intermediate states.


Once an isomer is chosen to be the initial state, the remaining challenge concerns overcoming the large angular momentum difference between the initial and final states throughout the lasing process. They need to be separated by at least E2 (in most cases $\gtrsim \mathrm{E3}$). nPP then will involve at least one upward transition with an energy gap $E_\mathrm{gap}\gtrsim \SI{1}{keV}$ in general.
To overcome this upward gap, it then demands that intensive $\gamma$-photons with an energy matching $E_\mathrm{gap}$ be supplied, which creates a practical problem. 

\subsection{Converter-type grasers}
Although there are a few cases of deformed nuclei where the intermediate states are separated by $E_\mathrm{gap}\lesssim \SI{1}{keV}$ and E1 transition, the typical energy gap between intermediate states is $\gtrsim \SI{10}{keV}$, which is too large to resonate with optical photons. Note that while this is not a problem in isomer pumping, as they can be filled by photons in the spectrum of $\gamma-$flash, the amplification process requires those $\gamma$-photons to be supplied with increasing intensity throughout the lasing process.
Thus, the strength of the graser can only be as strong as the integrated strength of the $\gamma-$flash (accumulated within an interval up to the spectrum width of the supplied laser), 
which is currently up to $\mathcal{P_{\gamma}}\approx \SI{e12}{Wcm^{-2}}$ for $E_{\gamma}\leq \SI{5}{MeV}$. In practice, the amplification process proceeds at the speed of light plus the stimulating time-cost for nPP\footnote{Since nPP involves multiple transitions between nuclear states before stimulated emission, the time cost is not likely negligible.}. Therefore, the $\gamma$-flash duration should also cover the total stimulated-emission time-cost---which involves nPP and is $\lesssim \SI{1}{fs}$ per event. Assuming each stimulated event costs \SI{0.5}{fs}, one could have $\sim 50$ events covered by the typical \SI{25}{fs} duration of $\gamma-$flash, which results in the limitation of a total number of $\sim 2^{50}\sim \SI{e15}{}$ aligned and coherently released $\gamma$-rays. Although this number well exceeds the current upper limit of the $\gamma$-photons can be generated in the $\gamma-$flash ($\lesssim \SI{e9}{}$ per eV interval for $E_{\gamma}\leq \SI{5}{MeV}$), it might serve as a bottleneck in the future. 
\begin{figure*}[t]
\includegraphics[width=\textwidth,clip=true]{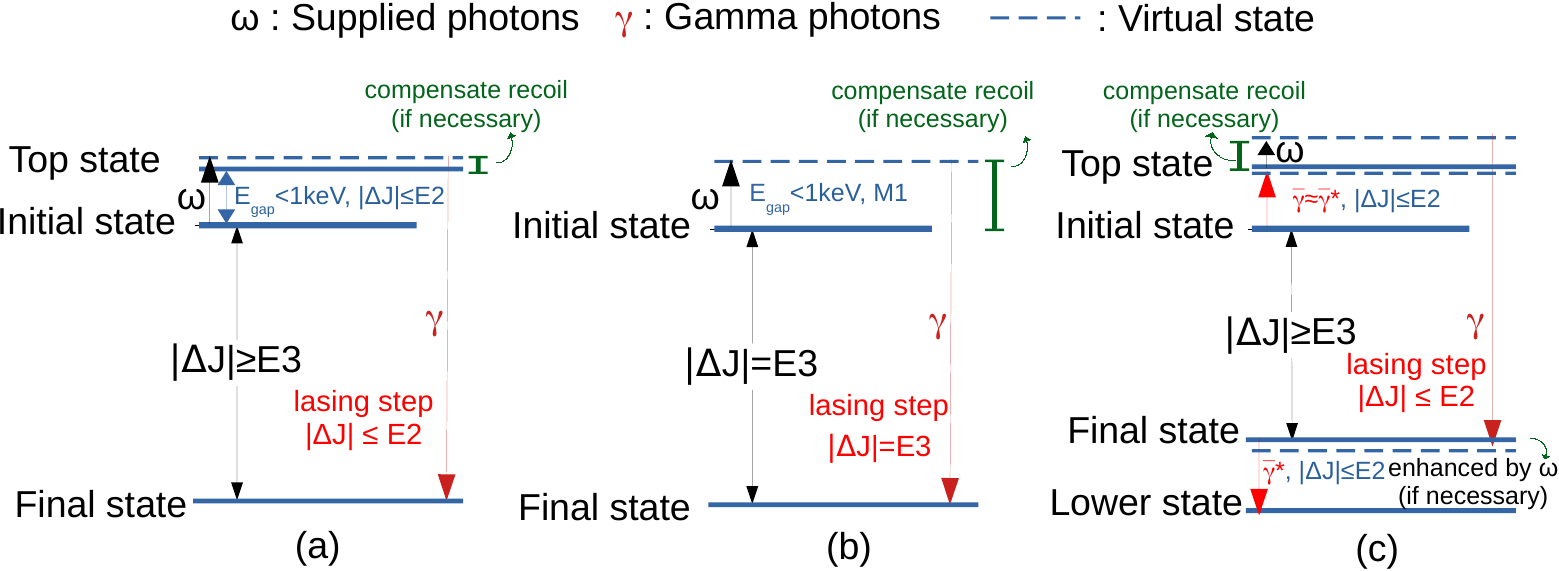}
\caption{Illustration of optical-photon-triggered grasers: (a) and (b); and self-feeding grasers: (c).}
\label{p3}
\end{figure*}

In summary, the above type of grasers can be applied to a wide range of nuclei, with transitions described in Fig.~\ref{fig1}. However, this type of graser serves as ``$\gamma-$converters", which gather the incoming $\gamma$-photons within a specific energy interval (depending on the supplied laser spectrum width) and convert them into a coherent output.

\subsection{Optical-photon-triggered grasers}
Grasers do not rely on continuous support from $\gamma-$flash demand specific energy-level arrangements in the nucleus. As a general requirement, the initial state must still be separated from its final state by at least E3 or M2 transition (which allows it to have a $t_{1/2}$ ranging from days to years) to allow sufficient pumping time. Under this condition, one can select a nucleus where the top state is within \SI{1}{keV} and E1 from the initial state, while the final state is separated from the top state within the E2 transition. Then one could supply \SI{e3}{} more intense optical beams which corresponds to $\mathcal{P_{\omega}}\gtrsim \SI{e21}{Wcm^{-2}}$ to maintain an effective stimulated emission cross-section $\gtrsim \SI{e-25}{cm^2}$ for the $\gamma-$ photons, so that grasing could occur. This possibility is illustrated as case (a) in Fig.~\ref{p3}. Another possibility (case (b) in Fig.~\ref{p3}) is just to start with an isomer that is separated by E3 from the final state. Then, the E3 transition might be enhanced directly by injecting photons with $\mathcal{P_{\omega}}\gtrsim \SI{e23}{Wcm^{-2}}$. In both cases, a high concentration of isomer state ($N_\mathrm{i}/N_\mathrm{tot}\geq1$ \%) is required to enable amplification against the photon removal effects. Note that both settings, when realized in solid targets, lead to a single-shot graser arrangement, as the strong optical pulse will destroy the medium. These systems are, in principle, possible in the immediate future by utilizing a succession of replaceable nano-wire targets on a shot-to-shot base.  

\subsection{Self-feeding grasers}
Finally, a setting demands an extraordinary arrangement of the nucleus's level structure. The simplest case of this ``self-feeding" scheme consists of a 4-level system as illustrated in Fig.~\ref{p3} (c), where the final state spontaneously decays fast enough (normally through E1 but can be accelerated by the supplied photons) to lower states with one of the releasing photons ($\equiv \bar{\gamma}^*$) matching the energy gap $\bar{\gamma}$ (within \SI{1}{keV} difference) and multipolarity between the initial state and the top state. Then, except for the first nPP, which is still stimulated by the $\gamma-$flash, the spontaneous emitted $\bar{\gamma}^*$ can fill the upward gap in the next nPP process. In this way, each stimulated nucleus releases a $\gamma$ and a $\bar{\gamma}^*$, with $\bar{\gamma}^*$ being re-absorbed and $\gamma$ being amplified. Throughout the periodic table, by restricting isomers to $t_{1/2}> \SI{30}{d}$,  radioactive $\mathrm{^{174}Lu}$ would be a prime candidate as it has the required level structure. However, in this case, both the $\bar{\gamma}^{*}$ and $\bar{\gamma}$ are of E3 transition, which makes it unfavorable. Possibilities involving a combination of two or more nuclei with $\bar{\gamma}^*$'s feeding each other, though difficult in target manufacturing, cannot be ruled out.

\subsection{Desired quality of the gain medium and the supplied photons}
\label{sec3f}
Typical high-power lasers have $\SI{0.1}{eV} \lesssim E_{\omega}\lesssim \SI{10}{eV}$ and bandwidth \SI{e-3}{eV} to \SI{e-1}{eV}. Thus, even without destroying the target, to have a $\sigma_\mathrm{eff}\gtrsim \SI{e-25}{cm^2}$, the required intensity of the supplied photons would make the majority of the solid targets non-transparent to the supplied photons which can only penetrate a skin depth within a few nm from the surface. In principle, with special target designs, this does not prohibit any linear-type, single-pulse graser. In practice, it requires the surface of the target to be smooth (with a roughness within a few nm) along the lasing direction.

As recoil losses up to $\sim \SI{5}{eV}$ can be compensated by the supplied photons, the gain medium of grasers is not restricted to the solid type.  
Moreover, if it turns out that supplying optical photons with $\mathcal{P}_{\omega}\gtrsim \SI{e18}{Wcm^{-2}}$ which destroys most of the solid targets anyway would be necessary, there might be no practical advantage (in the sense of just allowing grasing to occur) of favoring solid targets instead of gas or liquid targets. 

Alternatively, it could be advantageous to populate the isomers in a gas form and have them bound onto the surface of a solid target. Suppose isomers can be selected/concentrated~\cite{szilard1934chemical,l1973b,murnick1979applications,jacquinot1979hyperfine} during this process. In that case, the absolute number of isomers required can be greatly reduced, as distributing them very densely on the surface will be sufficient. The pumping time estimated in Sec.~\ref{sec2c} can be greatly reduced from $10^4$ shots down to as low as a single shot for reaching the same minimum concentration.    

Regarding the energy of the supplied photons. As discussed in Sec.~\ref{sec2a}, the key factor in retaining a $\sigma_\mathrm{eff}\approx \SI{e-20}{cm^2}$ is $\mathcal{P}/(\Delta E)^2$, where $\Delta E$ is the greater of the supplied photon energy or the natural width of the intermediate state. As the natural width of the nuclear intermediate state $\lesssim \SI{e-3}{eV}$ to \SI{e-4}{eV}, it is of interest to investigate photon sources with $E_{\omega}\lesssim10^{-3}$ eV, i.e., within the THz range. Using photons belonging to this energy range as the main source of the supplied photons allows a much better resonance between the virtual and intermediate states. An intensity as low as $\mathcal{P}_{\omega}\approx \SI{e14}{Wcm^{-2}}$ would serve the same role as $\mathcal{P}_{\omega}\gtrsim \SI{e20}{Wcm^{-2}}$ for optical photons in the aspect of enhancing the effective stimulated emission cross-section. Recent advances in this research direction are of very promising nature~\cite{Fedorov2020,PhysRevLett.130.146702,Malaca2023}. 
A potential additional benefit of a THz photon source is that they might allow a higher threshold in intensity, i.e., $\mathcal{P}_{\omega}\gtrsim \SI{e18}{Wcm^{-2}}$ before damaging the targets. 

\section{Controlling the multi-photon process}
\label{sec4}
Note that the multi-photon process can trigger both isomer pumping and stimulated emission depending on the intensity of the photons and other conditions. The main control is the intensity of the $\gamma$- and the $\omega$-photons, which can be divided into three regimes.

In the low-intensity regime, all transitions are dominated by 1PP, which corresponds to $\mathcal{P}_{\omega}$ and $\mathcal{P}_{\gamma}\ll \SI{e9}{Wcm^{-2}}$. Under this condition, conventional step-wise pumping/de-excitation occurs. The creation or depletion process toward the desired states is generally slow.

In the medium-intensity regime, where $\SI{e9}{Wcm^{-2}} \lesssim\mathcal{P}_{\gamma}\lesssim \SI{e12}{Wcm^{-2}}$ and $\SI{e11}{Wcm^{-2}}  \lesssim\mathcal{P}_{\omega}\lesssim \SI{e18}{Wcm^{-2}}$, isomer pumping and depletion start to be enhanced by nPP. Whether pumping or depletion would occur (or be more significant) depends on the state's population. It is also subjected to fine-tuning based on the specific level structure coupled with the intensities and energy distribution of the photons, provided that one has detailed control of them. 

In the strong regime, which serves as a necessary but not a sufficient condition for grasing to occur, one requires ${P}_{\gamma}\gtrsim 
\SI{e11}{Wcm^{-2}}$ and $\mathcal{P}_{\omega}\gtrsim \SI{e14}{Wcm^{-2}}$. Importantly, additional conditions must be met.
First, the minimum concentration described by Eq.~(\ref{equal}) must be met. Thus, it depends on specific choices of target and nuclei together with $\mathcal{P}_{\omega}$ and $\mathcal{P}_{\gamma}$.
In general, $N_\mathrm{i}/N_\mathrm{tot}\gtrsim \SI{e-5}{}$ will be required. With high-lying states above the initial state with each jump within $E_\mathrm{gap}\lesssim \SI{5}{MeV}$ and $\Delta J^{\pi}\leq 2$, the strong-regime will enable converter-type grasers.
For optical/infrared photons supplied more than $\mathcal{P}_{\omega}\gtrsim \SI{e18}{Wcm^{-2}}$, and with either condition (a) or (b) as described in Fig.~\ref{p3}, optical-photon-pumped grasers can occur where the $\gamma-$flash is only used as the first trigger. Finally, the ``self-feeding" type of graser is, in principle, possible and requires the least supplies from optical/infrared and $\gamma-$sources. However, suitable combinations of elements have yet to be found.  

For non-M\"{o}{\ss}bauer nuclei, a suitable energy and alignment of the supplied photons are also required. Otherwise, even with all of the favorable conditions, only isomer creation and depletion will occur according to their populations. Those stricter conditions can be advantageous, as they offer extra controls to select the process desired.   

\section{Summary}
\label{sec5}
We investigated the feasibility of grasing and the implementation of graser systems through a novel approach combining the $\gamma-$flash generated by laser-matter interactions and optical/infrared photons supplied by intense laser sources.
Our key idea, which is essentially governed by a simple combinatorial enhancement due to ultra-intense beams, might open a completely new venue to probe phenomena that were previously inaccessible in nuclear, particle, and condensed matter physics.
Such an approach is justified now due to the emergence of strong, high-power laser systems. The dense and continuous spectrum of the $\gamma$-photons expected from laser-plasma interaction of an HPLS in the PW region has the potential to serve as the game-changer in this field, as it enables multi-photon processes and can be facilitated to overcome the graser dilemma as well as many other known obscurities. 
Specifically, we showed that a multi-photon process could solve problems regarding the small stimulated cross-section, recoil, or any line shift, line broadening, and stimulated emission across large multipolarities. 

At present, the first experimental realization of a graser system will likely utilize isomers separated within an E4 transition from the final state. It will mostly be a single-shot device, as the currently available $\gamma-$flashes require optical/infrared photons to be supplied with an intensity $ \gtrsim \SI{e18}{Wcm^{-2}}$, which causes target-destruction afterward.  
Any improvement or optimization in either target-design or laser-matter interaction scheme is therefore highly desirable if it improves the intensity of the $\gamma-$flash under \SI{5}{MeV} and significantly reduces the burden of the supplied photons. 

Our investigations also suggest that developing intensive far-infrared beams ($E_{\omega}\lesssim \SI{e-2}{eV}$) and using them as the main supplied photon source would be of interest. This then enables virtual states to resonate better with intermediate states and enhances the effective stimulated emission cross-section drastically.

We emphasize that our proposed graser will be of single-pulse nature only because of the lack of reflective materials. Nevertheless, our scheme is very general and can be applied to various nuclei. 




\vspace{0.5cm}
\begin{acknowledgments}
We thank our colleagues S. Ionescu, P. Tomassini, V. Horny and M. Cernaianu for
useful discussions and suggestions. This work was supported by the
the Extreme Light Infrastructure Nuclear Physics (ELI-NP) Phase II, a project co-financed by the Romanian Government and the European Union through the European Regional Development Fund - the Competitiveness Operational Programme (1/07.07.2016, COP, ID 1334);  the Romanian Ministry of Research and Innovation: PN23210105 (Phase 2, the Program Nucleu); and the ELI-RO grant Proiectul ELI12/16.10.2020 of the Romanian Government.
We acknowledge PRACE for awarding us access to Karolina at IT4Innovations, Czechia under project number EHPC-BEN-2023B05-023 (DD-23-83); IT4Innovations at Czech National Supercomputing Center under project number OPEN24-21 1892; Ministry of Education, Youth and Sports of the Czech Republic through the e-INFRA CZ (ID:90140) and CINECA through PRACE-ICEI standard call 2022 (P.I. Paolo Tomassini).
\end{acknowledgments}

\bibliography{graser_2pa} 

\begin{thebibliography}{128}%
\makeatletter
\providecommand \@ifxundefined [1]{%
 \@ifx{#1\undefined}
}%
\providecommand \@ifnum [1]{%
 \ifnum #1\expandafter \@firstoftwo
 \else \expandafter \@secondoftwo
 \fi
}%
\providecommand \@ifx [1]{%
 \ifx #1\expandafter \@firstoftwo
 \else \expandafter \@secondoftwo
 \fi
}%
\providecommand \natexlab [1]{#1}%
\providecommand \enquote  [1]{``#1''}%
\providecommand \bibnamefont  [1]{#1}%
\providecommand \bibfnamefont [1]{#1}%
\providecommand \citenamefont [1]{#1}%
\providecommand \href@noop [0]{\@secondoftwo}%
\providecommand \href [0]{\begingroup \@sanitize@url \@href}%
\providecommand \@href[1]{\@@startlink{#1}\@@href}%
\providecommand \@@href[1]{\endgroup#1\@@endlink}%
\providecommand \@sanitize@url [0]{\catcode `\\12\catcode `\$12\catcode `\&12\catcode `\#12\catcode `\^12\catcode `\_12\catcode `\%12\relax}%
\providecommand \@@startlink[1]{}%
\providecommand \@@endlink[0]{}%
\providecommand \url  [0]{\begingroup\@sanitize@url \@url }%
\providecommand \@url [1]{\endgroup\@href {#1}{\urlprefix }}%
\providecommand \urlprefix  [0]{URL }%
\providecommand \Eprint [0]{\href }%
\providecommand \doibase [0]{http://dx.doi.org/}%
\providecommand \selectlanguage [0]{\@gobble}%
\providecommand \bibinfo  [0]{\@secondoftwo}%
\providecommand \bibfield  [0]{\@secondoftwo}%
\providecommand \translation [1]{[#1]}%
\providecommand \BibitemOpen [0]{}%
\providecommand \bibitemStop [0]{}%
\providecommand \bibitemNoStop [0]{.\EOS\space}%
\providecommand \EOS [0]{\spacefactor3000\relax}%
\providecommand \BibitemShut  [1]{\csname bibitem#1\endcsname}%
\let\auto@bib@innerbib\@empty
\bibitem [{\citenamefont {Baldwin}\ \emph {et~al.}(1981)\citenamefont {Baldwin}, \citenamefont {Solem},\ and\ \citenamefont {Gol'danskii}}]{RevModPhys.53.687}%
  \BibitemOpen
  \bibfield  {author} {\bibinfo {author} {\bibfnamefont {G.~C.}\ \bibnamefont {Baldwin}}, \bibinfo {author} {\bibfnamefont {J.~C.}\ \bibnamefont {Solem}}, \ and\ \bibinfo {author} {\bibfnamefont {V.~I.}\ \bibnamefont {Gol'danskii}},\ }\href {\doibase 10.1103/RevModPhys.53.687} {\bibfield  {journal} {\bibinfo  {journal} {Rev. Mod. Phys.}\ }\textbf {\bibinfo {volume} {53}},\ \bibinfo {pages} {687} (\bibinfo {year} {1981})}\BibitemShut {NoStop}%
\bibitem [{\citenamefont {Baldwin}\ and\ \citenamefont {Solem}(1997)}]{RevModPhys.69.1085}%
  \BibitemOpen
  \bibfield  {author} {\bibinfo {author} {\bibfnamefont {G.~C.}\ \bibnamefont {Baldwin}}\ and\ \bibinfo {author} {\bibfnamefont {J.~C.}\ \bibnamefont {Solem}},\ }\href {\doibase 10.1103/RevModPhys.69.1085} {\bibfield  {journal} {\bibinfo  {journal} {Rev. Mod. Phys.}\ }\textbf {\bibinfo {volume} {69}},\ \bibinfo {pages} {1085} (\bibinfo {year} {1997})}\BibitemShut {NoStop}%
\bibitem [{\citenamefont {Tanaka}\ \emph {et~al.}(2020)\citenamefont {Tanaka}, \citenamefont {Spohr}, \citenamefont {Balabanski}, \citenamefont {Balascuta}, \citenamefont {Capponi}, \citenamefont {Cernaianu}, \citenamefont {Cuciuc}, \citenamefont {Cucoanes}, \citenamefont {Dancus}, \citenamefont {Dhal}, \citenamefont {Diaconescu}, \citenamefont {Doria}, \citenamefont {Ghenuche}, \citenamefont {Ghita}, \citenamefont {Kisyov}, \citenamefont {Nastasa}, \citenamefont {Ong}, \citenamefont {Rotaru}, \citenamefont {Sangwan}, \citenamefont {S\"{o}derstr\"{o}m}, \citenamefont {Stutman}, \citenamefont {Suliman}, \citenamefont {Tesileanu}, \citenamefont {Tudor}, \citenamefont {Tsoneva}, \citenamefont {Ur}, \citenamefont {Ursescu},\ and\ \citenamefont {Zamfir}}]{Tanaka2020}%
  \BibitemOpen
  \bibfield  {author} {\bibinfo {author} {\bibfnamefont {K.~A.}\ \bibnamefont {Tanaka}}, \bibinfo {author} {\bibfnamefont {K.~M.}\ \bibnamefont {Spohr}}, \bibinfo {author} {\bibfnamefont {D.~L.}\ \bibnamefont {Balabanski}}, \bibinfo {author} {\bibfnamefont {S.}~\bibnamefont {Balascuta}}, \bibinfo {author} {\bibfnamefont {L.}~\bibnamefont {Capponi}}, \bibinfo {author} {\bibfnamefont {M.~O.}\ \bibnamefont {Cernaianu}}, \bibinfo {author} {\bibfnamefont {M.}~\bibnamefont {Cuciuc}}, \bibinfo {author} {\bibfnamefont {A.}~\bibnamefont {Cucoanes}}, \bibinfo {author} {\bibfnamefont {I.}~\bibnamefont {Dancus}}, \bibinfo {author} {\bibfnamefont {A.}~\bibnamefont {Dhal}}, \bibinfo {author} {\bibfnamefont {B.}~\bibnamefont {Diaconescu}}, \bibinfo {author} {\bibfnamefont {D.}~\bibnamefont {Doria}}, \bibinfo {author} {\bibfnamefont {P.}~\bibnamefont {Ghenuche}}, \bibinfo {author} {\bibfnamefont {D.~G.}\ \bibnamefont {Ghita}}, \bibinfo {author} {\bibfnamefont {S.}~\bibnamefont {Kisyov}}, \bibinfo {author} {\bibfnamefont
  {V.}~\bibnamefont {Nastasa}}, \bibinfo {author} {\bibfnamefont {J.~F.}\ \bibnamefont {Ong}}, \bibinfo {author} {\bibfnamefont {F.}~\bibnamefont {Rotaru}}, \bibinfo {author} {\bibfnamefont {D.}~\bibnamefont {Sangwan}}, \bibinfo {author} {\bibfnamefont {P.-A.}\ \bibnamefont {S\"{o}derstr\"{o}m}}, \bibinfo {author} {\bibfnamefont {D.}~\bibnamefont {Stutman}}, \bibinfo {author} {\bibfnamefont {G.}~\bibnamefont {Suliman}}, \bibinfo {author} {\bibfnamefont {O.}~\bibnamefont {Tesileanu}}, \bibinfo {author} {\bibfnamefont {L.}~\bibnamefont {Tudor}}, \bibinfo {author} {\bibfnamefont {N.}~\bibnamefont {Tsoneva}}, \bibinfo {author} {\bibfnamefont {C.~A.}\ \bibnamefont {Ur}}, \bibinfo {author} {\bibfnamefont {D.}~\bibnamefont {Ursescu}}, \ and\ \bibinfo {author} {\bibfnamefont {N.~V.}\ \bibnamefont {Zamfir}},\ }\href {\doibase 10.1063/1.5093535} {\bibfield  {journal} {\bibinfo  {journal} {Matter and Radiation at Extremes}\ }\textbf {\bibinfo {volume} {5}} (\bibinfo {year} {2020}),\ 10.1063/1.5093535}\BibitemShut
  {NoStop}%
\bibitem [{\citenamefont {Doria}\ \emph {et~al.}(2020)\citenamefont {Doria}, \citenamefont {Cernaianu}, \citenamefont {Ghenuche}, \citenamefont {Stutman}, \citenamefont {Tanaka}, \citenamefont {Ticos},\ and\ \citenamefont {Ur}}]{Doria2020}%
  \BibitemOpen
  \bibfield  {author} {\bibinfo {author} {\bibfnamefont {D.}~\bibnamefont {Doria}}, \bibinfo {author} {\bibfnamefont {M.}~\bibnamefont {Cernaianu}}, \bibinfo {author} {\bibfnamefont {P.}~\bibnamefont {Ghenuche}}, \bibinfo {author} {\bibfnamefont {D.}~\bibnamefont {Stutman}}, \bibinfo {author} {\bibfnamefont {K.}~\bibnamefont {Tanaka}}, \bibinfo {author} {\bibfnamefont {C.}~\bibnamefont {Ticos}}, \ and\ \bibinfo {author} {\bibfnamefont {C.}~\bibnamefont {Ur}},\ }\href {\doibase 10.1088/1748-0221/15/09/C09053} {\bibfield  {journal} {\bibinfo  {journal} {Journal of Instrumentation}\ }\textbf {\bibinfo {volume} {15}},\ \bibinfo {pages} {C09053} (\bibinfo {year} {2020})}\BibitemShut {NoStop}%
\bibitem [{\citenamefont {Radier}\ \emph {et~al.}(2022)\citenamefont {Radier}, \citenamefont {Chalus}, \citenamefont {Charbonneau}, \citenamefont {Thambirajah}, \citenamefont {Deschamps}, \citenamefont {David}, \citenamefont {Barbe}, \citenamefont {Etter}, \citenamefont {Matras}, \citenamefont {Ricaud}, \citenamefont {Leroux}, \citenamefont {Richard}, \citenamefont {Lureau}, \citenamefont {Baleanu}, \citenamefont {Banici}, \citenamefont {Gradinariu}, \citenamefont {Caldararu}, \citenamefont {Capiteanu}, \citenamefont {Naziru}, \citenamefont {Diaconescu}, \citenamefont {Iancu}, \citenamefont {Dabu}, \citenamefont {Ursescu}, \citenamefont {Dancus}, \citenamefont {Ur}, \citenamefont {Tanaka},\ and\ \citenamefont {Zamfir}}]{Radier2022}%
  \BibitemOpen
  \bibfield  {author} {\bibinfo {author} {\bibfnamefont {C.}~\bibnamefont {Radier}}, \bibinfo {author} {\bibfnamefont {O.}~\bibnamefont {Chalus}}, \bibinfo {author} {\bibfnamefont {M.}~\bibnamefont {Charbonneau}}, \bibinfo {author} {\bibfnamefont {S.}~\bibnamefont {Thambirajah}}, \bibinfo {author} {\bibfnamefont {G.}~\bibnamefont {Deschamps}}, \bibinfo {author} {\bibfnamefont {S.}~\bibnamefont {David}}, \bibinfo {author} {\bibfnamefont {J.}~\bibnamefont {Barbe}}, \bibinfo {author} {\bibfnamefont {E.}~\bibnamefont {Etter}}, \bibinfo {author} {\bibfnamefont {G.}~\bibnamefont {Matras}}, \bibinfo {author} {\bibfnamefont {S.}~\bibnamefont {Ricaud}}, \bibinfo {author} {\bibfnamefont {V.}~\bibnamefont {Leroux}}, \bibinfo {author} {\bibfnamefont {C.}~\bibnamefont {Richard}}, \bibinfo {author} {\bibfnamefont {F.}~\bibnamefont {Lureau}}, \bibinfo {author} {\bibfnamefont {A.}~\bibnamefont {Baleanu}}, \bibinfo {author} {\bibfnamefont {R.}~\bibnamefont {Banici}}, \bibinfo {author} {\bibfnamefont {A.}~\bibnamefont
  {Gradinariu}}, \bibinfo {author} {\bibfnamefont {C.}~\bibnamefont {Caldararu}}, \bibinfo {author} {\bibfnamefont {C.}~\bibnamefont {Capiteanu}}, \bibinfo {author} {\bibfnamefont {A.}~\bibnamefont {Naziru}}, \bibinfo {author} {\bibfnamefont {B.}~\bibnamefont {Diaconescu}}, \bibinfo {author} {\bibfnamefont {V.}~\bibnamefont {Iancu}}, \bibinfo {author} {\bibfnamefont {R.}~\bibnamefont {Dabu}}, \bibinfo {author} {\bibfnamefont {D.}~\bibnamefont {Ursescu}}, \bibinfo {author} {\bibfnamefont {I.}~\bibnamefont {Dancus}}, \bibinfo {author} {\bibfnamefont {C.~A.}\ \bibnamefont {Ur}}, \bibinfo {author} {\bibfnamefont {K.~A.}\ \bibnamefont {Tanaka}}, \ and\ \bibinfo {author} {\bibfnamefont {N.~V.}\ \bibnamefont {Zamfir}},\ }\href {\doibase 10.1017/hpl.2022.11} {\bibfield  {journal} {\bibinfo  {journal} {High Power Laser Science and Engineering}\ }\textbf {\bibinfo {volume} {10}} (\bibinfo {year} {2022}),\ 10.1017/hpl.2022.11}\BibitemShut {NoStop}%
\bibitem [{\citenamefont {Wang}\ \emph {et~al.}(2021{\natexlab{a}})\citenamefont {Wang}, \citenamefont {Zhou}, \citenamefont {Liu},\ and\ \citenamefont {Wang}}]{PhysRevLett.127.052501}%
  \BibitemOpen
  \bibfield  {author} {\bibinfo {author} {\bibfnamefont {W.}~\bibnamefont {Wang}}, \bibinfo {author} {\bibfnamefont {J.}~\bibnamefont {Zhou}}, \bibinfo {author} {\bibfnamefont {B.}~\bibnamefont {Liu}}, \ and\ \bibinfo {author} {\bibfnamefont {X.}~\bibnamefont {Wang}},\ }\href {\doibase 10.1103/PhysRevLett.127.052501} {\bibfield  {journal} {\bibinfo  {journal} {Phys. Rev. Lett.}\ }\textbf {\bibinfo {volume} {127}},\ \bibinfo {pages} {052501} (\bibinfo {year} {2021}{\natexlab{a}})}\BibitemShut {NoStop}%
\bibitem [{\citenamefont {Lv}\ \emph {et~al.}(2019)\citenamefont {Lv}, \citenamefont {Duan},\ and\ \citenamefont {Liu}}]{PhysRevC.100.064610}%
  \BibitemOpen
  \bibfield  {author} {\bibinfo {author} {\bibfnamefont {W.}~\bibnamefont {Lv}}, \bibinfo {author} {\bibfnamefont {H.}~\bibnamefont {Duan}}, \ and\ \bibinfo {author} {\bibfnamefont {J.}~\bibnamefont {Liu}},\ }\href {\doibase 10.1103/PhysRevC.100.064610} {\bibfield  {journal} {\bibinfo  {journal} {Phys. Rev. C}\ }\textbf {\bibinfo {volume} {100}},\ \bibinfo {pages} {064610} (\bibinfo {year} {2019})}\BibitemShut {NoStop}%
\bibitem [{\citenamefont {Ghinescu}\ and\ \citenamefont {Delion}(2020)}]{PhysRevC.101.044304}%
  \BibitemOpen
  \bibfield  {author} {\bibinfo {author} {\bibfnamefont {S.~A.}\ \bibnamefont {Ghinescu}}\ and\ \bibinfo {author} {\bibfnamefont {D.~S.}\ \bibnamefont {Delion}},\ }\href {\doibase 10.1103/PhysRevC.101.044304} {\bibfield  {journal} {\bibinfo  {journal} {Phys. Rev. C}\ }\textbf {\bibinfo {volume} {101}},\ \bibinfo {pages} {044304} (\bibinfo {year} {2020})}\BibitemShut {NoStop}%
\bibitem [{\citenamefont {von~der Wense}\ \emph {et~al.}(2020)\citenamefont {von~der Wense}, \citenamefont {Bilous}, \citenamefont {Seiferle}, \citenamefont {Stellmer}, \citenamefont {Weitenberg}, \citenamefont {Thirolf}, \citenamefont {P{\'a}lffy},\ and\ \citenamefont {Kazakov}}]{von2020theory}%
  \BibitemOpen
  \bibfield  {author} {\bibinfo {author} {\bibfnamefont {L.}~\bibnamefont {von~der Wense}}, \bibinfo {author} {\bibfnamefont {P.~V.}\ \bibnamefont {Bilous}}, \bibinfo {author} {\bibfnamefont {B.}~\bibnamefont {Seiferle}}, \bibinfo {author} {\bibfnamefont {S.}~\bibnamefont {Stellmer}}, \bibinfo {author} {\bibfnamefont {J.}~\bibnamefont {Weitenberg}}, \bibinfo {author} {\bibfnamefont {P.~G.}\ \bibnamefont {Thirolf}}, \bibinfo {author} {\bibfnamefont {A.}~\bibnamefont {P{\'a}lffy}}, \ and\ \bibinfo {author} {\bibfnamefont {G.}~\bibnamefont {Kazakov}},\ }\href@noop {} {\bibfield  {journal} {\bibinfo  {journal} {The European Physical Journal A}\ }\textbf {\bibinfo {volume} {56}},\ \bibinfo {pages} {1} (\bibinfo {year} {2020})}\BibitemShut {NoStop}%
\bibitem [{\citenamefont {Wang}(2022)}]{PhysRevC.106.024606}%
  \BibitemOpen
  \bibfield  {author} {\bibinfo {author} {\bibfnamefont {X.}~\bibnamefont {Wang}},\ }\href {\doibase 10.1103/PhysRevC.106.024606} {\bibfield  {journal} {\bibinfo  {journal} {Phys. Rev. C}\ }\textbf {\bibinfo {volume} {106}},\ \bibinfo {pages} {024606} (\bibinfo {year} {2022})}\BibitemShut {NoStop}%
\bibitem [{\citenamefont {Bekx}\ \emph {et~al.}(2022)\citenamefont {Bekx}, \citenamefont {Lindsey}, \citenamefont {Glenzer},\ and\ \citenamefont {Schlesinger}}]{PhysRevC.105.054001}%
  \BibitemOpen
  \bibfield  {author} {\bibinfo {author} {\bibfnamefont {J.~J.}\ \bibnamefont {Bekx}}, \bibinfo {author} {\bibfnamefont {M.~L.}\ \bibnamefont {Lindsey}}, \bibinfo {author} {\bibfnamefont {S.~H.}\ \bibnamefont {Glenzer}}, \ and\ \bibinfo {author} {\bibfnamefont {K.-G.}\ \bibnamefont {Schlesinger}},\ }\href {\doibase 10.1103/PhysRevC.105.054001} {\bibfield  {journal} {\bibinfo  {journal} {Phys. Rev. C}\ }\textbf {\bibinfo {volume} {105}},\ \bibinfo {pages} {054001} (\bibinfo {year} {2022})}\BibitemShut {NoStop}%
\bibitem [{\citenamefont {Qi}\ \emph {et~al.}(2019)\citenamefont {Qi}, \citenamefont {Li}, \citenamefont {Xu}, \citenamefont {Fu},\ and\ \citenamefont {Wang}}]{PhysRevC.99.044610}%
  \BibitemOpen
  \bibfield  {author} {\bibinfo {author} {\bibfnamefont {J.}~\bibnamefont {Qi}}, \bibinfo {author} {\bibfnamefont {T.}~\bibnamefont {Li}}, \bibinfo {author} {\bibfnamefont {R.}~\bibnamefont {Xu}}, \bibinfo {author} {\bibfnamefont {L.}~\bibnamefont {Fu}}, \ and\ \bibinfo {author} {\bibfnamefont {X.}~\bibnamefont {Wang}},\ }\href {\doibase 10.1103/PhysRevC.99.044610} {\bibfield  {journal} {\bibinfo  {journal} {Phys. Rev. C}\ }\textbf {\bibinfo {volume} {99}},\ \bibinfo {pages} {044610} (\bibinfo {year} {2019})}\BibitemShut {NoStop}%
\bibitem [{\citenamefont {Queisser}\ and\ \citenamefont {Sch\"utzhold}(2019)}]{PhysRevC.100.041601}%
  \BibitemOpen
  \bibfield  {author} {\bibinfo {author} {\bibfnamefont {F.}~\bibnamefont {Queisser}}\ and\ \bibinfo {author} {\bibfnamefont {R.}~\bibnamefont {Sch\"utzhold}},\ }\href {\doibase 10.1103/PhysRevC.100.041601} {\bibfield  {journal} {\bibinfo  {journal} {Phys. Rev. C}\ }\textbf {\bibinfo {volume} {100}},\ \bibinfo {pages} {041601} (\bibinfo {year} {2019})}\BibitemShut {NoStop}%
\bibitem [{\citenamefont {Li}\ and\ \citenamefont {Wang}(2021)}]{Li2021}%
  \BibitemOpen
  \bibfield  {author} {\bibinfo {author} {\bibfnamefont {T.}~\bibnamefont {Li}}\ and\ \bibinfo {author} {\bibfnamefont {X.}~\bibnamefont {Wang}},\ }\href {\doibase 10.1088/1361-6471/ac1712} {\bibfield  {journal} {\bibinfo  {journal} {Journal of Physics G: Nuclear and Particle Physics}\ }\textbf {\bibinfo {volume} {48}},\ \bibinfo {pages} {095105} (\bibinfo {year} {2021})}\BibitemShut {NoStop}%
\bibitem [{\citenamefont {Liu}\ \emph {et~al.}(2021)\citenamefont {Liu}, \citenamefont {Duan}, \citenamefont {Ye},\ and\ \citenamefont {Liu}}]{PhysRevC.104.044614}%
  \BibitemOpen
  \bibfield  {author} {\bibinfo {author} {\bibfnamefont {S.}~\bibnamefont {Liu}}, \bibinfo {author} {\bibfnamefont {H.}~\bibnamefont {Duan}}, \bibinfo {author} {\bibfnamefont {D.}~\bibnamefont {Ye}}, \ and\ \bibinfo {author} {\bibfnamefont {J.}~\bibnamefont {Liu}},\ }\href {\doibase 10.1103/PhysRevC.104.044614} {\bibfield  {journal} {\bibinfo  {journal} {Phys. Rev. C}\ }\textbf {\bibinfo {volume} {104}},\ \bibinfo {pages} {044614} (\bibinfo {year} {2021})}\BibitemShut {NoStop}%
\bibitem [{\citenamefont {Lv}\ \emph {et~al.}(2022)\citenamefont {Lv}, \citenamefont {Wu}, \citenamefont {Duan}, \citenamefont {Liu},\ and\ \citenamefont {Liu}}]{Lv2022}%
  \BibitemOpen
  \bibfield  {author} {\bibinfo {author} {\bibfnamefont {W.}~\bibnamefont {Lv}}, \bibinfo {author} {\bibfnamefont {B.}~\bibnamefont {Wu}}, \bibinfo {author} {\bibfnamefont {H.}~\bibnamefont {Duan}}, \bibinfo {author} {\bibfnamefont {S.}~\bibnamefont {Liu}}, \ and\ \bibinfo {author} {\bibfnamefont {J.}~\bibnamefont {Liu}},\ }\href {\doibase 10.1140/epja/s10050-022-00697-8} {\bibfield  {journal} {\bibinfo  {journal} {The European Physical Journal A}\ }\textbf {\bibinfo {volume} {58}} (\bibinfo {year} {2022}),\ 10.1140/epja/s10050-022-00697-8}\BibitemShut {NoStop}%
\bibitem [{\citenamefont {Cheng}\ \emph {et~al.}(2023)\citenamefont {Cheng}, \citenamefont {Zhang}, \citenamefont {Xiao}, \citenamefont {Deng},\ and\ \citenamefont {Yu}}]{cheng2023laserassisted}%
  \BibitemOpen
  \bibfield  {author} {\bibinfo {author} {\bibfnamefont {J.-H.}\ \bibnamefont {Cheng}}, \bibinfo {author} {\bibfnamefont {W.-Y.}\ \bibnamefont {Zhang}}, \bibinfo {author} {\bibfnamefont {Q.}~\bibnamefont {Xiao}}, \bibinfo {author} {\bibfnamefont {J.-G.}\ \bibnamefont {Deng}}, \ and\ \bibinfo {author} {\bibfnamefont {T.-P.}\ \bibnamefont {Yu}},\ }\href@noop {} {\enquote {\bibinfo {title} {Laser-assisted deformed $\alpha$ decay of the ground state even-even nuclei},}\ } (\bibinfo {year} {2023}),\ \Eprint {http://arxiv.org/abs/2307.02095} {arXiv:2307.02095 [nucl-th]} \BibitemShut {NoStop}%
\bibitem [{\citenamefont {Xu}\ \emph {et~al.}(2023)\citenamefont {Xu}, \citenamefont {Tang}, \citenamefont {Wang}, \citenamefont {Li}, \citenamefont {Li}, \citenamefont {Cappellaro},\ and\ \citenamefont {Li}}]{PhysRevA.108.L021502}%
  \BibitemOpen
  \bibfield  {author} {\bibinfo {author} {\bibfnamefont {H.}~\bibnamefont {Xu}}, \bibinfo {author} {\bibfnamefont {H.}~\bibnamefont {Tang}}, \bibinfo {author} {\bibfnamefont {G.}~\bibnamefont {Wang}}, \bibinfo {author} {\bibfnamefont {C.}~\bibnamefont {Li}}, \bibinfo {author} {\bibfnamefont {B.}~\bibnamefont {Li}}, \bibinfo {author} {\bibfnamefont {P.}~\bibnamefont {Cappellaro}}, \ and\ \bibinfo {author} {\bibfnamefont {J.}~\bibnamefont {Li}},\ }\href {\doibase 10.1103/PhysRevA.108.L021502} {\bibfield  {journal} {\bibinfo  {journal} {Phys. Rev. A}\ }\textbf {\bibinfo {volume} {108}},\ \bibinfo {pages} {L021502} (\bibinfo {year} {2023})}\BibitemShut {NoStop}%
\bibitem [{\citenamefont {Spohr}\ \emph {et~al.}(2023)\citenamefont {Spohr}, \citenamefont {Doria}, \citenamefont {Baran}, \citenamefont {Cernaianu}, \citenamefont {Ghenuche}, \citenamefont {Nastasa}, \citenamefont {O’Donnell}, \citenamefont {S{\"o}derstr{\"o}m}, \citenamefont {Tudor},\ and\ \citenamefont {Ur}}]{Spohr:2023koj}%
  \BibitemOpen
  \bibfield  {author} {\bibinfo {author} {\bibfnamefont {K.}~\bibnamefont {Spohr}}, \bibinfo {author} {\bibfnamefont {D.}~\bibnamefont {Doria}}, \bibinfo {author} {\bibfnamefont {V.}~\bibnamefont {Baran}}, \bibinfo {author} {\bibfnamefont {M.}~\bibnamefont {Cernaianu}}, \bibinfo {author} {\bibfnamefont {P.}~\bibnamefont {Ghenuche}}, \bibinfo {author} {\bibfnamefont {V.}~\bibnamefont {Nastasa}}, \bibinfo {author} {\bibfnamefont {D.}~\bibnamefont {O’Donnell}}, \bibinfo {author} {\bibfnamefont {P.-A.}\ \bibnamefont {S{\"o}derstr{\"o}m}}, \bibinfo {author} {\bibfnamefont {L.}~\bibnamefont {Tudor}}, \ and\ \bibinfo {author} {\bibfnamefont {C.}~\bibnamefont {Ur}},\ }\href@noop {} {\bibfield  {journal} {\bibinfo  {journal} {The European Physical Journal A}\ }\textbf {\bibinfo {volume} {59}},\ \bibinfo {pages} {281} (\bibinfo {year} {2023})}\BibitemShut {NoStop}%
\bibitem [{\citenamefont {Wu}\ \emph {et~al.}(2023)\citenamefont {Wu}, \citenamefont {Fan}, \citenamefont {Ye}, \citenamefont {Ye}, \citenamefont {Gao}, \citenamefont {Yu}, \citenamefont {Xu}, \citenamefont {Zhang},\ and\ \citenamefont {Liu}}]{Wu:2023vti}%
  \BibitemOpen
  \bibfield  {author} {\bibinfo {author} {\bibfnamefont {B.}~\bibnamefont {Wu}}, \bibinfo {author} {\bibfnamefont {Z.}~\bibnamefont {Fan}}, \bibinfo {author} {\bibfnamefont {D.}~\bibnamefont {Ye}}, \bibinfo {author} {\bibfnamefont {T.}~\bibnamefont {Ye}}, \bibinfo {author} {\bibfnamefont {C.}~\bibnamefont {Gao}}, \bibinfo {author} {\bibfnamefont {C.}~\bibnamefont {Yu}}, \bibinfo {author} {\bibfnamefont {X.}~\bibnamefont {Xu}}, \bibinfo {author} {\bibfnamefont {C.}~\bibnamefont {Zhang}}, \ and\ \bibinfo {author} {\bibfnamefont {J.}~\bibnamefont {Liu}},\ }\href@noop {} {\  (\bibinfo {year} {2023})},\ \Eprint {http://arxiv.org/abs/2312.16777} {arXiv:2312.16777 [nucl-th]} \BibitemShut {NoStop}%
\bibitem [{\citenamefont {Yang}\ \emph {et~al.}(2024)\citenamefont {Yang}, \citenamefont {Spohr}, \citenamefont {Cernaianu}, \citenamefont {Doria}, \citenamefont {Ghenuche},\ and\ \citenamefont {Horny}}]{our_iso}%
  \BibitemOpen
  \bibfield  {author} {\bibinfo {author} {\bibfnamefont {C.~J.}\ \bibnamefont {Yang}}, \bibinfo {author} {\bibfnamefont {K.~M.}\ \bibnamefont {Spohr}}, \bibinfo {author} {\bibfnamefont {M.}~\bibnamefont {Cernaianu}}, \bibinfo {author} {\bibfnamefont {D.}~\bibnamefont {Doria}}, \bibinfo {author} {\bibfnamefont {P.}~\bibnamefont {Ghenuche}}, \ and\ \bibinfo {author} {\bibfnamefont {V.}~\bibnamefont {Horny}},\ }\href@noop {} {\  (\bibinfo {year} {2024})},\ \Eprint {http://arxiv.org/abs/2404.07909} {arXiv:2404.07909 [nucl-th]} \BibitemShut {NoStop}%
\bibitem [{\citenamefont {{Einstein}}(1916)}]{1916DPhyG..18..318E}%
  \BibitemOpen
  \bibfield  {author} {\bibinfo {author} {\bibfnamefont {A.}~\bibnamefont {{Einstein}}},\ }\href@noop {} {\bibfield  {journal} {\bibinfo  {journal} {Deutsche Physikalische Gesellschaft}\ }\textbf {\bibinfo {volume} {18}},\ \bibinfo {pages} {318} (\bibinfo {year} {1916})}\BibitemShut {NoStop}%
\bibitem [{\citenamefont {Baldwin}\ \emph {et~al.}(1962)\citenamefont {Baldwin}, \citenamefont {Neissel},\ and\ \citenamefont {Tonks}}]{baldwin1962}%
  \BibitemOpen
  \bibfield  {author} {\bibinfo {author} {\bibfnamefont {G.~C.}\ \bibnamefont {Baldwin}}, \bibinfo {author} {\bibfnamefont {J.~P.}\ \bibnamefont {Neissel}}, \ and\ \bibinfo {author} {\bibfnamefont {L.}~\bibnamefont {Tonks}},\ }\href@noop {} {} (\bibinfo {year} {1962})\BibitemShut {NoStop}%
\bibitem [{\citenamefont {Mughabghab}\ and\ \citenamefont {Garber}(1973)}]{Mughabghab1973}%
  \BibitemOpen
  \bibfield  {author} {\bibinfo {author} {\bibfnamefont {S.}~\bibnamefont {Mughabghab}}\ and\ \bibinfo {author} {\bibfnamefont {D.}~\bibnamefont {Garber}},\ }\href {\doibase 10.2172/4335199} {\emph {\bibinfo {title} {Neutron cross sections. Volume I. Resonance parameters}}},\ \bibinfo {type} {Tech. Rep.}\ (\bibinfo {year} {1973})\BibitemShut {NoStop}%
\bibitem [{\citenamefont {Baldwin}\ \emph {et~al.}(1963)\citenamefont {Baldwin}, \citenamefont {Neissel}, \citenamefont {Tonks}, \citenamefont {Vali},\ and\ \citenamefont {Vali}}]{1444442}%
  \BibitemOpen
  \bibfield  {author} {\bibinfo {author} {\bibfnamefont {G.}~\bibnamefont {Baldwin}}, \bibinfo {author} {\bibfnamefont {J.}~\bibnamefont {Neissel}}, \bibinfo {author} {\bibfnamefont {L.}~\bibnamefont {Tonks}}, \bibinfo {author} {\bibfnamefont {V.}~\bibnamefont {Vali}}, \ and\ \bibinfo {author} {\bibfnamefont {W.}~\bibnamefont {Vali}},\ }\href {\doibase 10.1109/PROC.1963.2512} {\bibfield  {journal} {\bibinfo  {journal} {Proceedings of the IEEE}\ }\textbf {\bibinfo {volume} {51}},\ \bibinfo {pages} {1247} (\bibinfo {year} {1963})}\BibitemShut {NoStop}%
\bibitem [{\citenamefont {Rivlin}(1963)}]{r1963a}%
  \BibitemOpen
  \bibfield  {author} {\bibinfo {author} {\bibfnamefont {L.~A.}\ \bibnamefont {Rivlin}},\ }\href@noop {} {\bibfield  {journal} {\bibinfo  {journal} {Vopr. Radioelektron. 6, 42–49}\ } (\bibinfo {year} {1963})}\BibitemShut {NoStop}%
\bibitem [{\citenamefont {Gol'danskii}\ and\ \citenamefont {Yu.}(1973)}]{g1973a}%
  \BibitemOpen
  \bibfield  {author} {\bibinfo {author} {\bibfnamefont {V.~I.}\ \bibnamefont {Gol'danskii}}\ and\ \bibinfo {author} {\bibfnamefont {K.}~\bibnamefont {Yu.}},\ }\href@noop {} {\bibfield  {journal} {\bibinfo  {journal} {Zh. Eksp. Teor. Fiz. 64, 90–107}\ } (\bibinfo {year} {1973})}\BibitemShut {NoStop}%
\bibitem [{\citenamefont {Letokhov}(1973)}]{l1973b}%
  \BibitemOpen
  \bibfield  {author} {\bibinfo {author} {\bibfnamefont {V.~S.}\ \bibnamefont {Letokhov}},\ }\href@noop {} {\bibfield  {journal} {\bibinfo  {journal} {Zh. Eksp. Teor. Fiz. 64, 1555–1567}\ } (\bibinfo {year} {1973})}\BibitemShut {NoStop}%
\bibitem [{\citenamefont {Marcuse}(1963)}]{1444212}%
  \BibitemOpen
  \bibfield  {author} {\bibinfo {author} {\bibfnamefont {D.}~\bibnamefont {Marcuse}},\ }\href {\doibase 10.1109/PROC.1963.2282} {\bibfield  {journal} {\bibinfo  {journal} {Proceedings of the IEEE}\ }\textbf {\bibinfo {volume} {51}},\ \bibinfo {pages} {849} (\bibinfo {year} {1963})}\BibitemShut {NoStop}%
\bibitem [{\citenamefont {Byrne}\ \emph {et~al.}(1974)\citenamefont {Byrne}, \citenamefont {Peters},\ and\ \citenamefont {Allen}}]{Byrne1974}%
  \BibitemOpen
  \bibfield  {author} {\bibinfo {author} {\bibfnamefont {J.}~\bibnamefont {Byrne}}, \bibinfo {author} {\bibfnamefont {G.~I.}\ \bibnamefont {Peters}}, \ and\ \bibinfo {author} {\bibfnamefont {L.}~\bibnamefont {Allen}},\ }\href {\doibase 10.1364/ao.13.002499} {\bibfield  {journal} {\bibinfo  {journal} {Applied Optics}\ }\textbf {\bibinfo {volume} {13}},\ \bibinfo {pages} {2499} (\bibinfo {year} {1974})}\BibitemShut {NoStop}%
\bibitem [{\citenamefont {Walker}\ and\ \citenamefont {Dracoulis}(1999)}]{Walker1999}%
  \BibitemOpen
  \bibfield  {author} {\bibinfo {author} {\bibfnamefont {P.}~\bibnamefont {Walker}}\ and\ \bibinfo {author} {\bibfnamefont {G.}~\bibnamefont {Dracoulis}},\ }\href {\doibase 10.1038/19911} {\bibfield  {journal} {\bibinfo  {journal} {Nature}\ }\textbf {\bibinfo {volume} {399}},\ \bibinfo {pages} {35} (\bibinfo {year} {1999})}\BibitemShut {NoStop}%
\bibitem [{\citenamefont {Karamian}\ and\ \citenamefont {Carroll}(2007)}]{karamian2007prospects}%
  \BibitemOpen
  \bibfield  {author} {\bibinfo {author} {\bibfnamefont {S.}~\bibnamefont {Karamian}}\ and\ \bibinfo {author} {\bibfnamefont {J.}~\bibnamefont {Carroll}},\ }\href@noop {} {\bibfield  {journal} {\bibinfo  {journal} {Laser physics}\ }\textbf {\bibinfo {volume} {17}},\ \bibinfo {pages} {80} (\bibinfo {year} {2007})}\BibitemShut {NoStop}%
\bibitem [{\citenamefont {Hayes}\ \emph {et~al.}(2006)\citenamefont {Hayes}, \citenamefont {Cline}, \citenamefont {Wu}, \citenamefont {Ai}, \citenamefont {Amro}, \citenamefont {Beausang}, \citenamefont {Casten}, \citenamefont {Gerl}, \citenamefont {Hecht}, \citenamefont {Heinz}, \citenamefont {Hughes}, \citenamefont {Janssens}, \citenamefont {Lister}, \citenamefont {Macchiavelli}, \citenamefont {Meyer}, \citenamefont {Moore}, \citenamefont {Napiorkowski}, \citenamefont {Pardo}, \citenamefont {Schlegel}, \citenamefont {Seweryniak}, \citenamefont {Simon}, \citenamefont {Srebrny}, \citenamefont {Teng}, \citenamefont {Vetter},\ and\ \citenamefont {Wollersheim}}]{PhysRevLett.96.042505}%
  \BibitemOpen
  \bibfield  {author} {\bibinfo {author} {\bibfnamefont {A.~B.}\ \bibnamefont {Hayes}}, \bibinfo {author} {\bibfnamefont {D.}~\bibnamefont {Cline}}, \bibinfo {author} {\bibfnamefont {C.~Y.}\ \bibnamefont {Wu}}, \bibinfo {author} {\bibfnamefont {J.}~\bibnamefont {Ai}}, \bibinfo {author} {\bibfnamefont {H.}~\bibnamefont {Amro}}, \bibinfo {author} {\bibfnamefont {C.}~\bibnamefont {Beausang}}, \bibinfo {author} {\bibfnamefont {R.~F.}\ \bibnamefont {Casten}}, \bibinfo {author} {\bibfnamefont {J.}~\bibnamefont {Gerl}}, \bibinfo {author} {\bibfnamefont {A.~A.}\ \bibnamefont {Hecht}}, \bibinfo {author} {\bibfnamefont {A.}~\bibnamefont {Heinz}}, \bibinfo {author} {\bibfnamefont {R.}~\bibnamefont {Hughes}}, \bibinfo {author} {\bibfnamefont {R.~V.~F.}\ \bibnamefont {Janssens}}, \bibinfo {author} {\bibfnamefont {C.~J.}\ \bibnamefont {Lister}}, \bibinfo {author} {\bibfnamefont {A.~O.}\ \bibnamefont {Macchiavelli}}, \bibinfo {author} {\bibfnamefont {D.~A.}\ \bibnamefont {Meyer}}, \bibinfo {author} {\bibfnamefont {E.~F.}\
  \bibnamefont {Moore}}, \bibinfo {author} {\bibfnamefont {P.}~\bibnamefont {Napiorkowski}}, \bibinfo {author} {\bibfnamefont {R.~C.}\ \bibnamefont {Pardo}}, \bibinfo {author} {\bibfnamefont {C.}~\bibnamefont {Schlegel}}, \bibinfo {author} {\bibfnamefont {D.}~\bibnamefont {Seweryniak}}, \bibinfo {author} {\bibfnamefont {M.~W.}\ \bibnamefont {Simon}}, \bibinfo {author} {\bibfnamefont {J.}~\bibnamefont {Srebrny}}, \bibinfo {author} {\bibfnamefont {R.}~\bibnamefont {Teng}}, \bibinfo {author} {\bibfnamefont {K.}~\bibnamefont {Vetter}}, \ and\ \bibinfo {author} {\bibfnamefont {H.~J.}\ \bibnamefont {Wollersheim}},\ }\href {\doibase 10.1103/PhysRevLett.96.042505} {\bibfield  {journal} {\bibinfo  {journal} {Phys. Rev. Lett.}\ }\textbf {\bibinfo {volume} {96}},\ \bibinfo {pages} {042505} (\bibinfo {year} {2006})}\BibitemShut {NoStop}%
\bibitem [{\citenamefont {Georgiev}\ \emph {et~al.}(2006)\citenamefont {Georgiev}, \citenamefont {Stefanescu}, \citenamefont {Balabanski}, \citenamefont {Butler}, \citenamefont {Cederk\"{a}ll}, \citenamefont {Davinson}, \citenamefont {Delahaye}, \citenamefont {Fedosseev}, \citenamefont {Fraile}, \citenamefont {Franchoo}, \citenamefont {Gladnishki}, \citenamefont {Heyde}, \citenamefont {Huyse}, \citenamefont {Ivanov}, \citenamefont {Iwanicki}, \citenamefont {Kr\"{o}ll}, \citenamefont {K\"{o}ster}, \citenamefont {Lagoyannis}, \citenamefont {Bianco}, \citenamefont {de~Maesschalck}, \citenamefont {Saltarelli}, \citenamefont {Sieber}, \citenamefont {Smirnova}, \citenamefont {van Duppen}, \citenamefont {Warr}, \citenamefont {Wenander},\ and\ \citenamefont {van~de Walle}}]{rex2006first}%
  \BibitemOpen
  \bibfield  {author} {\bibinfo {author} {\bibfnamefont {G.}~\bibnamefont {Georgiev}}, \bibinfo {author} {\bibfnamefont {I.}~\bibnamefont {Stefanescu}}, \bibinfo {author} {\bibfnamefont {D.~I.}\ \bibnamefont {Balabanski}}, \bibinfo {author} {\bibfnamefont {P.}~\bibnamefont {Butler}}, \bibinfo {author} {\bibfnamefont {J.}~\bibnamefont {Cederk\"{a}ll}}, \bibinfo {author} {\bibfnamefont {T.}~\bibnamefont {Davinson}}, \bibinfo {author} {\bibfnamefont {P.}~\bibnamefont {Delahaye}}, \bibinfo {author} {\bibfnamefont {V.~N.}\ \bibnamefont {Fedosseev}}, \bibinfo {author} {\bibfnamefont {L.~M.}\ \bibnamefont {Fraile}}, \bibinfo {author} {\bibfnamefont {S.}~\bibnamefont {Franchoo}}, \bibinfo {author} {\bibfnamefont {K.}~\bibnamefont {Gladnishki}}, \bibinfo {author} {\bibfnamefont {K.}~\bibnamefont {Heyde}}, \bibinfo {author} {\bibfnamefont {M.}~\bibnamefont {Huyse}}, \bibinfo {author} {\bibfnamefont {O.}~\bibnamefont {Ivanov}}, \bibinfo {author} {\bibfnamefont {J.}~\bibnamefont {Iwanicki}}, \bibinfo {author}
  {\bibfnamefont {T.}~\bibnamefont {Kr\"{o}ll}}, \bibinfo {author} {\bibfnamefont {U.}~\bibnamefont {K\"{o}ster}}, \bibinfo {author} {\bibfnamefont {A.}~\bibnamefont {Lagoyannis}}, \bibinfo {author} {\bibfnamefont {G.~L.}\ \bibnamefont {Bianco}}, \bibinfo {author} {\bibfnamefont {A.}~\bibnamefont {de~Maesschalck}}, \bibinfo {author} {\bibfnamefont {A.}~\bibnamefont {Saltarelli}}, \bibinfo {author} {\bibfnamefont {T.}~\bibnamefont {Sieber}}, \bibinfo {author} {\bibfnamefont {N.}~\bibnamefont {Smirnova}}, \bibinfo {author} {\bibfnamefont {P.}~\bibnamefont {van Duppen}}, \bibinfo {author} {\bibfnamefont {N.}~\bibnamefont {Warr}}, \bibinfo {author} {\bibfnamefont {F.}~\bibnamefont {Wenander}}, \ and\ \bibinfo {author} {\bibfnamefont {J.}~\bibnamefont {van~de Walle}},\ }\href {\doibase 10.1142/s0218301306005095} {\bibfield  {journal} {\bibinfo  {journal} {International Journal of Modern Physics E}\ }\textbf {\bibinfo {volume} {15}},\ \bibinfo {pages} {1505–1512} (\bibinfo {year} {2006})}\BibitemShut {NoStop}%
\bibitem [{\citenamefont {Feng}\ \emph {et~al.}(2022)\citenamefont {Feng}, \citenamefont {Wang}, \citenamefont {Fu}, \citenamefont {Chen}, \citenamefont {Tan}, \citenamefont {Li}, \citenamefont {Wang}, \citenamefont {Li}, \citenamefont {Zhang}, \citenamefont {Ma},\ and\ \citenamefont {Zhang}}]{Feng2022}%
  \BibitemOpen
  \bibfield  {author} {\bibinfo {author} {\bibfnamefont {J.}~\bibnamefont {Feng}}, \bibinfo {author} {\bibfnamefont {W.}~\bibnamefont {Wang}}, \bibinfo {author} {\bibfnamefont {C.}~\bibnamefont {Fu}}, \bibinfo {author} {\bibfnamefont {L.}~\bibnamefont {Chen}}, \bibinfo {author} {\bibfnamefont {J.}~\bibnamefont {Tan}}, \bibinfo {author} {\bibfnamefont {Y.}~\bibnamefont {Li}}, \bibinfo {author} {\bibfnamefont {J.}~\bibnamefont {Wang}}, \bibinfo {author} {\bibfnamefont {Y.}~\bibnamefont {Li}}, \bibinfo {author} {\bibfnamefont {G.}~\bibnamefont {Zhang}}, \bibinfo {author} {\bibfnamefont {Y.}~\bibnamefont {Ma}}, \ and\ \bibinfo {author} {\bibfnamefont {J.}~\bibnamefont {Zhang}},\ }\href {\doibase 10.1103/physrevlett.128.052501} {\bibfield  {journal} {\bibinfo  {journal} {Physical Review Letters}\ }\textbf {\bibinfo {volume} {128}} (\bibinfo {year} {2022}),\ 10.1103/physrevlett.128.052501}\BibitemShut {NoStop}%
\bibitem [{\citenamefont {Feng}\ \emph {et~al.}(2023)\citenamefont {Feng}, \citenamefont {Li}, \citenamefont {Tan}, \citenamefont {Wang}, \citenamefont {Li}, \citenamefont {Zhang}, \citenamefont {Meng}, \citenamefont {Ge}, \citenamefont {Liu}, \citenamefont {Yan}, \citenamefont {Fu}, \citenamefont {Chen},\ and\ \citenamefont {Zhang}}]{Feng2023}%
  \BibitemOpen
  \bibfield  {author} {\bibinfo {author} {\bibfnamefont {J.}~\bibnamefont {Feng}}, \bibinfo {author} {\bibfnamefont {Y.}~\bibnamefont {Li}}, \bibinfo {author} {\bibfnamefont {J.}~\bibnamefont {Tan}}, \bibinfo {author} {\bibfnamefont {W.}~\bibnamefont {Wang}}, \bibinfo {author} {\bibfnamefont {Y.}~\bibnamefont {Li}}, \bibinfo {author} {\bibfnamefont {X.}~\bibnamefont {Zhang}}, \bibinfo {author} {\bibfnamefont {Y.}~\bibnamefont {Meng}}, \bibinfo {author} {\bibfnamefont {X.}~\bibnamefont {Ge}}, \bibinfo {author} {\bibfnamefont {F.}~\bibnamefont {Liu}}, \bibinfo {author} {\bibfnamefont {W.}~\bibnamefont {Yan}}, \bibinfo {author} {\bibfnamefont {C.}~\bibnamefont {Fu}}, \bibinfo {author} {\bibfnamefont {L.}~\bibnamefont {Chen}}, \ and\ \bibinfo {author} {\bibfnamefont {J.}~\bibnamefont {Zhang}},\ }\href {\doibase 10.1002/lpor.202300514} {\bibfield  {journal} {\bibinfo  {journal} {Laser \& Photonics Reviews}\ } (\bibinfo {year} {2023}),\ 10.1002/lpor.202300514}\BibitemShut {NoStop}%
\bibitem [{\citenamefont {Winterberg}(1986)}]{Winterberg1986}%
  \BibitemOpen
  \bibfield  {author} {\bibinfo {author} {\bibfnamefont {F.}~\bibnamefont {Winterberg}},\ }in\ \href {\doibase 10.1063/1.35822} {\emph {\bibinfo {booktitle} {{AIP} Conference Proceedings}}}\ (\bibinfo  {publisher} {{AIP}},\ \bibinfo {year} {1986})\BibitemShut {NoStop}%
\bibitem [{\citenamefont {Collins}\ \emph {et~al.}(1979{\natexlab{a}})\citenamefont {Collins}, \citenamefont {Olariu}, \citenamefont {Petrascu},\ and\ \citenamefont {Popescu}}]{PhysRevLett.42.1397}%
  \BibitemOpen
  \bibfield  {author} {\bibinfo {author} {\bibfnamefont {C.~B.}\ \bibnamefont {Collins}}, \bibinfo {author} {\bibfnamefont {S.}~\bibnamefont {Olariu}}, \bibinfo {author} {\bibfnamefont {M.}~\bibnamefont {Petrascu}}, \ and\ \bibinfo {author} {\bibfnamefont {I.}~\bibnamefont {Popescu}},\ }\href {\doibase 10.1103/PhysRevLett.42.1397} {\bibfield  {journal} {\bibinfo  {journal} {Phys. Rev. Lett.}\ }\textbf {\bibinfo {volume} {42}},\ \bibinfo {pages} {1397} (\bibinfo {year} {1979}{\natexlab{a}})}\BibitemShut {NoStop}%
\bibitem [{\citenamefont {Collins}\ \emph {et~al.}(1979{\natexlab{b}})\citenamefont {Collins}, \citenamefont {Olariu}, \citenamefont {Petrascu},\ and\ \citenamefont {Popescu}}]{PhysRevC.20.1942}%
  \BibitemOpen
  \bibfield  {author} {\bibinfo {author} {\bibfnamefont {C.~B.}\ \bibnamefont {Collins}}, \bibinfo {author} {\bibfnamefont {S.}~\bibnamefont {Olariu}}, \bibinfo {author} {\bibfnamefont {M.}~\bibnamefont {Petrascu}}, \ and\ \bibinfo {author} {\bibfnamefont {I.}~\bibnamefont {Popescu}},\ }\href {\doibase 10.1103/PhysRevC.20.1942} {\bibfield  {journal} {\bibinfo  {journal} {Phys. Rev. C}\ }\textbf {\bibinfo {volume} {20}},\ \bibinfo {pages} {1942} (\bibinfo {year} {1979}{\natexlab{b}})}\BibitemShut {NoStop}%
\bibitem [{\citenamefont {Olariu}\ \emph {et~al.}(1981{\natexlab{a}})\citenamefont {Olariu}, \citenamefont {Popescu},\ and\ \citenamefont {Collins}}]{PhysRevC.23.50}%
  \BibitemOpen
  \bibfield  {author} {\bibinfo {author} {\bibfnamefont {S.}~\bibnamefont {Olariu}}, \bibinfo {author} {\bibfnamefont {I.}~\bibnamefont {Popescu}}, \ and\ \bibinfo {author} {\bibfnamefont {C.~B.}\ \bibnamefont {Collins}},\ }\href {\doibase 10.1103/PhysRevC.23.50} {\bibfield  {journal} {\bibinfo  {journal} {Phys. Rev. C}\ }\textbf {\bibinfo {volume} {23}},\ \bibinfo {pages} {50} (\bibinfo {year} {1981}{\natexlab{a}})}\BibitemShut {NoStop}%
\bibitem [{\citenamefont {Olariu}\ \emph {et~al.}(1981{\natexlab{b}})\citenamefont {Olariu}, \citenamefont {Popescu},\ and\ \citenamefont {Collins}}]{PhysRevC.23.1007}%
  \BibitemOpen
  \bibfield  {author} {\bibinfo {author} {\bibfnamefont {S.}~\bibnamefont {Olariu}}, \bibinfo {author} {\bibfnamefont {I.}~\bibnamefont {Popescu}}, \ and\ \bibinfo {author} {\bibfnamefont {C.~B.}\ \bibnamefont {Collins}},\ }\href {\doibase 10.1103/PhysRevC.23.1007} {\bibfield  {journal} {\bibinfo  {journal} {Phys. Rev. C}\ }\textbf {\bibinfo {volume} {23}},\ \bibinfo {pages} {1007} (\bibinfo {year} {1981}{\natexlab{b}})}\BibitemShut {NoStop}%
\bibitem [{\citenamefont {Collins}\ \emph {et~al.}(1982)\citenamefont {Collins}, \citenamefont {Lee}, \citenamefont {Shemwell}, \citenamefont {DePaola}, \citenamefont {Olariu},\ and\ \citenamefont {Popescu}}]{Collins1982}%
  \BibitemOpen
  \bibfield  {author} {\bibinfo {author} {\bibfnamefont {C.~B.}\ \bibnamefont {Collins}}, \bibinfo {author} {\bibfnamefont {F.~W.}\ \bibnamefont {Lee}}, \bibinfo {author} {\bibfnamefont {D.~M.}\ \bibnamefont {Shemwell}}, \bibinfo {author} {\bibfnamefont {B.~D.}\ \bibnamefont {DePaola}}, \bibinfo {author} {\bibfnamefont {S.}~\bibnamefont {Olariu}}, \ and\ \bibinfo {author} {\bibfnamefont {I.~I.}\ \bibnamefont {Popescu}},\ }\href {\doibase 10.1063/1.331291} {\bibfield  {journal} {\bibinfo  {journal} {Journal of Applied Physics}\ }\textbf {\bibinfo {volume} {53}},\ \bibinfo {pages} {4645} (\bibinfo {year} {1982})}\BibitemShut {NoStop}%
\bibitem [{\citenamefont {G\"{o}ppert-Mayer}(1931)}]{GM}%
  \BibitemOpen
  \bibfield  {author} {\bibinfo {author} {\bibfnamefont {M.}~\bibnamefont {G\"{o}ppert-Mayer}},\ }\href {\doibase 10.1002/andp.19314010303} {\bibfield  {journal} {\bibinfo  {journal} {Annalen der Physik}\ }\textbf {\bibinfo {volume} {401}},\ \bibinfo {pages} {273} (\bibinfo {year} {1931})}\BibitemShut {NoStop}%
\bibitem [{\citenamefont {Lambropoulos}(1976)}]{LAMBROPOULOS197687}%
  \BibitemOpen
  \bibfield  {author} {\bibinfo {author} {\bibfnamefont {P.}~\bibnamefont {Lambropoulos}}\ }(\bibinfo  {publisher} {Academic Press},\ \bibinfo {year} {1976})\ pp.\ \bibinfo {pages} {87--164}\BibitemShut {NoStop}%
\bibitem [{\citenamefont {Friedrich}(2017)}]{Friedrich2017}%
  \BibitemOpen
  \bibfield  {author} {\bibinfo {author} {\bibfnamefont {H.}~\bibnamefont {Friedrich}},\ }\href {\doibase 10.1007/978-3-319-47769-5} {\emph {\bibinfo {title} {Theoretical Atomic Physics}}}\ (\bibinfo  {publisher} {Springer International Publishing},\ \bibinfo {year} {2017})\BibitemShut {NoStop}%
\bibitem [{\citenamefont {Delone}\ and\ \citenamefont {Krainov}(2000)}]{Delone2000}%
  \BibitemOpen
  \bibfield  {author} {\bibinfo {author} {\bibfnamefont {N.~B.}\ \bibnamefont {Delone}}\ and\ \bibinfo {author} {\bibfnamefont {V.~P.}\ \bibnamefont {Krainov}},\ }\href {\doibase 10.1007/978-3-642-57208-1} {\emph {\bibinfo {title} {Multiphoton Processes in Atoms}}}\ (\bibinfo  {publisher} {Springer Berlin Heidelberg},\ \bibinfo {year} {2000})\BibitemShut {NoStop}%
\bibitem [{\citenamefont {Grechukhin}(1963)}]{Grechukhin1963}%
  \BibitemOpen
  \bibfield  {author} {\bibinfo {author} {\bibfnamefont {D.}~\bibnamefont {Grechukhin}},\ }\href {\doibase 10.1016/0029-5582(63)90873-3} {\bibfield  {journal} {\bibinfo  {journal} {Nuclear Physics}\ }\textbf {\bibinfo {volume} {47}},\ \bibinfo {pages} {273–292} (\bibinfo {year} {1963})}\BibitemShut {NoStop}%
\bibitem [{\citenamefont {Eichler}(1974)}]{PhysRevA.9.1762}%
  \BibitemOpen
  \bibfield  {author} {\bibinfo {author} {\bibfnamefont {J.}~\bibnamefont {Eichler}},\ }\href {\doibase 10.1103/PhysRevA.9.1762} {\bibfield  {journal} {\bibinfo  {journal} {Phys. Rev. A}\ }\textbf {\bibinfo {volume} {9}},\ \bibinfo {pages} {1762} (\bibinfo {year} {1974})}\BibitemShut {NoStop}%
\bibitem [{\citenamefont {Friar}\ and\ \citenamefont {Rosen}(1974)}]{Friar1974}%
  \BibitemOpen
  \bibfield  {author} {\bibinfo {author} {\bibfnamefont {J.}~\bibnamefont {Friar}}\ and\ \bibinfo {author} {\bibfnamefont {M.}~\bibnamefont {Rosen}},\ }\href {\doibase 10.1016/0003-4916(74)90038-4} {\bibfield  {journal} {\bibinfo  {journal} {Annals of Physics}\ }\textbf {\bibinfo {volume} {87}},\ \bibinfo {pages} {289–326} (\bibinfo {year} {1974})}\BibitemShut {NoStop}%
\bibitem [{\citenamefont {Friar}(1975)}]{Friar1975}%
  \BibitemOpen
  \bibfield  {author} {\bibinfo {author} {\bibfnamefont {J.}~\bibnamefont {Friar}},\ }\href {\doibase 10.1016/0003-4916(75)90049-4} {\bibfield  {journal} {\bibinfo  {journal} {Annals of Physics}\ }\textbf {\bibinfo {volume} {95}},\ \bibinfo {pages} {170–201} (\bibinfo {year} {1975})}\BibitemShut {NoStop}%
\bibitem [{\citenamefont {Kramp}\ \emph {et~al.}(1987)\citenamefont {Kramp}, \citenamefont {Habs}, \citenamefont {Kroth}, \citenamefont {Music}, \citenamefont {Schirmer}, \citenamefont {Schwalm},\ and\ \citenamefont {Broude}}]{Kramp1987}%
  \BibitemOpen
  \bibfield  {author} {\bibinfo {author} {\bibfnamefont {J.}~\bibnamefont {Kramp}}, \bibinfo {author} {\bibfnamefont {D.}~\bibnamefont {Habs}}, \bibinfo {author} {\bibfnamefont {R.}~\bibnamefont {Kroth}}, \bibinfo {author} {\bibfnamefont {M.}~\bibnamefont {Music}}, \bibinfo {author} {\bibfnamefont {J.}~\bibnamefont {Schirmer}}, \bibinfo {author} {\bibfnamefont {D.}~\bibnamefont {Schwalm}}, \ and\ \bibinfo {author} {\bibfnamefont {C.}~\bibnamefont {Broude}},\ }\href {\doibase 10.1016/0375-9474(87)90625-7} {\bibfield  {journal} {\bibinfo  {journal} {Nuclear Physics A}\ }\textbf {\bibinfo {volume} {474}},\ \bibinfo {pages} {412–450} (\bibinfo {year} {1987})}\BibitemShut {NoStop}%
\bibitem [{\citenamefont {Gorodetzky}\ \emph {et~al.}(1961)\citenamefont {Gorodetzky}, \citenamefont {Sutter}, \citenamefont {Armbruster}, \citenamefont {Chevallier}, \citenamefont {Mennrath}, \citenamefont {Scheibling},\ and\ \citenamefont {Yoccoz}}]{PhysRevLett.7.170}%
  \BibitemOpen
  \bibfield  {author} {\bibinfo {author} {\bibfnamefont {S.}~\bibnamefont {Gorodetzky}}, \bibinfo {author} {\bibfnamefont {G.}~\bibnamefont {Sutter}}, \bibinfo {author} {\bibfnamefont {R.}~\bibnamefont {Armbruster}}, \bibinfo {author} {\bibfnamefont {P.}~\bibnamefont {Chevallier}}, \bibinfo {author} {\bibfnamefont {P.}~\bibnamefont {Mennrath}}, \bibinfo {author} {\bibfnamefont {F.}~\bibnamefont {Scheibling}}, \ and\ \bibinfo {author} {\bibfnamefont {J.}~\bibnamefont {Yoccoz}},\ }\href {\doibase 10.1103/PhysRevLett.7.170} {\bibfield  {journal} {\bibinfo  {journal} {Phys. Rev. Lett.}\ }\textbf {\bibinfo {volume} {7}},\ \bibinfo {pages} {170} (\bibinfo {year} {1961})}\BibitemShut {NoStop}%
\bibitem [{\citenamefont {Alburger}\ and\ \citenamefont {Parker}(1964)}]{Alburger1964}%
  \BibitemOpen
  \bibfield  {author} {\bibinfo {author} {\bibfnamefont {D.~E.}\ \bibnamefont {Alburger}}\ and\ \bibinfo {author} {\bibfnamefont {P.~D.}\ \bibnamefont {Parker}},\ }\href {\doibase 10.1103/physrev.135.b294} {\bibfield  {journal} {\bibinfo  {journal} {Physical Review}\ }\textbf {\bibinfo {volume} {135}},\ \bibinfo {pages} {B294–B300} (\bibinfo {year} {1964})}\BibitemShut {NoStop}%
\bibitem [{\citenamefont {Harihar}\ \emph {et~al.}(1970)\citenamefont {Harihar}, \citenamefont {Ullman},\ and\ \citenamefont {Wu}}]{Harihar1970}%
  \BibitemOpen
  \bibfield  {author} {\bibinfo {author} {\bibfnamefont {P.}~\bibnamefont {Harihar}}, \bibinfo {author} {\bibfnamefont {J.~D.}\ \bibnamefont {Ullman}}, \ and\ \bibinfo {author} {\bibfnamefont {C.~S.}\ \bibnamefont {Wu}},\ }\href {\doibase 10.1103/physrevc.2.462} {\bibfield  {journal} {\bibinfo  {journal} {Physical Review C}\ }\textbf {\bibinfo {volume} {2}},\ \bibinfo {pages} {462–467} (\bibinfo {year} {1970})}\BibitemShut {NoStop}%
\bibitem [{\citenamefont {Nakayama}(1973)}]{Nakayama1973}%
  \BibitemOpen
  \bibfield  {author} {\bibinfo {author} {\bibfnamefont {Y.}~\bibnamefont {Nakayama}},\ }\href {\doibase 10.1103/physrevc.7.322} {\bibfield  {journal} {\bibinfo  {journal} {Physical Review C}\ }\textbf {\bibinfo {volume} {7}},\ \bibinfo {pages} {322–330} (\bibinfo {year} {1973})}\BibitemShut {NoStop}%
\bibitem [{\citenamefont {Vanderleeden}\ and\ \citenamefont {Jastram}(1970)}]{Vanderleeden:1970zn}%
  \BibitemOpen
  \bibfield  {author} {\bibinfo {author} {\bibfnamefont {J.~C.}\ \bibnamefont {Vanderleeden}}\ and\ \bibinfo {author} {\bibfnamefont {P.~S.}\ \bibnamefont {Jastram}},\ }\href {\doibase 10.1103/PhysRevC.1.1025} {\bibfield  {journal} {\bibinfo  {journal} {Phys. Rev. C}\ }\textbf {\bibinfo {volume} {1}},\ \bibinfo {pages} {1025} (\bibinfo {year} {1970})}\BibitemShut {NoStop}%
\bibitem [{\citenamefont {Schirmer}\ \emph {et~al.}(1984)\citenamefont {Schirmer}, \citenamefont {Habs}, \citenamefont {Kroth}, \citenamefont {Kwong}, \citenamefont {Schwalm}, \citenamefont {Zirnbauer},\ and\ \citenamefont {Broude}}]{PhysRevLett.53.1897}%
  \BibitemOpen
  \bibfield  {author} {\bibinfo {author} {\bibfnamefont {J.}~\bibnamefont {Schirmer}}, \bibinfo {author} {\bibfnamefont {D.}~\bibnamefont {Habs}}, \bibinfo {author} {\bibfnamefont {R.}~\bibnamefont {Kroth}}, \bibinfo {author} {\bibfnamefont {N.}~\bibnamefont {Kwong}}, \bibinfo {author} {\bibfnamefont {D.}~\bibnamefont {Schwalm}}, \bibinfo {author} {\bibfnamefont {M.}~\bibnamefont {Zirnbauer}}, \ and\ \bibinfo {author} {\bibfnamefont {C.}~\bibnamefont {Broude}},\ }\href {\doibase 10.1103/PhysRevLett.53.1897} {\bibfield  {journal} {\bibinfo  {journal} {Phys. Rev. Lett.}\ }\textbf {\bibinfo {volume} {53}},\ \bibinfo {pages} {1897} (\bibinfo {year} {1984})}\BibitemShut {NoStop}%
\bibitem [{\citenamefont {S\"oderstr\"om}\ \emph {et~al.}(2020)\citenamefont {S\"oderstr\"om} \emph {et~al.}}]{Soderstrom:2020iaz}%
  \BibitemOpen
  \bibfield  {author} {\bibinfo {author} {\bibfnamefont {P.~A.}\ \bibnamefont {S\"oderstr\"om}} \emph {et~al.},\ }\href {\doibase 10.1038/s41467-020-16787-4} {\bibfield  {journal} {\bibinfo  {journal} {Nature Commun.}\ }\textbf {\bibinfo {volume} {11}},\ \bibinfo {pages} {3242} (\bibinfo {year} {2020})},\ \Eprint {http://arxiv.org/abs/2001.00554} {arXiv:2001.00554 [nucl-ex]} \BibitemShut {NoStop}%
\bibitem [{\citenamefont {Freire-Fernández}\ \emph {et~al.}(2023)\citenamefont {Freire-Fernández}, \citenamefont {Korten}, \citenamefont {Chen}, \citenamefont {Litvinov}, \citenamefont {Litvinov}, \citenamefont {Sanjari}, \citenamefont {Weick}, \citenamefont {Akinci}, \citenamefont {Albers}, \citenamefont {Armstrong}, \citenamefont {Banerjee}, \citenamefont {Blaum}, \citenamefont {Brandau}, \citenamefont {Brown}, \citenamefont {Bruno}, \citenamefont {Carroll}, \citenamefont {Chen}, \citenamefont {Chiara}, \citenamefont {Cortes}, \citenamefont {Dellmann}, \citenamefont {Dillmann}, \citenamefont {Dmytriiev}, \citenamefont {Forstner}, \citenamefont {Geissel}, \citenamefont {Glorius}, \citenamefont {Görgen}, \citenamefont {Górska}, \citenamefont {Griffin}, \citenamefont {Gumberidze}, \citenamefont {Harayama}, \citenamefont {Hess}, \citenamefont {Hubbard}, \citenamefont {Ide}, \citenamefont {John}, \citenamefont {Joseph}, \citenamefont {Jurado}, \citenamefont {Kalaydjieva}, \citenamefont {Kanika},
  \citenamefont {Kondev}, \citenamefont {Koseoglou}, \citenamefont {Kosir}, \citenamefont {Kozhuharov}, \citenamefont {Kulikov}, \citenamefont {Leckenby}, \citenamefont {Lorenz}, \citenamefont {Marsh}, \citenamefont {Mistry}, \citenamefont {Ozawa}, \citenamefont {Pietralla}, \citenamefont {Podolyák}, \citenamefont {Polettini}, \citenamefont {Sguazzin}, \citenamefont {Sidhu}, \citenamefont {Steck}, \citenamefont {Stöhlker}, \citenamefont {Swartz}, \citenamefont {Vesic}, \citenamefont {Walker}, \citenamefont {Yamaguchi},\ and\ \citenamefont {Zidarova}}]{freirefernández2023measurement}%
  \BibitemOpen
  \bibfield  {author} {\bibinfo {author} {\bibfnamefont {D.}~\bibnamefont {Freire-Fernández}}, \bibinfo {author} {\bibfnamefont {W.}~\bibnamefont {Korten}}, \bibinfo {author} {\bibfnamefont {R.~J.}\ \bibnamefont {Chen}}, \bibinfo {author} {\bibfnamefont {S.}~\bibnamefont {Litvinov}}, \bibinfo {author} {\bibfnamefont {Y.~A.}\ \bibnamefont {Litvinov}}, \bibinfo {author} {\bibfnamefont {M.~S.}\ \bibnamefont {Sanjari}}, \bibinfo {author} {\bibfnamefont {H.}~\bibnamefont {Weick}}, \bibinfo {author} {\bibfnamefont {F.~C.}\ \bibnamefont {Akinci}}, \bibinfo {author} {\bibfnamefont {H.~M.}\ \bibnamefont {Albers}}, \bibinfo {author} {\bibfnamefont {M.}~\bibnamefont {Armstrong}}, \bibinfo {author} {\bibfnamefont {A.}~\bibnamefont {Banerjee}}, \bibinfo {author} {\bibfnamefont {K.}~\bibnamefont {Blaum}}, \bibinfo {author} {\bibfnamefont {C.}~\bibnamefont {Brandau}}, \bibinfo {author} {\bibfnamefont {B.~A.}\ \bibnamefont {Brown}}, \bibinfo {author} {\bibfnamefont {C.~G.}\ \bibnamefont {Bruno}}, \bibinfo {author}
  {\bibfnamefont {J.~J.}\ \bibnamefont {Carroll}}, \bibinfo {author} {\bibfnamefont {X.}~\bibnamefont {Chen}}, \bibinfo {author} {\bibfnamefont {C.~J.}\ \bibnamefont {Chiara}}, \bibinfo {author} {\bibfnamefont {M.~L.}\ \bibnamefont {Cortes}}, \bibinfo {author} {\bibfnamefont {S.~F.}\ \bibnamefont {Dellmann}}, \bibinfo {author} {\bibfnamefont {I.}~\bibnamefont {Dillmann}}, \bibinfo {author} {\bibfnamefont {D.}~\bibnamefont {Dmytriiev}}, \bibinfo {author} {\bibfnamefont {O.}~\bibnamefont {Forstner}}, \bibinfo {author} {\bibfnamefont {H.}~\bibnamefont {Geissel}}, \bibinfo {author} {\bibfnamefont {J.}~\bibnamefont {Glorius}}, \bibinfo {author} {\bibfnamefont {A.}~\bibnamefont {Görgen}}, \bibinfo {author} {\bibfnamefont {M.}~\bibnamefont {Górska}}, \bibinfo {author} {\bibfnamefont {C.~J.}\ \bibnamefont {Griffin}}, \bibinfo {author} {\bibfnamefont {A.}~\bibnamefont {Gumberidze}}, \bibinfo {author} {\bibfnamefont {S.}~\bibnamefont {Harayama}}, \bibinfo {author} {\bibfnamefont {R.}~\bibnamefont {Hess}}, \bibinfo
  {author} {\bibfnamefont {N.}~\bibnamefont {Hubbard}}, \bibinfo {author} {\bibfnamefont {K.~E.}\ \bibnamefont {Ide}}, \bibinfo {author} {\bibfnamefont {P.~R.}\ \bibnamefont {John}}, \bibinfo {author} {\bibfnamefont {R.}~\bibnamefont {Joseph}}, \bibinfo {author} {\bibfnamefont {B.}~\bibnamefont {Jurado}}, \bibinfo {author} {\bibfnamefont {D.}~\bibnamefont {Kalaydjieva}}, \bibinfo {author} {\bibfnamefont {K.}~\bibnamefont {Kanika}}, \bibinfo {author} {\bibfnamefont {F.~G.}\ \bibnamefont {Kondev}}, \bibinfo {author} {\bibfnamefont {P.}~\bibnamefont {Koseoglou}}, \bibinfo {author} {\bibfnamefont {G.}~\bibnamefont {Kosir}}, \bibinfo {author} {\bibfnamefont {C.}~\bibnamefont {Kozhuharov}}, \bibinfo {author} {\bibfnamefont {I.}~\bibnamefont {Kulikov}}, \bibinfo {author} {\bibfnamefont {G.}~\bibnamefont {Leckenby}}, \bibinfo {author} {\bibfnamefont {B.}~\bibnamefont {Lorenz}}, \bibinfo {author} {\bibfnamefont {J.}~\bibnamefont {Marsh}}, \bibinfo {author} {\bibfnamefont {A.}~\bibnamefont {Mistry}}, \bibinfo {author}
  {\bibfnamefont {A.}~\bibnamefont {Ozawa}}, \bibinfo {author} {\bibfnamefont {N.}~\bibnamefont {Pietralla}}, \bibinfo {author} {\bibfnamefont {Z.}~\bibnamefont {Podolyák}}, \bibinfo {author} {\bibfnamefont {M.}~\bibnamefont {Polettini}}, \bibinfo {author} {\bibfnamefont {M.}~\bibnamefont {Sguazzin}}, \bibinfo {author} {\bibfnamefont {R.~S.}\ \bibnamefont {Sidhu}}, \bibinfo {author} {\bibfnamefont {M.}~\bibnamefont {Steck}}, \bibinfo {author} {\bibfnamefont {T.}~\bibnamefont {Stöhlker}}, \bibinfo {author} {\bibfnamefont {J.~A.}\ \bibnamefont {Swartz}}, \bibinfo {author} {\bibfnamefont {J.}~\bibnamefont {Vesic}}, \bibinfo {author} {\bibfnamefont {P.~M.}\ \bibnamefont {Walker}}, \bibinfo {author} {\bibfnamefont {T.}~\bibnamefont {Yamaguchi}}, \ and\ \bibinfo {author} {\bibfnamefont {R.}~\bibnamefont {Zidarova}},\ }\href@noop {} {\enquote {\bibinfo {title} {Measurement of the isolated nuclear two-photon decay in $^{72}\mathrm{Ge}$},}\ } (\bibinfo {year} {2023}),\ \Eprint {http://arxiv.org/abs/2312.11313}
  {arXiv:2312.11313 [nucl-ex]} \BibitemShut {NoStop}%
\bibitem [{\citenamefont {Yang}(2020)}]{Yang:2019hkn}%
  \BibitemOpen
  \bibfield  {author} {\bibinfo {author} {\bibfnamefont {C.~J.}\ \bibnamefont {Yang}},\ }\href {\doibase 10.1140/epja/s10050-020-00104-0} {\bibfield  {journal} {\bibinfo  {journal} {Eur. Phys. J. A}\ }\textbf {\bibinfo {volume} {56}},\ \bibinfo {pages} {96} (\bibinfo {year} {2020})},\ \Eprint {http://arxiv.org/abs/1905.12510} {arXiv:1905.12510 [nucl-th]} \BibitemShut {NoStop}%
\bibitem [{\citenamefont {Yang}\ \emph {et~al.}(2023)\citenamefont {Yang}, \citenamefont {Ekstr\"om}, \citenamefont {Forss\'en}, \citenamefont {Hagen}, \citenamefont {Rupak},\ and\ \citenamefont {van Kolck}}]{Yang:2021vxa}%
  \BibitemOpen
  \bibfield  {author} {\bibinfo {author} {\bibfnamefont {C.~J.}\ \bibnamefont {Yang}}, \bibinfo {author} {\bibfnamefont {A.}~\bibnamefont {Ekstr\"om}}, \bibinfo {author} {\bibfnamefont {C.}~\bibnamefont {Forss\'en}}, \bibinfo {author} {\bibfnamefont {G.}~\bibnamefont {Hagen}}, \bibinfo {author} {\bibfnamefont {G.}~\bibnamefont {Rupak}}, \ and\ \bibinfo {author} {\bibfnamefont {U.}~\bibnamefont {van Kolck}},\ }\href {\doibase 10.1140/epja/s10050-023-01149-7} {\bibfield  {journal} {\bibinfo  {journal} {Eur. Phys. J. A}\ }\textbf {\bibinfo {volume} {59}},\ \bibinfo {pages} {233} (\bibinfo {year} {2023})},\ \Eprint {http://arxiv.org/abs/2109.13303} {arXiv:2109.13303 [nucl-th]} \BibitemShut {NoStop}%
\bibitem [{\citenamefont {Becker}\ \emph {et~al.}(1984)\citenamefont {Becker}, \citenamefont {Schlicher},\ and\ \citenamefont {Scully}}]{BECKER1984441}%
  \BibitemOpen
  \bibfield  {author} {\bibinfo {author} {\bibfnamefont {W.}~\bibnamefont {Becker}}, \bibinfo {author} {\bibfnamefont {R.}~\bibnamefont {Schlicher}}, \ and\ \bibinfo {author} {\bibfnamefont {M.}~\bibnamefont {Scully}},\ }\href {\doibase https://doi.org/10.1016/0375-9601(84)90989-7} {\bibfield  {journal} {\bibinfo  {journal} {Physics Letters A}\ }\textbf {\bibinfo {volume} {106}},\ \bibinfo {pages} {441} (\bibinfo {year} {1984})}\BibitemShut {NoStop}%
\bibitem [{\citenamefont {Blatt}\ and\ \citenamefont {Weisskopf}(1952)}]{Blatt:1952ije}%
  \BibitemOpen
  \bibfield  {author} {\bibinfo {author} {\bibfnamefont {J.~M.}\ \bibnamefont {Blatt}}\ and\ \bibinfo {author} {\bibfnamefont {V.~F.}\ \bibnamefont {Weisskopf}},\ }\href {\doibase 10.1007/978-1-4612-9959-2} {\emph {\bibinfo {title} {{Theoretical nuclear physics}}}}\ (\bibinfo  {publisher} {Springer},\ \bibinfo {address} {New York},\ \bibinfo {year} {1952})\BibitemShut {NoStop}%
\bibitem [{\citenamefont {Taira}\ \emph {et~al.}(2017)\citenamefont {Taira}, \citenamefont {Hayakawa},\ and\ \citenamefont {Katoh}}]{Taira2017}%
  \BibitemOpen
  \bibfield  {author} {\bibinfo {author} {\bibfnamefont {Y.}~\bibnamefont {Taira}}, \bibinfo {author} {\bibfnamefont {T.}~\bibnamefont {Hayakawa}}, \ and\ \bibinfo {author} {\bibfnamefont {M.}~\bibnamefont {Katoh}},\ }\href {\doibase 10.1038/s41598-017-05187-2} {\bibfield  {journal} {\bibinfo  {journal} {Scientific Reports}\ }\textbf {\bibinfo {volume} {7}} (\bibinfo {year} {2017}),\ 10.1038/s41598-017-05187-2}\BibitemShut {NoStop}%
\bibitem [{\citenamefont {Maruyama}\ \emph {et~al.}(2022)\citenamefont {Maruyama}, \citenamefont {Hayakawa}, \citenamefont {Kajino},\ and\ \citenamefont {Cheoun}}]{Maruyama2022}%
  \BibitemOpen
  \bibfield  {author} {\bibinfo {author} {\bibfnamefont {T.}~\bibnamefont {Maruyama}}, \bibinfo {author} {\bibfnamefont {T.}~\bibnamefont {Hayakawa}}, \bibinfo {author} {\bibfnamefont {T.}~\bibnamefont {Kajino}}, \ and\ \bibinfo {author} {\bibfnamefont {M.-K.}\ \bibnamefont {Cheoun}},\ }\href {\doibase 10.1016/j.physletb.2021.136779} {\bibfield  {journal} {\bibinfo  {journal} {Physics Letters B}\ }\textbf {\bibinfo {volume} {826}},\ \bibinfo {pages} {136779} (\bibinfo {year} {2022})}\BibitemShut {NoStop}%
\bibitem [{\citenamefont {Shen}\ \emph {et~al.}(2019)\citenamefont {Shen}, \citenamefont {Wang}, \citenamefont {Xie}, \citenamefont {Min}, \citenamefont {Fu}, \citenamefont {Liu}, \citenamefont {Gong},\ and\ \citenamefont {Yuan}}]{Shen2019}%
  \BibitemOpen
  \bibfield  {author} {\bibinfo {author} {\bibfnamefont {Y.}~\bibnamefont {Shen}}, \bibinfo {author} {\bibfnamefont {X.}~\bibnamefont {Wang}}, \bibinfo {author} {\bibfnamefont {Z.}~\bibnamefont {Xie}}, \bibinfo {author} {\bibfnamefont {C.}~\bibnamefont {Min}}, \bibinfo {author} {\bibfnamefont {X.}~\bibnamefont {Fu}}, \bibinfo {author} {\bibfnamefont {Q.}~\bibnamefont {Liu}}, \bibinfo {author} {\bibfnamefont {M.}~\bibnamefont {Gong}}, \ and\ \bibinfo {author} {\bibfnamefont {X.}~\bibnamefont {Yuan}},\ }\href {\doibase 10.1038/s41377-019-0194-2} {\bibfield  {journal} {\bibinfo  {journal} {Light: Science \& Applications}\ }\textbf {\bibinfo {volume} {8}} (\bibinfo {year} {2019}),\ 10.1038/s41377-019-0194-2}\BibitemShut {NoStop}%
\bibitem [{\citenamefont {Peele}\ \emph {et~al.}(2002)\citenamefont {Peele}, \citenamefont {McMahon}, \citenamefont {Paterson}, \citenamefont {Tran}, \citenamefont {Mancuso}, \citenamefont {Nugent}, \citenamefont {Hayes}, \citenamefont {Harvey}, \citenamefont {Lai},\ and\ \citenamefont {McNulty}}]{Peele2002}%
  \BibitemOpen
  \bibfield  {author} {\bibinfo {author} {\bibfnamefont {A.~G.}\ \bibnamefont {Peele}}, \bibinfo {author} {\bibfnamefont {P.~J.}\ \bibnamefont {McMahon}}, \bibinfo {author} {\bibfnamefont {D.}~\bibnamefont {Paterson}}, \bibinfo {author} {\bibfnamefont {C.~Q.}\ \bibnamefont {Tran}}, \bibinfo {author} {\bibfnamefont {A.~P.}\ \bibnamefont {Mancuso}}, \bibinfo {author} {\bibfnamefont {K.~A.}\ \bibnamefont {Nugent}}, \bibinfo {author} {\bibfnamefont {J.~P.}\ \bibnamefont {Hayes}}, \bibinfo {author} {\bibfnamefont {E.}~\bibnamefont {Harvey}}, \bibinfo {author} {\bibfnamefont {B.}~\bibnamefont {Lai}}, \ and\ \bibinfo {author} {\bibfnamefont {I.}~\bibnamefont {McNulty}},\ }\href {\doibase 10.1364/ol.27.001752} {\bibfield  {journal} {\bibinfo  {journal} {Optics Letters}\ }\textbf {\bibinfo {volume} {27}},\ \bibinfo {pages} {1752} (\bibinfo {year} {2002})}\BibitemShut {NoStop}%
\bibitem [{\citenamefont {Terhalle}\ \emph {et~al.}(2011)\citenamefont {Terhalle}, \citenamefont {Langner}, \citenamefont {P\"{a}iv\"{a}nranta}, \citenamefont {Guzenko}, \citenamefont {David},\ and\ \citenamefont {Ekinci}}]{Terhalle2011}%
  \BibitemOpen
  \bibfield  {author} {\bibinfo {author} {\bibfnamefont {B.}~\bibnamefont {Terhalle}}, \bibinfo {author} {\bibfnamefont {A.}~\bibnamefont {Langner}}, \bibinfo {author} {\bibfnamefont {B.}~\bibnamefont {P\"{a}iv\"{a}nranta}}, \bibinfo {author} {\bibfnamefont {V.~A.}\ \bibnamefont {Guzenko}}, \bibinfo {author} {\bibfnamefont {C.}~\bibnamefont {David}}, \ and\ \bibinfo {author} {\bibfnamefont {Y.}~\bibnamefont {Ekinci}},\ }\href {\doibase 10.1364/ol.36.004143} {\bibfield  {journal} {\bibinfo  {journal} {Optics Letters}\ }\textbf {\bibinfo {volume} {36}},\ \bibinfo {pages} {4143} (\bibinfo {year} {2011})}\BibitemShut {NoStop}%
\bibitem [{\citenamefont {Gariepy}\ \emph {et~al.}(2014)\citenamefont {Gariepy}, \citenamefont {Leach}, \citenamefont {Kim}, \citenamefont {Hammond}, \citenamefont {Frumker}, \citenamefont {Boyd},\ and\ \citenamefont {Corkum}}]{PhysRevLett.113.153901}%
  \BibitemOpen
  \bibfield  {author} {\bibinfo {author} {\bibfnamefont {G.}~\bibnamefont {Gariepy}}, \bibinfo {author} {\bibfnamefont {J.}~\bibnamefont {Leach}}, \bibinfo {author} {\bibfnamefont {K.~T.}\ \bibnamefont {Kim}}, \bibinfo {author} {\bibfnamefont {T.~J.}\ \bibnamefont {Hammond}}, \bibinfo {author} {\bibfnamefont {E.}~\bibnamefont {Frumker}}, \bibinfo {author} {\bibfnamefont {R.~W.}\ \bibnamefont {Boyd}}, \ and\ \bibinfo {author} {\bibfnamefont {P.~B.}\ \bibnamefont {Corkum}},\ }\href {\doibase 10.1103/PhysRevLett.113.153901} {\bibfield  {journal} {\bibinfo  {journal} {Phys. Rev. Lett.}\ }\textbf {\bibinfo {volume} {113}},\ \bibinfo {pages} {153901} (\bibinfo {year} {2014})}\BibitemShut {NoStop}%
\bibitem [{\citenamefont {Hemsing}\ \emph {et~al.}(2013)\citenamefont {Hemsing}, \citenamefont {Knyazik}, \citenamefont {Dunning}, \citenamefont {Xiang}, \citenamefont {Marinelli}, \citenamefont {Hast},\ and\ \citenamefont {Rosenzweig}}]{Hemsing2013}%
  \BibitemOpen
  \bibfield  {author} {\bibinfo {author} {\bibfnamefont {E.}~\bibnamefont {Hemsing}}, \bibinfo {author} {\bibfnamefont {A.}~\bibnamefont {Knyazik}}, \bibinfo {author} {\bibfnamefont {M.}~\bibnamefont {Dunning}}, \bibinfo {author} {\bibfnamefont {D.}~\bibnamefont {Xiang}}, \bibinfo {author} {\bibfnamefont {A.}~\bibnamefont {Marinelli}}, \bibinfo {author} {\bibfnamefont {C.}~\bibnamefont {Hast}}, \ and\ \bibinfo {author} {\bibfnamefont {J.~B.}\ \bibnamefont {Rosenzweig}},\ }\href {\doibase 10.1038/nphys2712} {\bibfield  {journal} {\bibinfo  {journal} {Nature Physics}\ }\textbf {\bibinfo {volume} {9}},\ \bibinfo {pages} {549–553} (\bibinfo {year} {2013})}\BibitemShut {NoStop}%
\bibitem [{\citenamefont {Ababekri}\ \emph {et~al.}(2024)\citenamefont {Ababekri}, \citenamefont {Guo}, \citenamefont {Wan}, \citenamefont {Qiao}, \citenamefont {Li}, \citenamefont {Lv}, \citenamefont {Zhang}, \citenamefont {Zhou}, \citenamefont {Gu},\ and\ \citenamefont {Li}}]{Ababekri:2022mob}%
  \BibitemOpen
  \bibfield  {author} {\bibinfo {author} {\bibfnamefont {M.}~\bibnamefont {Ababekri}}, \bibinfo {author} {\bibfnamefont {R.-T.}\ \bibnamefont {Guo}}, \bibinfo {author} {\bibfnamefont {F.}~\bibnamefont {Wan}}, \bibinfo {author} {\bibfnamefont {B.}~\bibnamefont {Qiao}}, \bibinfo {author} {\bibfnamefont {Z.}~\bibnamefont {Li}}, \bibinfo {author} {\bibfnamefont {C.}~\bibnamefont {Lv}}, \bibinfo {author} {\bibfnamefont {B.}~\bibnamefont {Zhang}}, \bibinfo {author} {\bibfnamefont {W.}~\bibnamefont {Zhou}}, \bibinfo {author} {\bibfnamefont {Y.}~\bibnamefont {Gu}}, \ and\ \bibinfo {author} {\bibfnamefont {J.-X.}\ \bibnamefont {Li}},\ }\href {\doibase 10.1103/PhysRevD.109.016005} {\bibfield  {journal} {\bibinfo  {journal} {Phys. Rev. D}\ }\textbf {\bibinfo {volume} {109}},\ \bibinfo {pages} {016005} (\bibinfo {year} {2024})},\ \Eprint {http://arxiv.org/abs/2211.05467} {arXiv:2211.05467 [hep-ph]} \BibitemShut {NoStop}%
\bibitem [{\citenamefont {Taira}\ and\ \citenamefont {Katoh}(2018)}]{PhysRevA.98.052130}%
  \BibitemOpen
  \bibfield  {author} {\bibinfo {author} {\bibfnamefont {Y.}~\bibnamefont {Taira}}\ and\ \bibinfo {author} {\bibfnamefont {M.}~\bibnamefont {Katoh}},\ }\href {\doibase 10.1103/PhysRevA.98.052130} {\bibfield  {journal} {\bibinfo  {journal} {Phys. Rev. A}\ }\textbf {\bibinfo {volume} {98}},\ \bibinfo {pages} {052130} (\bibinfo {year} {2018})}\BibitemShut {NoStop}%
\bibitem [{\citenamefont {Zhu}\ \emph {et~al.}(2018)\citenamefont {Zhu}, \citenamefont {Yu}, \citenamefont {Chen}, \citenamefont {Weng},\ and\ \citenamefont {Sheng}}]{Zhu2018}%
  \BibitemOpen
  \bibfield  {author} {\bibinfo {author} {\bibfnamefont {X.-L.}\ \bibnamefont {Zhu}}, \bibinfo {author} {\bibfnamefont {T.-P.}\ \bibnamefont {Yu}}, \bibinfo {author} {\bibfnamefont {M.}~\bibnamefont {Chen}}, \bibinfo {author} {\bibfnamefont {S.-M.}\ \bibnamefont {Weng}}, \ and\ \bibinfo {author} {\bibfnamefont {Z.-M.}\ \bibnamefont {Sheng}},\ }\href {\doibase 10.1088/1367-2630/aad71a} {\bibfield  {journal} {\bibinfo  {journal} {New Journal of Physics}\ }\textbf {\bibinfo {volume} {20}},\ \bibinfo {pages} {083013} (\bibinfo {year} {2018})}\BibitemShut {NoStop}%
\bibitem [{\citenamefont {Feng}\ \emph {et~al.}(2019)\citenamefont {Feng}, \citenamefont {Qin}, \citenamefont {Geng}, \citenamefont {Yu}, \citenamefont {Wang}, \citenamefont {Wu}, \citenamefont {Yan}, \citenamefont {Ji},\ and\ \citenamefont {Shen}}]{Feng2019}%
  \BibitemOpen
  \bibfield  {author} {\bibinfo {author} {\bibfnamefont {B.}~\bibnamefont {Feng}}, \bibinfo {author} {\bibfnamefont {C.~Y.}\ \bibnamefont {Qin}}, \bibinfo {author} {\bibfnamefont {X.~S.}\ \bibnamefont {Geng}}, \bibinfo {author} {\bibfnamefont {Q.}~\bibnamefont {Yu}}, \bibinfo {author} {\bibfnamefont {W.~Q.}\ \bibnamefont {Wang}}, \bibinfo {author} {\bibfnamefont {Y.~T.}\ \bibnamefont {Wu}}, \bibinfo {author} {\bibfnamefont {X.}~\bibnamefont {Yan}}, \bibinfo {author} {\bibfnamefont {L.~L.}\ \bibnamefont {Ji}}, \ and\ \bibinfo {author} {\bibfnamefont {B.~F.}\ \bibnamefont {Shen}},\ }\href {\doibase 10.1038/s41598-019-55217-4} {\bibfield  {journal} {\bibinfo  {journal} {Scientific Reports}\ }\textbf {\bibinfo {volume} {9}} (\bibinfo {year} {2019}),\ 10.1038/s41598-019-55217-4}\BibitemShut {NoStop}%
\bibitem [{\citenamefont {Wang}\ \emph {et~al.}(2020{\natexlab{a}})\citenamefont {Wang}, \citenamefont {Li}, \citenamefont {Gan}, \citenamefont {Xie}, \citenamefont {Zhong}, \citenamefont {Zhou}, \citenamefont {Zhu}, \citenamefont {He},\ and\ \citenamefont {Qiao}}]{PhysRevApplied.14.014094}%
  \BibitemOpen
  \bibfield  {author} {\bibinfo {author} {\bibfnamefont {J.}~\bibnamefont {Wang}}, \bibinfo {author} {\bibfnamefont {X.}~\bibnamefont {Li}}, \bibinfo {author} {\bibfnamefont {L.}~\bibnamefont {Gan}}, \bibinfo {author} {\bibfnamefont {Y.}~\bibnamefont {Xie}}, \bibinfo {author} {\bibfnamefont {C.}~\bibnamefont {Zhong}}, \bibinfo {author} {\bibfnamefont {C.}~\bibnamefont {Zhou}}, \bibinfo {author} {\bibfnamefont {S.}~\bibnamefont {Zhu}}, \bibinfo {author} {\bibfnamefont {X.}~\bibnamefont {He}}, \ and\ \bibinfo {author} {\bibfnamefont {B.}~\bibnamefont {Qiao}},\ }\href {\doibase 10.1103/PhysRevApplied.14.014094} {\bibfield  {journal} {\bibinfo  {journal} {Phys. Rev. Appl.}\ }\textbf {\bibinfo {volume} {14}},\ \bibinfo {pages} {014094} (\bibinfo {year} {2020}{\natexlab{a}})}\BibitemShut {NoStop}%
\bibitem [{\citenamefont {Zhang}\ \emph {et~al.}(2023)\citenamefont {Zhang}, \citenamefont {Bake}, \citenamefont {Xiao}, \citenamefont {Sang},\ and\ \citenamefont {Xie}}]{Zhang2023}%
  \BibitemOpen
  \bibfield  {author} {\bibinfo {author} {\bibfnamefont {C.-W.}\ \bibnamefont {Zhang}}, \bibinfo {author} {\bibfnamefont {M.-A.}\ \bibnamefont {Bake}}, \bibinfo {author} {\bibfnamefont {H.}~\bibnamefont {Xiao}}, \bibinfo {author} {\bibfnamefont {H.-B.}\ \bibnamefont {Sang}}, \ and\ \bibinfo {author} {\bibfnamefont {B.-S.}\ \bibnamefont {Xie}},\ }\href {\doibase 10.1063/5.0136143} {\bibfield  {journal} {\bibinfo  {journal} {Physics of Plasmas}\ }\textbf {\bibinfo {volume} {30}} (\bibinfo {year} {2023}),\ 10.1063/5.0136143}\BibitemShut {NoStop}%
\bibitem [{\citenamefont {Liu}\ \emph {et~al.}(2016)\citenamefont {Liu}, \citenamefont {Shen}, \citenamefont {Zhang}, \citenamefont {Shi}, \citenamefont {Ji}, \citenamefont {Wang}, \citenamefont {Yi}, \citenamefont {Zhang}, \citenamefont {Xu}, \citenamefont {Pei},\ and\ \citenamefont {Xu}}]{Liu2016}%
  \BibitemOpen
  \bibfield  {author} {\bibinfo {author} {\bibfnamefont {C.}~\bibnamefont {Liu}}, \bibinfo {author} {\bibfnamefont {B.}~\bibnamefont {Shen}}, \bibinfo {author} {\bibfnamefont {X.}~\bibnamefont {Zhang}}, \bibinfo {author} {\bibfnamefont {Y.}~\bibnamefont {Shi}}, \bibinfo {author} {\bibfnamefont {L.}~\bibnamefont {Ji}}, \bibinfo {author} {\bibfnamefont {W.}~\bibnamefont {Wang}}, \bibinfo {author} {\bibfnamefont {L.}~\bibnamefont {Yi}}, \bibinfo {author} {\bibfnamefont {L.}~\bibnamefont {Zhang}}, \bibinfo {author} {\bibfnamefont {T.}~\bibnamefont {Xu}}, \bibinfo {author} {\bibfnamefont {Z.}~\bibnamefont {Pei}}, \ and\ \bibinfo {author} {\bibfnamefont {Z.}~\bibnamefont {Xu}},\ }\href {\doibase 10.1063/1.4963396} {\bibfield  {journal} {\bibinfo  {journal} {Physics of Plasmas}\ }\textbf {\bibinfo {volume} {23}} (\bibinfo {year} {2016}),\ 10.1063/1.4963396}\BibitemShut {NoStop}%
\bibitem [{\citenamefont {Zhang}\ \emph {et~al.}(2021)\citenamefont {Zhang}, \citenamefont {Zhao}, \citenamefont {Hu}, \citenamefont {Li}, \citenamefont {Lu}, \citenamefont {Cao}, \citenamefont {Zou}, \citenamefont {Sheng}, \citenamefont {Pegoraro}, \citenamefont {McKenna}, \citenamefont {Shao},\ and\ \citenamefont {Yu}}]{Zhang2021}%
  \BibitemOpen
  \bibfield  {author} {\bibinfo {author} {\bibfnamefont {H.}~\bibnamefont {Zhang}}, \bibinfo {author} {\bibfnamefont {J.}~\bibnamefont {Zhao}}, \bibinfo {author} {\bibfnamefont {Y.}~\bibnamefont {Hu}}, \bibinfo {author} {\bibfnamefont {Q.}~\bibnamefont {Li}}, \bibinfo {author} {\bibfnamefont {Y.}~\bibnamefont {Lu}}, \bibinfo {author} {\bibfnamefont {Y.}~\bibnamefont {Cao}}, \bibinfo {author} {\bibfnamefont {D.}~\bibnamefont {Zou}}, \bibinfo {author} {\bibfnamefont {Z.}~\bibnamefont {Sheng}}, \bibinfo {author} {\bibfnamefont {F.}~\bibnamefont {Pegoraro}}, \bibinfo {author} {\bibfnamefont {P.}~\bibnamefont {McKenna}}, \bibinfo {author} {\bibfnamefont {F.}~\bibnamefont {Shao}}, \ and\ \bibinfo {author} {\bibfnamefont {T.}~\bibnamefont {Yu}},\ }\href {\doibase 10.1017/hpl.2021.29} {\bibfield  {journal} {\bibinfo  {journal} {High Power Laser Science and Engineering}\ ,\ \bibinfo {pages} {1–24}} (\bibinfo {year} {2021})}\BibitemShut {NoStop}%
\bibitem [{\citenamefont {Liu}\ \emph {et~al.}(2020)\citenamefont {Liu}, \citenamefont {Salamin}, \citenamefont {Dou}, \citenamefont {Xu},\ and\ \citenamefont {Li}}]{Liu2020}%
  \BibitemOpen
  \bibfield  {author} {\bibinfo {author} {\bibfnamefont {Y.-Y.}\ \bibnamefont {Liu}}, \bibinfo {author} {\bibfnamefont {Y.~I.}\ \bibnamefont {Salamin}}, \bibinfo {author} {\bibfnamefont {Z.-K.}\ \bibnamefont {Dou}}, \bibinfo {author} {\bibfnamefont {Z.-F.}\ \bibnamefont {Xu}}, \ and\ \bibinfo {author} {\bibfnamefont {J.-X.}\ \bibnamefont {Li}},\ }\href {\doibase 10.1364/ol.45.000395} {\bibfield  {journal} {\bibinfo  {journal} {Optics Letters}\ }\textbf {\bibinfo {volume} {45}},\ \bibinfo {pages} {395} (\bibinfo {year} {2020})}\BibitemShut {NoStop}%
\bibitem [{\citenamefont {Jentschura}\ and\ \citenamefont {Serbo}(2011)}]{PhysRevLett.106.013001}%
  \BibitemOpen
  \bibfield  {author} {\bibinfo {author} {\bibfnamefont {U.~D.}\ \bibnamefont {Jentschura}}\ and\ \bibinfo {author} {\bibfnamefont {V.~G.}\ \bibnamefont {Serbo}},\ }\href {\doibase 10.1103/PhysRevLett.106.013001} {\bibfield  {journal} {\bibinfo  {journal} {Phys. Rev. Lett.}\ }\textbf {\bibinfo {volume} {106}},\ \bibinfo {pages} {013001} (\bibinfo {year} {2011})}\BibitemShut {NoStop}%
\bibitem [{\citenamefont {Chen}\ \emph {et~al.}(2018)\citenamefont {Chen}, \citenamefont {Li}, \citenamefont {Hatsagortsyan},\ and\ \citenamefont {Keitel}}]{PhysRevLett.121.074801}%
  \BibitemOpen
  \bibfield  {author} {\bibinfo {author} {\bibfnamefont {Y.-Y.}\ \bibnamefont {Chen}}, \bibinfo {author} {\bibfnamefont {J.-X.}\ \bibnamefont {Li}}, \bibinfo {author} {\bibfnamefont {K.~Z.}\ \bibnamefont {Hatsagortsyan}}, \ and\ \bibinfo {author} {\bibfnamefont {C.~H.}\ \bibnamefont {Keitel}},\ }\href {\doibase 10.1103/PhysRevLett.121.074801} {\bibfield  {journal} {\bibinfo  {journal} {Phys. Rev. Lett.}\ }\textbf {\bibinfo {volume} {121}},\ \bibinfo {pages} {074801} (\bibinfo {year} {2018})}\BibitemShut {NoStop}%
\bibitem [{\citenamefont {Hu}\ \emph {et~al.}(2021)\citenamefont {Hu}, \citenamefont {Zhao}, \citenamefont {Zhang}, \citenamefont {Lu}, \citenamefont {Wang}, \citenamefont {Hu}, \citenamefont {Shao},\ and\ \citenamefont {Yu}}]{Hu2021}%
  \BibitemOpen
  \bibfield  {author} {\bibinfo {author} {\bibfnamefont {Y.-T.}\ \bibnamefont {Hu}}, \bibinfo {author} {\bibfnamefont {J.}~\bibnamefont {Zhao}}, \bibinfo {author} {\bibfnamefont {H.}~\bibnamefont {Zhang}}, \bibinfo {author} {\bibfnamefont {Y.}~\bibnamefont {Lu}}, \bibinfo {author} {\bibfnamefont {W.-Q.}\ \bibnamefont {Wang}}, \bibinfo {author} {\bibfnamefont {L.-X.}\ \bibnamefont {Hu}}, \bibinfo {author} {\bibfnamefont {F.-Q.}\ \bibnamefont {Shao}}, \ and\ \bibinfo {author} {\bibfnamefont {T.-P.}\ \bibnamefont {Yu}},\ }\href {\doibase 10.1063/5.0028203} {\bibfield  {journal} {\bibinfo  {journal} {Applied Physics Letters}\ }\textbf {\bibinfo {volume} {118}} (\bibinfo {year} {2021}),\ 10.1063/5.0028203}\BibitemShut {NoStop}%
\bibitem [{\citenamefont {Younis}\ \emph {et~al.}(2022)\citenamefont {Younis}, \citenamefont {Hafizi},\ and\ \citenamefont {Gordon}}]{Younis2022}%
  \BibitemOpen
  \bibfield  {author} {\bibinfo {author} {\bibfnamefont {D.}~\bibnamefont {Younis}}, \bibinfo {author} {\bibfnamefont {B.}~\bibnamefont {Hafizi}}, \ and\ \bibinfo {author} {\bibfnamefont {D.~F.}\ \bibnamefont {Gordon}},\ }\href {\doibase 10.1063/5.0102909} {\bibfield  {journal} {\bibinfo  {journal} {Physics of Plasmas}\ }\textbf {\bibinfo {volume} {29}} (\bibinfo {year} {2022}),\ 10.1063/5.0102909}\BibitemShut {NoStop}%
\bibitem [{\citenamefont {Lu}\ \emph {et~al.}(2023)\citenamefont {Lu}, \citenamefont {Guo}, \citenamefont {Li}, \citenamefont {Ababekri}, \citenamefont {Chen}, \citenamefont {Fu}, \citenamefont {Lv}, \citenamefont {Xu}, \citenamefont {Kong}, \citenamefont {Niu},\ and\ \citenamefont {Li}}]{PhysRevLett.131.202502}%
  \BibitemOpen
  \bibfield  {author} {\bibinfo {author} {\bibfnamefont {Z.-W.}\ \bibnamefont {Lu}}, \bibinfo {author} {\bibfnamefont {L.}~\bibnamefont {Guo}}, \bibinfo {author} {\bibfnamefont {Z.-Z.}\ \bibnamefont {Li}}, \bibinfo {author} {\bibfnamefont {M.}~\bibnamefont {Ababekri}}, \bibinfo {author} {\bibfnamefont {F.-Q.}\ \bibnamefont {Chen}}, \bibinfo {author} {\bibfnamefont {C.}~\bibnamefont {Fu}}, \bibinfo {author} {\bibfnamefont {C.}~\bibnamefont {Lv}}, \bibinfo {author} {\bibfnamefont {R.}~\bibnamefont {Xu}}, \bibinfo {author} {\bibfnamefont {X.}~\bibnamefont {Kong}}, \bibinfo {author} {\bibfnamefont {Y.-F.}\ \bibnamefont {Niu}}, \ and\ \bibinfo {author} {\bibfnamefont {J.-X.}\ \bibnamefont {Li}},\ }\href {\doibase 10.1103/PhysRevLett.131.202502} {\bibfield  {journal} {\bibinfo  {journal} {Phys. Rev. Lett.}\ }\textbf {\bibinfo {volume} {131}},\ \bibinfo {pages} {202502} (\bibinfo {year} {2023})}\BibitemShut {NoStop}%
\bibitem [{\citenamefont {Balabanski}\ and\ \citenamefont {Luo}(2024)}]{Balabanski2024}%
  \BibitemOpen
  \bibfield  {author} {\bibinfo {author} {\bibfnamefont {D.~L.}\ \bibnamefont {Balabanski}}\ and\ \bibinfo {author} {\bibfnamefont {W.}~\bibnamefont {Luo}},\ }\href {\doibase 10.1140/epjs/s11734-024-01132-3} {\bibfield  {journal} {\bibinfo  {journal} {The European Physical Journal Special Topics}\ } (\bibinfo {year} {2024}),\ 10.1140/epjs/s11734-024-01132-3}\BibitemShut {NoStop}%
\bibitem [{\citenamefont {Wang}\ \emph {et~al.}(2020{\natexlab{b}})\citenamefont {Wang}, \citenamefont {Ribeyre}, \citenamefont {Gong}, \citenamefont {Jansen}, \citenamefont {d'Humi\`eres}, \citenamefont {Stutman}, \citenamefont {Toncian},\ and\ \citenamefont {Arefiev}}]{PhysRevApplied.13.054024}%
  \BibitemOpen
  \bibfield  {author} {\bibinfo {author} {\bibfnamefont {T.}~\bibnamefont {Wang}}, \bibinfo {author} {\bibfnamefont {X.}~\bibnamefont {Ribeyre}}, \bibinfo {author} {\bibfnamefont {Z.}~\bibnamefont {Gong}}, \bibinfo {author} {\bibfnamefont {O.}~\bibnamefont {Jansen}}, \bibinfo {author} {\bibfnamefont {E.}~\bibnamefont {d'Humi\`eres}}, \bibinfo {author} {\bibfnamefont {D.}~\bibnamefont {Stutman}}, \bibinfo {author} {\bibfnamefont {T.}~\bibnamefont {Toncian}}, \ and\ \bibinfo {author} {\bibfnamefont {A.}~\bibnamefont {Arefiev}},\ }\href {\doibase 10.1103/PhysRevApplied.13.054024} {\bibfield  {journal} {\bibinfo  {journal} {Phys. Rev. Appl.}\ }\textbf {\bibinfo {volume} {13}},\ \bibinfo {pages} {054024} (\bibinfo {year} {2020}{\natexlab{b}})}\BibitemShut {NoStop}%
\bibitem [{\citenamefont {Xue}\ \emph {et~al.}(2020)\citenamefont {Xue}, \citenamefont {Dou}, \citenamefont {Wan}, \citenamefont {Yu}, \citenamefont {Wang}, \citenamefont {Ren}, \citenamefont {Zhao}, \citenamefont {Zhao}, \citenamefont {Xu},\ and\ \citenamefont {Li}}]{Xue2020}%
  \BibitemOpen
  \bibfield  {author} {\bibinfo {author} {\bibfnamefont {K.}~\bibnamefont {Xue}}, \bibinfo {author} {\bibfnamefont {Z.-K.}\ \bibnamefont {Dou}}, \bibinfo {author} {\bibfnamefont {F.}~\bibnamefont {Wan}}, \bibinfo {author} {\bibfnamefont {T.-P.}\ \bibnamefont {Yu}}, \bibinfo {author} {\bibfnamefont {W.-M.}\ \bibnamefont {Wang}}, \bibinfo {author} {\bibfnamefont {J.-R.}\ \bibnamefont {Ren}}, \bibinfo {author} {\bibfnamefont {Q.}~\bibnamefont {Zhao}}, \bibinfo {author} {\bibfnamefont {Y.-T.}\ \bibnamefont {Zhao}}, \bibinfo {author} {\bibfnamefont {Z.-F.}\ \bibnamefont {Xu}}, \ and\ \bibinfo {author} {\bibfnamefont {J.-X.}\ \bibnamefont {Li}},\ }\href {\doibase 10.1063/5.0007734} {\bibfield  {journal} {\bibinfo  {journal} {Matter and Radiation at Extremes}\ }\textbf {\bibinfo {volume} {5}} (\bibinfo {year} {2020}),\ 10.1063/5.0007734}\BibitemShut {NoStop}%
\bibitem [{\citenamefont {Wang}\ \emph {et~al.}(2021{\natexlab{b}})\citenamefont {Wang}, \citenamefont {Blackman}, \citenamefont {Chin},\ and\ \citenamefont {Arefiev}}]{PhysRevE.104.045206}%
  \BibitemOpen
  \bibfield  {author} {\bibinfo {author} {\bibfnamefont {T.}~\bibnamefont {Wang}}, \bibinfo {author} {\bibfnamefont {D.}~\bibnamefont {Blackman}}, \bibinfo {author} {\bibfnamefont {K.}~\bibnamefont {Chin}}, \ and\ \bibinfo {author} {\bibfnamefont {A.}~\bibnamefont {Arefiev}},\ }\href {\doibase 10.1103/PhysRevE.104.045206} {\bibfield  {journal} {\bibinfo  {journal} {Phys. Rev. E}\ }\textbf {\bibinfo {volume} {104}},\ \bibinfo {pages} {045206} (\bibinfo {year} {2021}{\natexlab{b}})}\BibitemShut {NoStop}%
\bibitem [{\citenamefont {Heppe}\ and\ \citenamefont {Kumar}(2022)}]{Heppe2022}%
  \BibitemOpen
  \bibfield  {author} {\bibinfo {author} {\bibfnamefont {C.}~\bibnamefont {Heppe}}\ and\ \bibinfo {author} {\bibfnamefont {N.}~\bibnamefont {Kumar}},\ }\href {\doibase 10.3389/fphy.2022.987830} {\bibfield  {journal} {\bibinfo  {journal} {Frontiers in Physics}\ }\textbf {\bibinfo {volume} {10}} (\bibinfo {year} {2022}),\ 10.3389/fphy.2022.987830}\BibitemShut {NoStop}%
\bibitem [{\citenamefont {Shen}\ \emph {et~al.}(2024)\citenamefont {Shen}, \citenamefont {Pukhov},\ and\ \citenamefont {Qiao}}]{Shen2024}%
  \BibitemOpen
  \bibfield  {author} {\bibinfo {author} {\bibfnamefont {X.}~\bibnamefont {Shen}}, \bibinfo {author} {\bibfnamefont {A.}~\bibnamefont {Pukhov}}, \ and\ \bibinfo {author} {\bibfnamefont {B.}~\bibnamefont {Qiao}},\ }\href {\doibase 10.1038/s42005-024-01575-z} {\bibfield  {journal} {\bibinfo  {journal} {Communications Physics}\ }\textbf {\bibinfo {volume} {7}} (\bibinfo {year} {2024}),\ 10.1038/s42005-024-01575-z}\BibitemShut {NoStop}%
\bibitem [{\citenamefont {Yoon}\ \emph {et~al.}(2019)\citenamefont {Yoon}, \citenamefont {Jeon}, \citenamefont {Shin}, \citenamefont {Lee}, \citenamefont {Lee}, \citenamefont {Choi}, \citenamefont {Kim}, \citenamefont {Sung},\ and\ \citenamefont {Nam}}]{Yoon:19}%
  \BibitemOpen
  \bibfield  {author} {\bibinfo {author} {\bibfnamefont {J.~W.}\ \bibnamefont {Yoon}}, \bibinfo {author} {\bibfnamefont {C.}~\bibnamefont {Jeon}}, \bibinfo {author} {\bibfnamefont {J.}~\bibnamefont {Shin}}, \bibinfo {author} {\bibfnamefont {S.~K.}\ \bibnamefont {Lee}}, \bibinfo {author} {\bibfnamefont {H.~W.}\ \bibnamefont {Lee}}, \bibinfo {author} {\bibfnamefont {I.~W.}\ \bibnamefont {Choi}}, \bibinfo {author} {\bibfnamefont {H.~T.}\ \bibnamefont {Kim}}, \bibinfo {author} {\bibfnamefont {J.~H.}\ \bibnamefont {Sung}}, \ and\ \bibinfo {author} {\bibfnamefont {C.~H.}\ \bibnamefont {Nam}},\ }\href {\doibase 10.1364/OE.27.020412} {\bibfield  {journal} {\bibinfo  {journal} {Opt. Express}\ }\textbf {\bibinfo {volume} {27}},\ \bibinfo {pages} {20412} (\bibinfo {year} {2019})}\BibitemShut {NoStop}%
\bibitem [{\citenamefont {Yoon}\ \emph {et~al.}(2021)\citenamefont {Yoon}, \citenamefont {Kim}, \citenamefont {Choi}, \citenamefont {Sung}, \citenamefont {Lee}, \citenamefont {Lee},\ and\ \citenamefont {Nam}}]{Yoon:21}%
  \BibitemOpen
  \bibfield  {author} {\bibinfo {author} {\bibfnamefont {J.~W.}\ \bibnamefont {Yoon}}, \bibinfo {author} {\bibfnamefont {Y.~G.}\ \bibnamefont {Kim}}, \bibinfo {author} {\bibfnamefont {I.~W.}\ \bibnamefont {Choi}}, \bibinfo {author} {\bibfnamefont {J.~H.}\ \bibnamefont {Sung}}, \bibinfo {author} {\bibfnamefont {H.~W.}\ \bibnamefont {Lee}}, \bibinfo {author} {\bibfnamefont {S.~K.}\ \bibnamefont {Lee}}, \ and\ \bibinfo {author} {\bibfnamefont {C.~H.}\ \bibnamefont {Nam}},\ }\href {\doibase 10.1364/OPTICA.420520} {\bibfield  {journal} {\bibinfo  {journal} {Optica}\ }\textbf {\bibinfo {volume} {8}},\ \bibinfo {pages} {630} (\bibinfo {year} {2021})}\BibitemShut {NoStop}%
\bibitem [{\citenamefont {Li}\ \emph {et~al.}(2018)\citenamefont {Li}, \citenamefont {Gan}, \citenamefont {Yu}, \citenamefont {Wang}, \citenamefont {Liu}, \citenamefont {Guo}, \citenamefont {Xu}, \citenamefont {Xu}, \citenamefont {Hang}, \citenamefont {Xu}, \citenamefont {Wang}, \citenamefont {Huang}, \citenamefont {Cao}, \citenamefont {Yao}, \citenamefont {Zhang}, \citenamefont {Chen}, \citenamefont {Tang}, \citenamefont {Li}, \citenamefont {Liu}, \citenamefont {Li}, \citenamefont {He}, \citenamefont {Yin}, \citenamefont {Liang}, \citenamefont {Leng}, \citenamefont {Li},\ and\ \citenamefont {Xu}}]{Li:18}%
  \BibitemOpen
  \bibfield  {author} {\bibinfo {author} {\bibfnamefont {W.}~\bibnamefont {Li}}, \bibinfo {author} {\bibfnamefont {Z.}~\bibnamefont {Gan}}, \bibinfo {author} {\bibfnamefont {L.}~\bibnamefont {Yu}}, \bibinfo {author} {\bibfnamefont {C.}~\bibnamefont {Wang}}, \bibinfo {author} {\bibfnamefont {Y.}~\bibnamefont {Liu}}, \bibinfo {author} {\bibfnamefont {Z.}~\bibnamefont {Guo}}, \bibinfo {author} {\bibfnamefont {L.}~\bibnamefont {Xu}}, \bibinfo {author} {\bibfnamefont {M.}~\bibnamefont {Xu}}, \bibinfo {author} {\bibfnamefont {Y.}~\bibnamefont {Hang}}, \bibinfo {author} {\bibfnamefont {Y.}~\bibnamefont {Xu}}, \bibinfo {author} {\bibfnamefont {J.}~\bibnamefont {Wang}}, \bibinfo {author} {\bibfnamefont {P.}~\bibnamefont {Huang}}, \bibinfo {author} {\bibfnamefont {H.}~\bibnamefont {Cao}}, \bibinfo {author} {\bibfnamefont {B.}~\bibnamefont {Yao}}, \bibinfo {author} {\bibfnamefont {X.}~\bibnamefont {Zhang}}, \bibinfo {author} {\bibfnamefont {L.}~\bibnamefont {Chen}}, \bibinfo {author} {\bibfnamefont {Y.}~\bibnamefont
  {Tang}}, \bibinfo {author} {\bibfnamefont {S.}~\bibnamefont {Li}}, \bibinfo {author} {\bibfnamefont {X.}~\bibnamefont {Liu}}, \bibinfo {author} {\bibfnamefont {S.}~\bibnamefont {Li}}, \bibinfo {author} {\bibfnamefont {M.}~\bibnamefont {He}}, \bibinfo {author} {\bibfnamefont {D.}~\bibnamefont {Yin}}, \bibinfo {author} {\bibfnamefont {X.}~\bibnamefont {Liang}}, \bibinfo {author} {\bibfnamefont {Y.}~\bibnamefont {Leng}}, \bibinfo {author} {\bibfnamefont {R.}~\bibnamefont {Li}}, \ and\ \bibinfo {author} {\bibfnamefont {Z.}~\bibnamefont {Xu}},\ }\href {\doibase 10.1364/OL.43.005681} {\bibfield  {journal} {\bibinfo  {journal} {Opt. Lett.}\ }\textbf {\bibinfo {volume} {43}},\ \bibinfo {pages} {5681} (\bibinfo {year} {2018})}\BibitemShut {NoStop}%
\bibitem [{\citenamefont {Yu}\ \emph {et~al.}(2018)\citenamefont {Yu}, \citenamefont {Xu}, \citenamefont {Liu}, \citenamefont {Li}, \citenamefont {Li}, \citenamefont {Liu}, \citenamefont {Li}, \citenamefont {Wu}, \citenamefont {Yang}, \citenamefont {Yang}, \citenamefont {Wang}, \citenamefont {Lu}, \citenamefont {Leng}, \citenamefont {Li},\ and\ \citenamefont {Xu}}]{Yu:18}%
  \BibitemOpen
  \bibfield  {author} {\bibinfo {author} {\bibfnamefont {L.}~\bibnamefont {Yu}}, \bibinfo {author} {\bibfnamefont {Y.}~\bibnamefont {Xu}}, \bibinfo {author} {\bibfnamefont {Y.}~\bibnamefont {Liu}}, \bibinfo {author} {\bibfnamefont {Y.}~\bibnamefont {Li}}, \bibinfo {author} {\bibfnamefont {S.}~\bibnamefont {Li}}, \bibinfo {author} {\bibfnamefont {Z.}~\bibnamefont {Liu}}, \bibinfo {author} {\bibfnamefont {W.}~\bibnamefont {Li}}, \bibinfo {author} {\bibfnamefont {F.}~\bibnamefont {Wu}}, \bibinfo {author} {\bibfnamefont {X.}~\bibnamefont {Yang}}, \bibinfo {author} {\bibfnamefont {Y.}~\bibnamefont {Yang}}, \bibinfo {author} {\bibfnamefont {C.}~\bibnamefont {Wang}}, \bibinfo {author} {\bibfnamefont {X.}~\bibnamefont {Lu}}, \bibinfo {author} {\bibfnamefont {Y.}~\bibnamefont {Leng}}, \bibinfo {author} {\bibfnamefont {R.}~\bibnamefont {Li}}, \ and\ \bibinfo {author} {\bibfnamefont {Z.}~\bibnamefont {Xu}},\ }\href {\doibase 10.1364/OE.26.002625} {\bibfield  {journal} {\bibinfo  {journal} {Opt. Express}\ }\textbf
  {\bibinfo {volume} {26}},\ \bibinfo {pages} {2625} (\bibinfo {year} {2018})}\BibitemShut {NoStop}%
\bibitem [{\citenamefont {Ur}\ \emph {et~al.}(2015)\citenamefont {Ur}, \citenamefont {Balabanski}, \citenamefont {Cata-Danil}, \citenamefont {Gales}, \citenamefont {Morjan}, \citenamefont {Tesileanu}, \citenamefont {Ursescu}, \citenamefont {Ursu},\ and\ \citenamefont {Zamfir}}]{ur2015eli}%
  \BibitemOpen
  \bibfield  {author} {\bibinfo {author} {\bibfnamefont {C.}~\bibnamefont {Ur}}, \bibinfo {author} {\bibfnamefont {D.}~\bibnamefont {Balabanski}}, \bibinfo {author} {\bibfnamefont {G.}~\bibnamefont {Cata-Danil}}, \bibinfo {author} {\bibfnamefont {S.}~\bibnamefont {Gales}}, \bibinfo {author} {\bibfnamefont {I.}~\bibnamefont {Morjan}}, \bibinfo {author} {\bibfnamefont {O.}~\bibnamefont {Tesileanu}}, \bibinfo {author} {\bibfnamefont {D.}~\bibnamefont {Ursescu}}, \bibinfo {author} {\bibfnamefont {I.}~\bibnamefont {Ursu}}, \ and\ \bibinfo {author} {\bibfnamefont {N.}~\bibnamefont {Zamfir}},\ }\href@noop {} {\bibfield  {journal} {\bibinfo  {journal} {Nuclear Instruments and Methods in Physics Research Section B: Beam Interactions with Materials and Atoms}\ }\textbf {\bibinfo {volume} {355}},\ \bibinfo {pages} {198} (\bibinfo {year} {2015})}\BibitemShut {NoStop}%
\bibitem [{\citenamefont {Borghesi}\ \emph {et~al.}(2006)\citenamefont {Borghesi}, \citenamefont {Fuchs}, \citenamefont {Bulanov}, \citenamefont {MacKinnon}, \citenamefont {Patel},\ and\ \citenamefont {Roth}}]{Borghesi2006}%
  \BibitemOpen
  \bibfield  {author} {\bibinfo {author} {\bibfnamefont {M.}~\bibnamefont {Borghesi}}, \bibinfo {author} {\bibfnamefont {J.}~\bibnamefont {Fuchs}}, \bibinfo {author} {\bibfnamefont {S.~V.}\ \bibnamefont {Bulanov}}, \bibinfo {author} {\bibfnamefont {A.~J.}\ \bibnamefont {MacKinnon}}, \bibinfo {author} {\bibfnamefont {P.~K.}\ \bibnamefont {Patel}}, \ and\ \bibinfo {author} {\bibfnamefont {M.}~\bibnamefont {Roth}},\ }\href {\doibase 10.13182/fst06-a1159} {\bibfield  {journal} {\bibinfo  {journal} {Fusion Science and Technology}\ }\textbf {\bibinfo {volume} {49}},\ \bibinfo {pages} {412} (\bibinfo {year} {2006})}\BibitemShut {NoStop}%
\bibitem [{\citenamefont {Roth}\ \emph {et~al.}(2002)\citenamefont {Roth}, \citenamefont {Blazevic}, \citenamefont {Geissel}, \citenamefont {Schlegel}, \citenamefont {Cowan}, \citenamefont {Allen}, \citenamefont {Gauthier}, \citenamefont {Audebert}, \citenamefont {Fuchs}, \citenamefont {Meyer-ter Vehn} \emph {et~al.}}]{roth2002energetic}%
  \BibitemOpen
  \bibfield  {author} {\bibinfo {author} {\bibfnamefont {M.}~\bibnamefont {Roth}}, \bibinfo {author} {\bibfnamefont {A.}~\bibnamefont {Blazevic}}, \bibinfo {author} {\bibfnamefont {M.}~\bibnamefont {Geissel}}, \bibinfo {author} {\bibfnamefont {T.}~\bibnamefont {Schlegel}}, \bibinfo {author} {\bibfnamefont {T.}~\bibnamefont {Cowan}}, \bibinfo {author} {\bibfnamefont {M.}~\bibnamefont {Allen}}, \bibinfo {author} {\bibfnamefont {J.-C.}\ \bibnamefont {Gauthier}}, \bibinfo {author} {\bibfnamefont {P.}~\bibnamefont {Audebert}}, \bibinfo {author} {\bibfnamefont {J.}~\bibnamefont {Fuchs}}, \bibinfo {author} {\bibfnamefont {J.}~\bibnamefont {Meyer-ter Vehn}},  \emph {et~al.},\ }\href@noop {} {\bibfield  {journal} {\bibinfo  {journal} {Physical Review Special Topics-Accelerators and Beams}\ }\textbf {\bibinfo {volume} {5}},\ \bibinfo {pages} {061301} (\bibinfo {year} {2002})}\BibitemShut {NoStop}%
\bibitem [{\citenamefont {Logan}\ \emph {et~al.}(2006)\citenamefont {Logan}, \citenamefont {Bangerter}, \citenamefont {Callahan}, \citenamefont {Tabak}, \citenamefont {Roth}, \citenamefont {Perkins},\ and\ \citenamefont {Caporaso}}]{logan2006assessment}%
  \BibitemOpen
  \bibfield  {author} {\bibinfo {author} {\bibfnamefont {B.~G.}\ \bibnamefont {Logan}}, \bibinfo {author} {\bibfnamefont {R.~O.}\ \bibnamefont {Bangerter}}, \bibinfo {author} {\bibfnamefont {D.~A.}\ \bibnamefont {Callahan}}, \bibinfo {author} {\bibfnamefont {M.}~\bibnamefont {Tabak}}, \bibinfo {author} {\bibfnamefont {M.}~\bibnamefont {Roth}}, \bibinfo {author} {\bibfnamefont {L.~J.}\ \bibnamefont {Perkins}}, \ and\ \bibinfo {author} {\bibfnamefont {G.}~\bibnamefont {Caporaso}},\ }\href@noop {} {\bibfield  {journal} {\bibinfo  {journal} {Fusion science and technology}\ }\textbf {\bibinfo {volume} {49}},\ \bibinfo {pages} {399} (\bibinfo {year} {2006})}\BibitemShut {NoStop}%
\bibitem [{\citenamefont {Roth}(2008)}]{Roth2008}%
  \BibitemOpen
  \bibfield  {author} {\bibinfo {author} {\bibfnamefont {M.}~\bibnamefont {Roth}},\ }\href {\doibase 10.1088/0741-3335/51/1/014004} {\bibfield  {journal} {\bibinfo  {journal} {Plasma Physics and Controlled Fusion}\ }\textbf {\bibinfo {volume} {51}},\ \bibinfo {pages} {014004} (\bibinfo {year} {2008})}\BibitemShut {NoStop}%
\bibitem [{\citenamefont {Karamian}\ and\ \citenamefont {Carroll}(2011)}]{PhysRevC.83.024604}%
  \BibitemOpen
  \bibfield  {author} {\bibinfo {author} {\bibfnamefont {S.~A.}\ \bibnamefont {Karamian}}\ and\ \bibinfo {author} {\bibfnamefont {J.~J.}\ \bibnamefont {Carroll}},\ }\href {\doibase 10.1103/PhysRevC.83.024604} {\bibfield  {journal} {\bibinfo  {journal} {Phys. Rev. C}\ }\textbf {\bibinfo {volume} {83}},\ \bibinfo {pages} {024604} (\bibinfo {year} {2011})}\BibitemShut {NoStop}%
\bibitem [{\citenamefont {Sharpe}\ and\ \citenamefont {Schittm}(1959)}]{sharp}%
  \BibitemOpen
  \bibfield  {author} {\bibinfo {author} {\bibfnamefont {B.~S.}\ \bibnamefont {Sharpe}}\ and\ \bibinfo {author} {\bibfnamefont {R.~A.}\ \bibnamefont {Schittm}},\ }\href@noop {} {\bibfield  {journal} {\bibinfo  {journal} {Internal Report No. GA910, General Atomics Corp. , San Diego}\ } (\bibinfo {year} {1959})}\BibitemShut {NoStop}%
\bibitem [{\citenamefont {Vali}\ and\ \citenamefont {Vali}(1963)}]{1443607}%
  \BibitemOpen
  \bibfield  {author} {\bibinfo {author} {\bibfnamefont {V.}~\bibnamefont {Vali}}\ and\ \bibinfo {author} {\bibfnamefont {W.}~\bibnamefont {Vali}},\ }\href {\doibase 10.1109/PROC.1963.1677} {\bibfield  {journal} {\bibinfo  {journal} {Proceedings of the IEEE}\ }\textbf {\bibinfo {volume} {51}},\ \bibinfo {pages} {182} (\bibinfo {year} {1963})}\BibitemShut {NoStop}%
\bibitem [{\citenamefont {Billings}\ \emph {et~al.}(1953)\citenamefont {Billings}, \citenamefont {Hitchcock},\ and\ \citenamefont {Zelikoff}}]{billings1953photochemical}%
  \BibitemOpen
  \bibfield  {author} {\bibinfo {author} {\bibfnamefont {B.~H.}\ \bibnamefont {Billings}}, \bibinfo {author} {\bibfnamefont {W.}~\bibnamefont {Hitchcock}}, \ and\ \bibinfo {author} {\bibfnamefont {M.}~\bibnamefont {Zelikoff}},\ }\href@noop {} {\bibfield  {journal} {\bibinfo  {journal} {The Journal of Chemical Physics}\ }\textbf {\bibinfo {volume} {21}},\ \bibinfo {pages} {1762} (\bibinfo {year} {1953})}\BibitemShut {NoStop}%
\bibitem [{\citenamefont {Gunning}\ and\ \citenamefont {Strausz}(1963)}]{Gunning1963}%
  \BibitemOpen
  \bibfield  {author} {\bibinfo {author} {\bibfnamefont {H.~E.}\ \bibnamefont {Gunning}}\ and\ \bibinfo {author} {\bibfnamefont {O.~P.}\ \bibnamefont {Strausz}},\ }\href {\doibase 10.1002/9780470133316.ch7} {\enquote {\bibinfo {title} {Isotopic effects and the mechanism of energy transfer in mercury photosensitization},}\ } (\bibinfo {year} {1963})\BibitemShut {NoStop}%
\bibitem [{\citenamefont {Liuti}\ \emph {et~al.}(1966)\citenamefont {Liuti}, \citenamefont {Dondes},\ and\ \citenamefont {Harteck}}]{liuti1966isotopic}%
  \BibitemOpen
  \bibfield  {author} {\bibinfo {author} {\bibfnamefont {G.}~\bibnamefont {Liuti}}, \bibinfo {author} {\bibfnamefont {S.}~\bibnamefont {Dondes}}, \ and\ \bibinfo {author} {\bibfnamefont {P.}~\bibnamefont {Harteck}},\ }\href@noop {} {\bibfield  {journal} {\bibinfo  {journal} {The Journal of Chemical Physics}\ }\textbf {\bibinfo {volume} {44}},\ \bibinfo {pages} {4052} (\bibinfo {year} {1966})}\BibitemShut {NoStop}%
\bibitem [{\citenamefont {Tsoneva}\ \emph {et~al.}(2000)\citenamefont {Tsoneva}, \citenamefont {Stoyanov}, \citenamefont {Gangrsky}, \citenamefont {Ponomarev}, \citenamefont {Balabanov},\ and\ \citenamefont {Tonchev}}]{PhysRevC.61.044303}%
  \BibitemOpen
  \bibfield  {author} {\bibinfo {author} {\bibfnamefont {N.}~\bibnamefont {Tsoneva}}, \bibinfo {author} {\bibfnamefont {C.}~\bibnamefont {Stoyanov}}, \bibinfo {author} {\bibfnamefont {Y.~P.}\ \bibnamefont {Gangrsky}}, \bibinfo {author} {\bibfnamefont {V.~Y.}\ \bibnamefont {Ponomarev}}, \bibinfo {author} {\bibfnamefont {N.~P.}\ \bibnamefont {Balabanov}}, \ and\ \bibinfo {author} {\bibfnamefont {A.~P.}\ \bibnamefont {Tonchev}},\ }\href {\doibase 10.1103/PhysRevC.61.044303} {\bibfield  {journal} {\bibinfo  {journal} {Phys. Rev. C}\ }\textbf {\bibinfo {volume} {61}},\ \bibinfo {pages} {044303} (\bibinfo {year} {2000})}\BibitemShut {NoStop}%
\bibitem [{\citenamefont {Spohr}\ \emph {et~al.}(2006)\citenamefont {Spohr}, \citenamefont {Chapman}, \citenamefont {Hanvey}, \citenamefont {Ledingham}, \citenamefont {Mccanny}, \citenamefont {Mckenna}, \citenamefont {Robson},\ and\ \citenamefont {Shaw}}]{Spohr2006}%
  \BibitemOpen
  \bibfield  {author} {\bibinfo {author} {\bibfnamefont {K.}~\bibnamefont {Spohr}}, \bibinfo {author} {\bibfnamefont {R.}~\bibnamefont {Chapman}}, \bibinfo {author} {\bibfnamefont {S.}~\bibnamefont {Hanvey}}, \bibinfo {author} {\bibfnamefont {K.}~\bibnamefont {Ledingham}}, \bibinfo {author} {\bibfnamefont {T.}~\bibnamefont {Mccanny}}, \bibinfo {author} {\bibfnamefont {P.}~\bibnamefont {Mckenna}}, \bibinfo {author} {\bibfnamefont {L.}~\bibnamefont {Robson}}, \ and\ \bibinfo {author} {\bibfnamefont {M.}~\bibnamefont {Shaw}},\ }\href {\doibase 10.1080/09500340600895631} {\bibfield  {journal} {\bibinfo  {journal} {Journal of Modern Optics}\ }\textbf {\bibinfo {volume} {53}},\ \bibinfo {pages} {2633} (\bibinfo {year} {2006})}\BibitemShut {NoStop}%
\bibitem [{\citenamefont {Kbokhlov}\ and\ \citenamefont {Il'inskii}(1973)}]{kho1973}%
  \BibitemOpen
  \bibfield  {author} {\bibinfo {author} {\bibfnamefont {B.~V.}\ \bibnamefont {Kbokhlov}}\ and\ \bibinfo {author} {\bibfnamefont {Y.~A.}\ \bibnamefont {Il'inskii}},\ }\href@noop {} {\bibfield  {journal} {\bibinfo  {journal} {Nonlinear Processes in Optics, edited by R. E. Soloukhin and G. V. Krivoschokov (USSR Acad. Sci., Siberian Branch, Novosibir sk)}\ } (\bibinfo {year} {1973})}\BibitemShut {NoStop}%
\bibitem [{\citenamefont {Il'inskii}\ and\ \citenamefont {Khokhlov}(1974{\natexlab{a}})}]{ill1974a}%
  \BibitemOpen
  \bibfield  {author} {\bibinfo {author} {\bibfnamefont {Y.~A.}\ \bibnamefont {Il'inskii}}\ and\ \bibinfo {author} {\bibfnamefont {R.~V.}\ \bibnamefont {Khokhlov}},\ }\href@noop {} {\bibfield  {journal} {\bibinfo  {journal} {Sov. Phys. JETP}\ }\textbf {\bibinfo {volume} {38}},\ \bibinfo {pages} {809} (\bibinfo {year} {1974}{\natexlab{a}})}\BibitemShut {NoStop}%
\bibitem [{\citenamefont {Il'inskii}\ and\ \citenamefont {Khokhlov}(1974{\natexlab{b}})}]{ill1974b}%
  \BibitemOpen
  \bibfield  {author} {\bibinfo {author} {\bibfnamefont {Y.~A.}\ \bibnamefont {Il'inskii}}\ and\ \bibinfo {author} {\bibfnamefont {R.~V.}\ \bibnamefont {Khokhlov}},\ }\href@noop {} {\bibfield  {journal} {\bibinfo  {journal} {Sov. Phys. Usp.}\ }\textbf {\bibinfo {volume} {16}},\ \bibinfo {pages} {565} (\bibinfo {year} {1974}{\natexlab{b}})}\BibitemShut {NoStop}%
\bibitem [{\citenamefont {Gol'danskii}\ \emph {et~al.}(1974{\natexlab{a}})\citenamefont {Gol'danskii}, \citenamefont {Karyagin},\ and\ \citenamefont {Namiot}}]{go1974a}%
  \BibitemOpen
  \bibfield  {author} {\bibinfo {author} {\bibfnamefont {U.~I.}\ \bibnamefont {Gol'danskii}}, \bibinfo {author} {\bibfnamefont {S.~V.}\ \bibnamefont {Karyagin}}, \ and\ \bibinfo {author} {\bibfnamefont {V.~A.}\ \bibnamefont {Namiot}},\ }\href@noop {} {\bibfield  {journal} {\bibinfo  {journal} {Zh. Eksp. Teor. Fiz. Pis'ma Red}\ }\textbf {\bibinfo {volume} {19}},\ \bibinfo {pages} {625} (\bibinfo {year} {1974}{\natexlab{a}})}\BibitemShut {NoStop}%
\bibitem [{\citenamefont {Gol'danskii}\ \emph {et~al.}(1974{\natexlab{b}})\citenamefont {Gol'danskii}, \citenamefont {Karyagin},\ and\ \citenamefont {Namiot}}]{go1974b}%
  \BibitemOpen
  \bibfield  {author} {\bibinfo {author} {\bibfnamefont {U.~I.}\ \bibnamefont {Gol'danskii}}, \bibinfo {author} {\bibfnamefont {S.~V.}\ \bibnamefont {Karyagin}}, \ and\ \bibinfo {author} {\bibfnamefont {V.~A.}\ \bibnamefont {Namiot}},\ }\href@noop {} {\bibfield  {journal} {\bibinfo  {journal} {J. Phys. (Paris)}\ }\textbf {\bibinfo {volume} {35}},\ \bibinfo {pages} {192} (\bibinfo {year} {1974}{\natexlab{b}})}\BibitemShut {NoStop}%
\bibitem [{\citenamefont {Namiot}(1973)}]{na1973}%
  \BibitemOpen
  \bibfield  {author} {\bibinfo {author} {\bibfnamefont {V.~A.}\ \bibnamefont {Namiot}},\ }\href@noop {} {\bibfield  {journal} {\bibinfo  {journal} {Zh. Eksp. Teor. Fiz. Pis'Ina Red.}\ }\textbf {\bibinfo {volume} {18}},\ \bibinfo {pages} {369} (\bibinfo {year} {1973})}\BibitemShut {NoStop}%
\bibitem [{\citenamefont {Kagan}(1973)}]{kagan1975}%
  \BibitemOpen
  \bibfield  {author} {\bibinfo {author} {\bibfnamefont {Y.~M.}\ \bibnamefont {Kagan}},\ }\href@noop {} {\bibfield  {journal} {\bibinfo  {journal} {Proceedings of the Intenzational Conference on Mossbauer Spectroscopy, Cracozo, Poland, 1975, edited by A. Hrynkiewicz and J. Sawicka, (Acad. GorniczoHufnicza, Cracow)}\ }\textbf {\bibinfo {volume} {2}},\ \bibinfo {pages} {17} (\bibinfo {year} {1973})}\BibitemShut {NoStop}%
\bibitem [{\citenamefont {Karyagin}(1976)}]{kar1976}%
  \BibitemOpen
  \bibfield  {author} {\bibinfo {author} {\bibfnamefont {S.~V.}\ \bibnamefont {Karyagin}},\ }\href@noop {} {\bibfield  {journal} {\bibinfo  {journal} {Sov. Phys. —Tech. Phys. Lett.}\ }\textbf {\bibinfo {volume} {2}},\ \bibinfo {pages} {196} (\bibinfo {year} {1976})}\BibitemShut {NoStop}%
\bibitem [{\citenamefont {Kagan}(1977)}]{kar1977}%
  \BibitemOpen
  \bibfield  {author} {\bibinfo {author} {\bibfnamefont {Y.~M.}\ \bibnamefont {Kagan}},\ }\href@noop {} {\bibfield  {journal} {\bibinfo  {journal} {Proceedings of the International Conference on Mossbaue~ Spectmscopy, Bucharest, Roman- , edited by D. Barb and D. Tarina (Central Institute of Physics, Bucharest)}\ }\textbf {\bibinfo {volume} {2}},\ \bibinfo {pages} {1} (\bibinfo {year} {1977})}\BibitemShut {NoStop}%
\bibitem [{\citenamefont {Lesko}\ \emph {et~al.}(2021)\citenamefont {Lesko}, \citenamefont {Timmers}, \citenamefont {Xing}, \citenamefont {Kowligy}, \citenamefont {Lind},\ and\ \citenamefont {Diddams}}]{Lesko2021}%
  \BibitemOpen
  \bibfield  {author} {\bibinfo {author} {\bibfnamefont {D.~M.~B.}\ \bibnamefont {Lesko}}, \bibinfo {author} {\bibfnamefont {H.}~\bibnamefont {Timmers}}, \bibinfo {author} {\bibfnamefont {S.}~\bibnamefont {Xing}}, \bibinfo {author} {\bibfnamefont {A.}~\bibnamefont {Kowligy}}, \bibinfo {author} {\bibfnamefont {A.~J.}\ \bibnamefont {Lind}}, \ and\ \bibinfo {author} {\bibfnamefont {S.~A.}\ \bibnamefont {Diddams}},\ }\href {\doibase 10.1038/s41566-021-00778-y} {\bibfield  {journal} {\bibinfo  {journal} {Nature Photonics}\ }\textbf {\bibinfo {volume} {15}},\ \bibinfo {pages} {281} (\bibinfo {year} {2021})}\BibitemShut {NoStop}%
\bibitem [{\citenamefont {Ozawa}\ and\ \citenamefont {Udem}(2021)}]{Ozawa2021}%
  \BibitemOpen
  \bibfield  {author} {\bibinfo {author} {\bibfnamefont {A.}~\bibnamefont {Ozawa}}\ and\ \bibinfo {author} {\bibfnamefont {T.}~\bibnamefont {Udem}},\ }\href {\doibase 10.1038/s41566-021-00788-w} {\bibfield  {journal} {\bibinfo  {journal} {Nature Photonics}\ }\textbf {\bibinfo {volume} {15}},\ \bibinfo {pages} {247} (\bibinfo {year} {2021})}\BibitemShut {NoStop}%
\bibitem [{\citenamefont {Baldwin}(1974)}]{Baldwin1974}%
  \BibitemOpen
  \bibfield  {author} {\bibinfo {author} {\bibfnamefont {G.~C.}\ \bibnamefont {Baldwin}},\ }\href {\doibase 10.1007/978-1-4684-8416-8_23} {\bibfield  {journal} {\bibinfo  {journal} {Laser Interaction and Related Plasma Phenomena}\ ,\ \bibinfo {pages} {875}} (\bibinfo {year} {1974})}\BibitemShut {NoStop}%
\bibitem [{\citenamefont {Allen}\ and\ \citenamefont {Peters}(1971)}]{allen1971amplified}%
  \BibitemOpen
  \bibfield  {author} {\bibinfo {author} {\bibfnamefont {L.}~\bibnamefont {Allen}}\ and\ \bibinfo {author} {\bibfnamefont {G.}~\bibnamefont {Peters}},\ }\href@noop {} {\bibfield  {journal} {\bibinfo  {journal} {Journal of Physics A: General Physics}\ }\textbf {\bibinfo {volume} {4}},\ \bibinfo {pages} {564} (\bibinfo {year} {1971})}\BibitemShut {NoStop}%
\bibitem [{\citenamefont {M{\"o}{\ss}bauer}(1958)}]{Mssbauer1958}%
  \BibitemOpen
  \bibfield  {author} {\bibinfo {author} {\bibfnamefont {R.~L.}\ \bibnamefont {M{\"o}{\ss}bauer}},\ }\href {\doibase 10.1007/bf01344210} {\bibfield  {journal} {\bibinfo  {journal} {Zeitschrift fur Physik}\ }\textbf {\bibinfo {volume} {151}},\ \bibinfo {pages} {124} (\bibinfo {year} {1958})}\BibitemShut {NoStop}%
\bibitem [{\citenamefont {Krasnov}\ and\ \citenamefont {Shaparev}(1975)}]{Krasnov1975}%
  \BibitemOpen
  \bibfield  {author} {\bibinfo {author} {\bibfnamefont {I.~V.}\ \bibnamefont {Krasnov}}\ and\ \bibinfo {author} {\bibfnamefont {N.~Y.}\ \bibnamefont {Shaparev}},\ }\href {\doibase 10.1070/qe1975v005n12abeh012158} {\bibfield  {journal} {\bibinfo  {journal} {Soviet Journal of Quantum Electronics}\ }\textbf {\bibinfo {volume} {5}},\ \bibinfo {pages} {1420} (\bibinfo {year} {1975})}\BibitemShut {NoStop}%
\bibitem [{\citenamefont {Szilard}\ and\ \citenamefont {Chalmers}(1934)}]{szilard1934chemical}%
  \BibitemOpen
  \bibfield  {author} {\bibinfo {author} {\bibfnamefont {L.}~\bibnamefont {Szilard}}\ and\ \bibinfo {author} {\bibfnamefont {T.}~\bibnamefont {Chalmers}},\ }\href@noop {} {\bibfield  {journal} {\bibinfo  {journal} {Nature}\ }\textbf {\bibinfo {volume} {134}},\ \bibinfo {pages} {462} (\bibinfo {year} {1934})}\BibitemShut {NoStop}%
\bibitem [{\citenamefont {Murnick}\ and\ \citenamefont {Feld}(1979)}]{murnick1979applications}%
  \BibitemOpen
  \bibfield  {author} {\bibinfo {author} {\bibfnamefont {D.~E.}\ \bibnamefont {Murnick}}\ and\ \bibinfo {author} {\bibfnamefont {M.~S.}\ \bibnamefont {Feld}},\ }\href@noop {} {\bibfield  {journal} {\bibinfo  {journal} {Annual Review of Nuclear and Particle Science}\ }\textbf {\bibinfo {volume} {29}},\ \bibinfo {pages} {411} (\bibinfo {year} {1979})}\BibitemShut {NoStop}%
\bibitem [{\citenamefont {Jacquinot}\ and\ \citenamefont {Klapisch}(1979)}]{jacquinot1979hyperfine}%
  \BibitemOpen
  \bibfield  {author} {\bibinfo {author} {\bibfnamefont {P.}~\bibnamefont {Jacquinot}}\ and\ \bibinfo {author} {\bibfnamefont {R.}~\bibnamefont {Klapisch}},\ }\href@noop {} {\bibfield  {journal} {\bibinfo  {journal} {Reports on Progress in Physics}\ }\textbf {\bibinfo {volume} {42}},\ \bibinfo {pages} {773} (\bibinfo {year} {1979})}\BibitemShut {NoStop}%
\bibitem [{\citenamefont {Fedorov}\ and\ \citenamefont {Tzortzakis}(2020)}]{Fedorov2020}%
  \BibitemOpen
  \bibfield  {author} {\bibinfo {author} {\bibfnamefont {V.~Y.}\ \bibnamefont {Fedorov}}\ and\ \bibinfo {author} {\bibfnamefont {S.}~\bibnamefont {Tzortzakis}},\ }\href {\doibase 10.1038/s41377-020-00423-3} {\bibfield  {journal} {\bibinfo  {journal} {Light: Science \& Applications}\ }\textbf {\bibinfo {volume} {9}} (\bibinfo {year} {2020}),\ 10.1038/s41377-020-00423-3}\BibitemShut {NoStop}%
\bibitem [{\citenamefont {Yao}\ \emph {et~al.}(2023)\citenamefont {Yao}, \citenamefont {Gui}, \citenamefont {Rao}, \citenamefont {Zhang}, \citenamefont {Lu},\ and\ \citenamefont {Hu}}]{PhysRevLett.130.146702}%
  \BibitemOpen
  \bibfield  {author} {\bibinfo {author} {\bibfnamefont {B.}~\bibnamefont {Yao}}, \bibinfo {author} {\bibfnamefont {Y.~S.}\ \bibnamefont {Gui}}, \bibinfo {author} {\bibfnamefont {J.~W.}\ \bibnamefont {Rao}}, \bibinfo {author} {\bibfnamefont {Y.~H.}\ \bibnamefont {Zhang}}, \bibinfo {author} {\bibfnamefont {W.}~\bibnamefont {Lu}}, \ and\ \bibinfo {author} {\bibfnamefont {C.-M.}\ \bibnamefont {Hu}},\ }\href {\doibase 10.1103/PhysRevLett.130.146702} {\bibfield  {journal} {\bibinfo  {journal} {Phys. Rev. Lett.}\ }\textbf {\bibinfo {volume} {130}},\ \bibinfo {pages} {146702} (\bibinfo {year} {2023})}\BibitemShut {NoStop}%
\bibitem [{\citenamefont {Malaca}\ \emph {et~al.}(2023)\citenamefont {Malaca}, \citenamefont {Pardal}, \citenamefont {Ramsey}, \citenamefont {Pierce}, \citenamefont {Weichman}, \citenamefont {Andriyash}, \citenamefont {Mori}, \citenamefont {Palastro}, \citenamefont {Fonseca},\ and\ \citenamefont {Vieira}}]{Malaca2023}%
  \BibitemOpen
  \bibfield  {author} {\bibinfo {author} {\bibfnamefont {B.}~\bibnamefont {Malaca}}, \bibinfo {author} {\bibfnamefont {M.}~\bibnamefont {Pardal}}, \bibinfo {author} {\bibfnamefont {D.}~\bibnamefont {Ramsey}}, \bibinfo {author} {\bibfnamefont {J.~R.}\ \bibnamefont {Pierce}}, \bibinfo {author} {\bibfnamefont {K.}~\bibnamefont {Weichman}}, \bibinfo {author} {\bibfnamefont {I.~A.}\ \bibnamefont {Andriyash}}, \bibinfo {author} {\bibfnamefont {W.~B.}\ \bibnamefont {Mori}}, \bibinfo {author} {\bibfnamefont {J.~P.}\ \bibnamefont {Palastro}}, \bibinfo {author} {\bibfnamefont {R.~A.}\ \bibnamefont {Fonseca}}, \ and\ \bibinfo {author} {\bibfnamefont {J.}~\bibnamefont {Vieira}},\ }\href {\doibase 10.1038/s41566-023-01311-z} {\bibfield  {journal} {\bibinfo  {journal} {Nature Photonics}\ } (\bibinfo {year} {2023}),\ 10.1038/s41566-023-01311-z}\BibitemShut {NoStop}%
\end{thebibliography}%
\bibliographystyle{apsrev4-1}

\end{document}